\documentclass{ucetd}

\usepackage[T1]{fontenc}
\usepackage{subcaption,graphicx}
\usepackage[numbers]{natbib}
\usepackage{mathtools}  
\usepackage{amssymb}    
\usepackage{amsthm}
\usepackage{float}
\usepackage{multirow} 
\usepackage{algorithm}
\usepackage{algpseudocode}

\newcommand{\angstrom}{\text{\normalfont\AA}}
\newcommand{\thesistitle}{Leveraging Deep Generative model for computational protein design and optimization}
\newcommand{\thesisauthor}{Boqiao Lai}
\department{Toyota Technological Institute at Chicago}
\degree{Doctor of Philosophy in Computer Science}
\date{AUGUST, 2024}

\title{\thesistitle}
\author{\thesisauthor}

\dedication{To XXX}
\epigraph{\textit{“I have approximate answers and possible beliefs in different degrees of certainty about different things, but I'm not absolutely sure of anything.”}\\ -Richard Feynman}

\usepackage{doi}
\usepackage{xurl}
\hypersetup{bookmarksnumbered,
            unicode,
            linktoc=all,
            pdftitle={\thesistitle},
            pdfauthor={\thesisauthor},
            pdfsubject={},                                
            pdfborder={0 0 0}}
\makeatletter
\let\ORG@hyper@linkstart\hyper@linkstart
\protected\def\hyper@linkstart#1#2{%
  \lowercase{\ORG@hyper@linkstart{#1}{#2}}}
\makeatother

\begin{document}
\maketitle
\makecopyright
\makeepigraph
\tableofcontents
\listoffigures
\listoftables

\acknowledgments
The completion of this thesis marks the end of a challenging yet rewarding journey, one that would not have been possible without the support and guidance of many individuals.

First and foremost, I extend my sincere gratitude to my advisor, Dr. Jinbo Xu. Your insightful guidance, unwavering support, and intellectual rigor have been instrumental in shaping both this work and my growth as a researcher. Your ability to push me beyond my perceived limits while offering patience and encouragement has been truly invaluable.

To my committee members, Dr. Aly Khan and Dr. Avrim Blum: your expertise, constructive feedback, and challenging questions have significantly elevated the quality of this research. I am deeply appreciative of the time and effort you have invested in my academic development.

I owe a debt of gratitude to my collaborators, Matthew McPartlon and Hugh Yeh. Your contributions, innovative ideas, and our spirited discussions have enriched this work immeasurably. The synergy of our collaboration has been a highlight of this research process.

On a personal note, I want to express my heartfelt thanks to my fiancée, Cindy Zhang. Your unwavering support, understanding, and love have been my anchor throughout this journey. Your belief in me, especially during the most challenging times, has been a source of strength and motivation.

To my parents: your unconditional love, countless sacrifices, and constant encouragement have been the foundation upon which all my achievements stand. Your support has allowed me to pursue my dreams, and for that, I am eternally grateful.

I would be remiss not to mention Hibiki and Cooper, whose feline companionship during long hours of research and writing provided a sense of comfort and routine.

Lastly, I want to acknowledge the friendship and support of Gavin Young, Tom Li, Cathy Zhang, and Vijay Pillai. Your camaraderie, encouragement, and ability to provide perspective have been crucial in maintaining balance throughout this academic pursuit.

To everyone mentioned here, and to those whose names may not appear but whose impact has been felt: your collective support, wisdom, and encouragement have not only made this thesis possible but have also made the journey profoundly meaningful. Thank you.

\abstract
Proteins are the fundamental macromolecules that play diverse and crucial roles in all living matter and have tremendous implications in healthcare, manufacturing, and biotechnology. Their functions are largely determined by the sequences of amino acids that compose them and their unique three-dimensional structures when folded. The recent surge in highly accurate computational protein structure prediction tools has equipped scientists with the means to derive preliminary structural insights without the onerous costs of experimental structure determination. These breakthroughs hold profound promise for building robust and efficient \textit{in silico} protein design systems.

While the prospect of designing \textit{de novo} proteins with precise computational accuracy remains a grand challenge in biochemical engineering, conventional assembly-based and rational design methods often grapple with the expansive design space, resulting in suboptimal design success rates. Despite recently emerged deep learning-based models have shown promise in improving the efficiency of the computational protein design process, a significant gap persists between current design paradigms and their experimental realization. This thesis will investigate the potential of deep generative models in refining protein structure and sequence design methods, aiming to develop frameworks capable of crafting novel protein sequences with predetermined structures or specific functionalities. By harnessing extensive protein databases and cutting-edge neural architectures, this research aims to enhance precision and robustness in current protein design paradigms, potentially paving the way for advancements across various scientific fields.

The thesis is structured into three main Sections. The first section provides a comprehensive background on computational protein design, highlighting the current challenges faced by the scientific community. It then introduces the machine learning techniques that have been developed to address these challenges, focusing on those particularly relevant to our research. This section aims to establish the foundation necessary for understanding the novel approaches presented in the subsequent sections. The second section introduces a deep structure generative model capable of producing ensembles of high-quality structural variants. We demonstrate how these structure ensembles can facilitate robust and diverse \textit{de novo} protein design pipelines. This section showcases the power of our approach in expanding the possibilities of computational protein design beyond traditional methods. The third section presents an iterative design paradigm that leverages the models described in Section 2. We illustrate the versatility and effectiveness of this paradigm through its application to a diverse array of protein design and engineering tasks. These applications include unconditional \textit{de novo} structure generation, demonstrating our ability to expand current protein fold spaces. We also explore the design and optimization of a small-molecule activated fluorescent protein system, where we show improved \textit{in vitro} fitness and stability in the designed proteins. Furthermore, we present a motif-grounded protein binder design, transforming the important therapeutic target PD1 into distinct \textit{de novo} scaffolds with potential for enhanced stability. Lastly, we demonstrate structure-based \textit{de novo} CDR design for antibody engineering, utilizing an enhanced structure generative model optimized specifically for antibody design.

Through these applications, we demonstrate the broad utility and significant advancements our approach brings to the field of computational protein design. This thesis aims to contribute to the ongoing efforts to expand the capabilities of protein engineering, potentially opening new avenues for therapeutic development and biotechnological applications.

\mainmatter
\chapter{Background}
\section{Proteins}
\begin{figure}[h]
    \centering
    \includegraphics[width=0.5\linewidth]{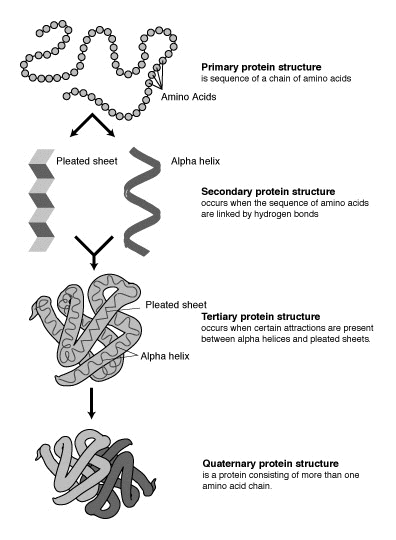}
    \caption{Illustration of the protein structure}
    \label{fig:protein-struct}
\end{figure}

Proteins are macromolecules that play vital roles in most biochemical processes. The central dogma of molecular biology established that proteins are the functional endpoint of the genetic code, made up of amino acids that carry out the biochemical interaction among relevant molecules \cite{crick1970central}. Proteins perform their function mostly when  folded into their three dimensional structures. Figure \ref{fig:protein-struct} showcased the four levels of proteins structures where the primary structure is the linear amino acid sequence in the protein chain; the secondary structure is the repeating local structure stabilized by hydrogen bonds where the two main types are alpha helices and beta sheets; the tertiary structure is the overall three dimension structure of a single protein chain; the quaternary structure is the global arrangement of multiple proteins chains in a multi-unit protein complex.

The function of a protein is ultimately linked to its structure. Specific structural motifs often correspond to particular functions, and changes in structure (due to mutations or environmental factors) can significantly impact a protein's function \cite{petsko2004protein}. While often depicted as static structures, proteins are dynamic molecules. They can undergo conformational changes that are often crucial to their function, such as in allosteric regulation or enzyme-substrate interactions\cite{henzler2007dynamic}. One of the main motivation behind this project is to explore the conformational structure space with deep generative models and how to use these model to improve current protein design paradigms. Understanding protein structure and function is fundamental to many areas of biochemical research, from basic molecular biology to drug design and biotechnology. As our ability to determine and predict protein structures improves, so does our capacity to understand and manipulate biological systems at the molecular level.
\section{Computational Protein Design}
Protein molecules are crucial in most biochemical processes, therefore, designing proteins with desired function and structure is of high interest to a wide range of scientists in the scientific community. Thus, the field presents a unique and significant challenge at the intersection of biochemistry, biophysics, and computer science \cite{huang2016coming}. Thanks to the rapid advances in the deep learning and machine learning field, computational scientists can leverage theses techniques to accelerate the field as a whole. 

Overall, the fundamental goal of protein design is to determine an amino acid sequence that will fold into a desired structure and perform a specific function. For most of the protein sequences, the combinatorial space of all the possible sequence composition is too vast to explore exhaustively. Conventional approaches to protein design and engineering includes methods like rational design and directed evolution\cite{schmid2001industrial,arnold2018directed}. Rational design aims to make specific and targeted changes to a protein by chemist to achieve the desired change in protein function. This approach is often labor intensive and low in success rate. Directed evolution designs protein by mimicking natural evolution with artificial selection pressures and does not require detailed understanding of the structure of the protein of interest. While this is a very powerful tool for protein engineering, there are many technical limitations such as limited sequence space exploration and reliance on \textit{in vitro} selection techniques. 

\begin{figure}[h]
    \centering
    \includegraphics[width=0.7\linewidth]{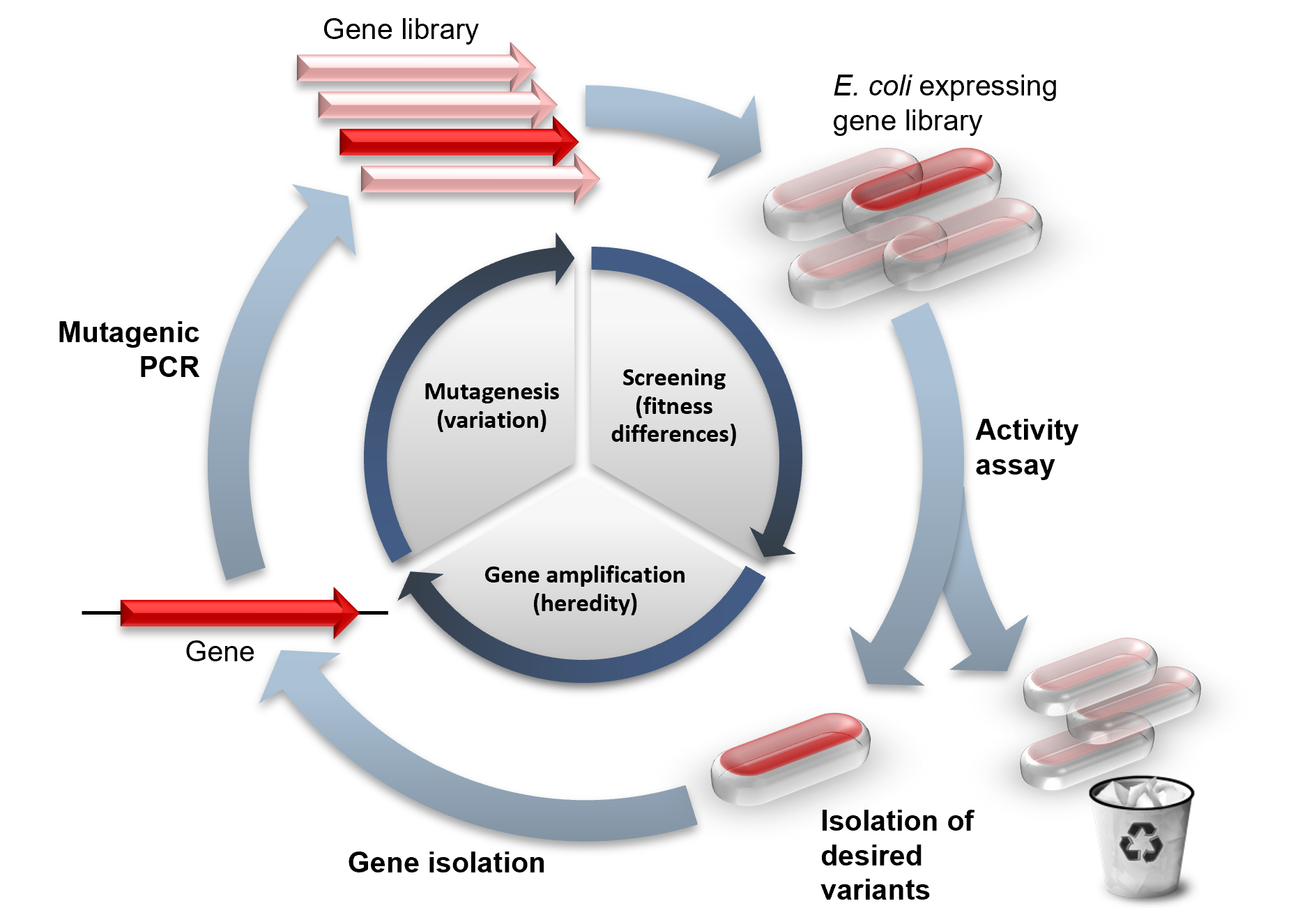}
    \caption{Illustration of in vitro directed evolution adopted from \cite{shafee2014evolvability} CC-BY 4.0}
    \label{fig:DE_cycle}
\end{figure}

Computational methods have become increasingly important in protein design due to their ability to explore vast sequence spaces efficiently. Central to early computational protein design is the use of energy functions to evaluate the stability and potential functionality of designed sequences \cite{alford2017rosetta}. These energy functions typically account for various interactions including van der Waals forces, electrostatic interactions, hydrogen bonding, and solvation effects. However, the folding of a protein is governed by numerous weak interactions, resulting in complex energy landscapes which are still challenging to explore efficiently \textit{in sillico}. To reduce the computational complexity, most methods use discrete side-chain conformations called rotamers. This approach significantly reduces the search space while still capturing the essential features of side-chain packing. Various algorithms are employed to search the sequence space, including Monte Carlo methods, genetic algorithms, dead-end elimination, and integer linear programming. Each of these approaches has its strengths and is suited to different aspects of the design problem\cite{gainza2016algorithms}.

\begin{figure}[H]
    \centering
    \includegraphics[width=1\linewidth]{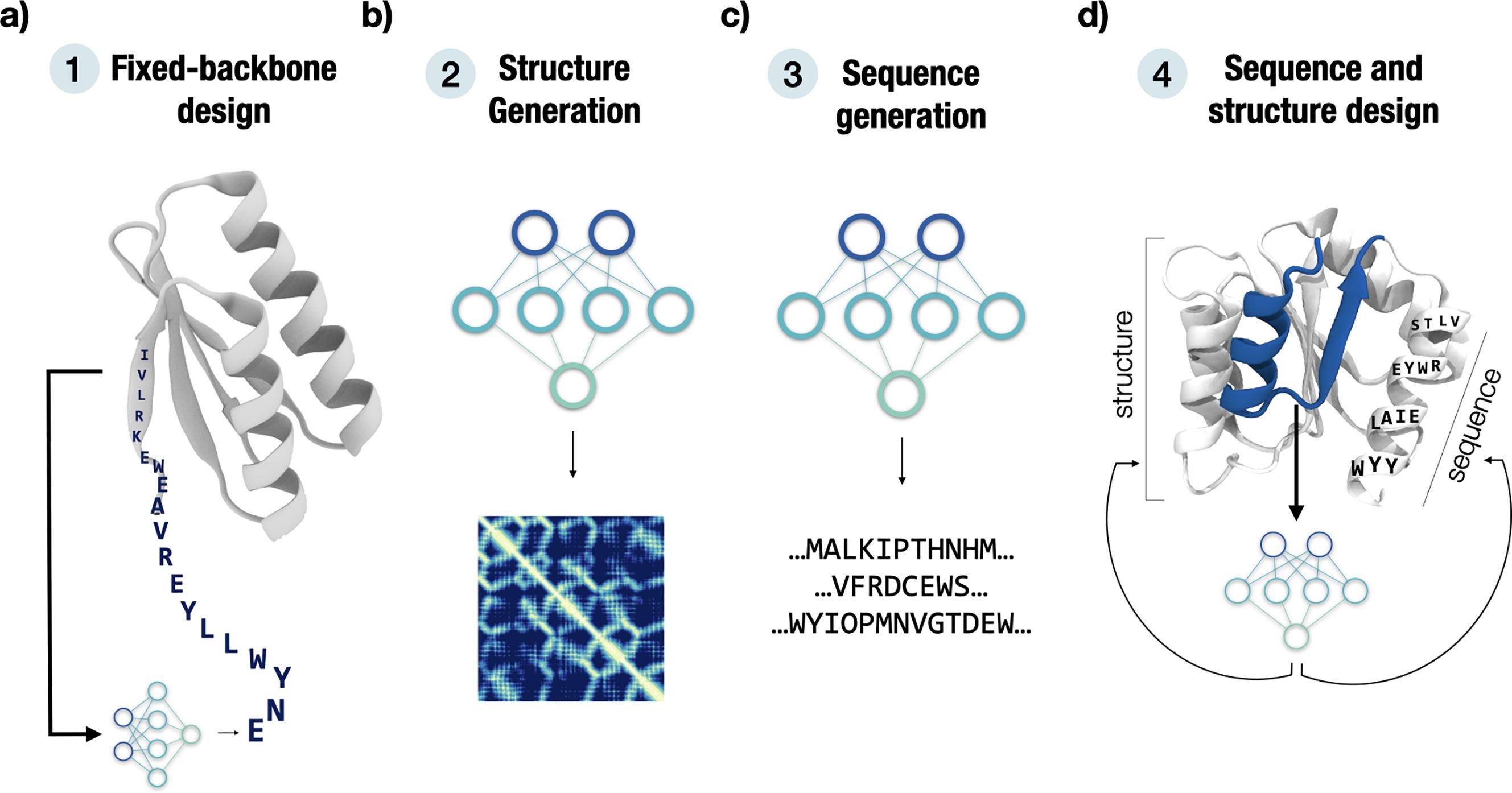}
    \caption{Illustration of different type of protein design tasks adopted from \cite{Ferruz2022.08.31.505981} CC-BY 4.0. a) Fixed backbone sequence design. b) Structure generation via constraints. c) Direct sequence generation from sequence-only models. d) Sequence-Structure co-design models.}
    \label{CPD_summary}
\end{figure}

The field of computational protein design has seen significant progress in recent years. For fixed backbone sequence design which seeks to recover the amino acid sequences that conform to the given backbone, methods such as 3D-CNN and geometric graph neural networks has shown promising successes \cite{anand2022protein,qi2020densecpd,jing2020learning,mcpartlon2022deep, dauparas2022robust,hsu2022learning} by leveraging the graphical and three dimensional representations of the backbone structures. For structure generation, methods such as the Generative adversarial network(GANs) and diffusion models are widely used to recover the either the two dimensional distance or orientation map \cite{anand2019fully,eguchi2022ig,wu2024protein}. More recent diffusion based methods can also generate the backbone structure directly \cite{trippe2022diffusion,watson2023novo}. 

\begin{figure}[H]
    \centering
    \includegraphics[width=0.99\linewidth]{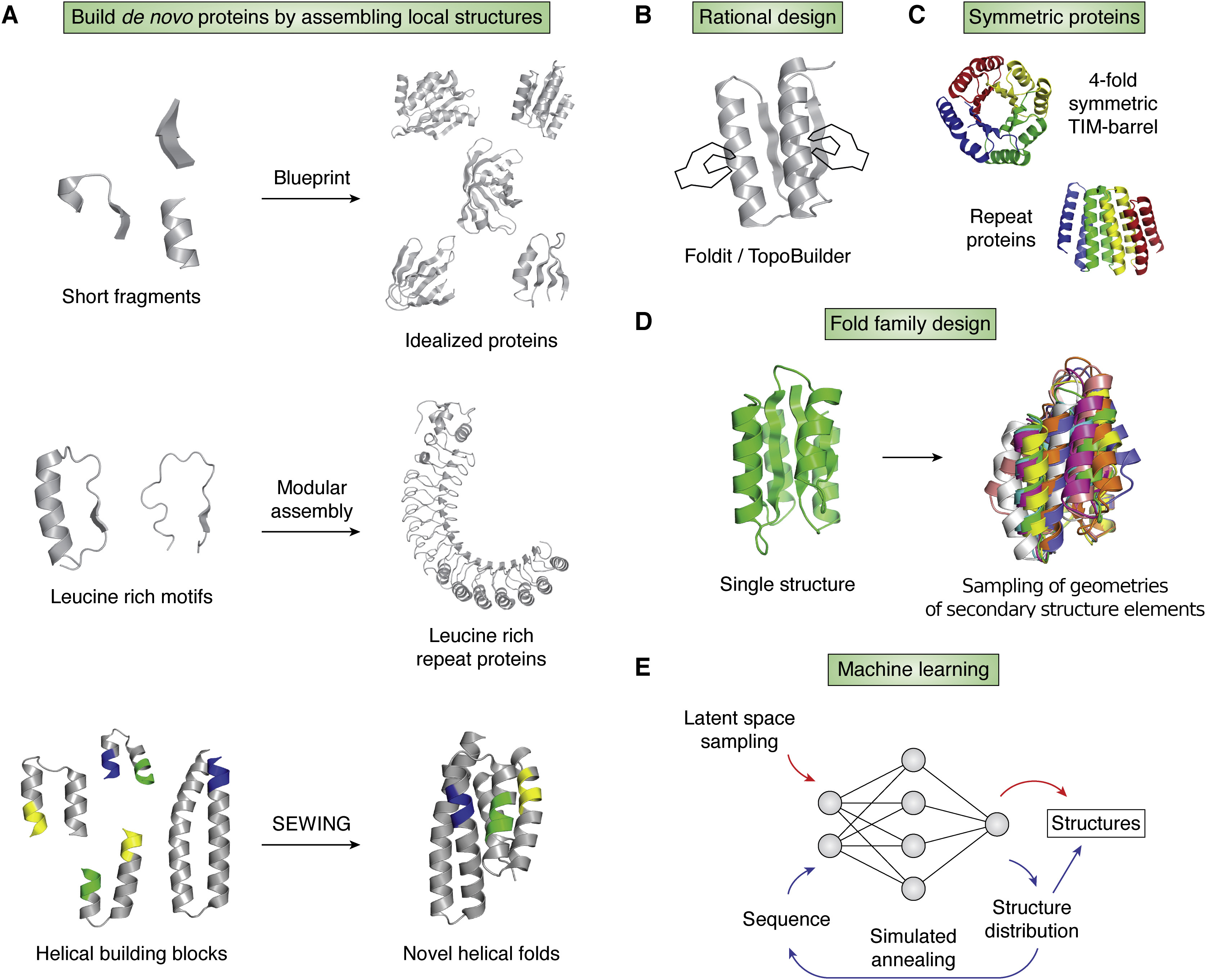}
    \caption{Methods for \textit{de novo} protein backbone generation adopted from \cite{pan2021recent} CC-BY 4.0. A) Assembly based backbone generation methods, the top shows  examples of blue print guided short fragment assembly design. The middle shows an examples of designs leveraging modular structure motifs such as Leucine-rich-motifs. The bottom shows fragment assembly using substructure graphs. B) Rational design methods such as TopoBuilder that leverages expert curated blueprints. C) Symmetric Protein design with repeats and symmetrical fold elements and repeat fragments. D) Family based fold design with family-wide geometry sampling. E) Machine learning methods generate \textit{de novo} protein structures and through various neural network architectures.}
    \label{fig:backbone_gen}
\end{figure}

The success of \textit{de novo} design of proteins heavily relies on the ability to generate high-quality designable protein backbone templates\cite{huang2016coming,kortemme2024novo,pan2021recent}, various strategies have been successfully employed for backbone structure generation which includes:\\
\textbf{Structure variations:} Redesigning existing backbone structures for new functions and simulate the backbone movement through molecular dynamics simulations.\cite{jiang2008novo,tinberg2013computational}. The disadvantage of this approach is apparent such that the process of generating variants with compute intensive methods such as molecular dynamic simulation limited its  applicability and stability. In addition, the accuracy and reliability of fast timescale simulation may not be adequate for the design purpose and therefore lead to optimal results.\\
\textbf{Family based design:} Fold family specific \textit{de novo} protein design such as helical bundles\cite{hill2000novo,pan2020expanding,basanta2020enumerative}. This approach designs protein with target function by adopting specific folds with desire functional structure motifs such as small molecule binding pockets where rational or computational fine-tunning of the active sites to achieve similar but novel function such as small molecule binders\cite{polizzi2020defined}, ion transport proteins\cite{joh2014novo}, and protein switches\cite{langan2019novo}. While this approach is powerful in designing functional variants that are neighboring to existing proteins, however, the flexibility of this approach is limited and structure diversity is often not part of the design objective.\\
\textbf{Assembly based design:}Assembly based backbone generation strategies leverages existing protein structure databases to find fragments and structure motifs that fits the design blueprint and assemble them into the desired typology. The first successful design of \textit{de novo} protein fold with assembly based methods were done over twenty years ago (Top7)\cite{kuhlman2003design}. Other assembly based methods leverages modular structure motifs such as leucine-rich-repeats\cite{park2015control} and helical blocks\cite{jacobs2016design} with substructure graphs to guide the assembly process. Despite the some notable successes, assembly based methods are still challenging due to the constraints presented by the structure motifs\cite{koga2012principles,marcos2017principles} and expert construction of viable blueprints\cite{lin2015control,marcos2018novo,yang2021bottom} render them hard to apply to practical protein design tasks. One of the challenges for assembly based \textit{de novo} protein design is to increase flexibility and expand the usable structure motifs.\\
\textbf{Machine learning and deep learning based design:}Machine learning based backbone generation models developed recently trained with structures from the PDB are used to generate \textit{de novo} protein structures. For example, a generative adversarial network(GAN) based method \cite{anand2019fully} can create protein structures by generating pairwise distance maps and decode with a pre-trained downstream neural networks. Another autoencoder base model\cite{eguchi2022ig} focused on immunoglobulins can generate immunoproteins by decoding latent representations. More recently, a model developed can build distance map by iterative refinement and network hallucination\cite{anishchenko2021novo} and repurposed a structure prediction model\cite{yang2020improved} to generate three dimensional protein structures.A recent study that utilized this model demonstrated its ability toe facilitate the design novel enzymes with competitive enzymatic activity\cite{yeh2023novo}. 
On the other hand, a slew of diffusion based models \cite{watson2023novo,lin2023generating,wu2024protein,ingraham2023illuminating} emerged for structure biology tasks aimed to address challenges presented by previous methods, specifically the ability to flexibly generate diverse and high-quality protein structures. These models have demonstrated success in designing novel monomeric proteins, protein binders, and enzyme active site scaffolds \cite{watson2023novo}, with some designs experimentally validated. Despite its success, there are still drawbacks from diffusion based structure models which often demand large amount of compute and hard to incorporate physical constraints into the generative process. It is also worth noting that the success of functional protein design still heavily relies on extensive laboratory screening.\\

\begin{figure}[H]
    \centering
    \includegraphics[width=0.99\linewidth]{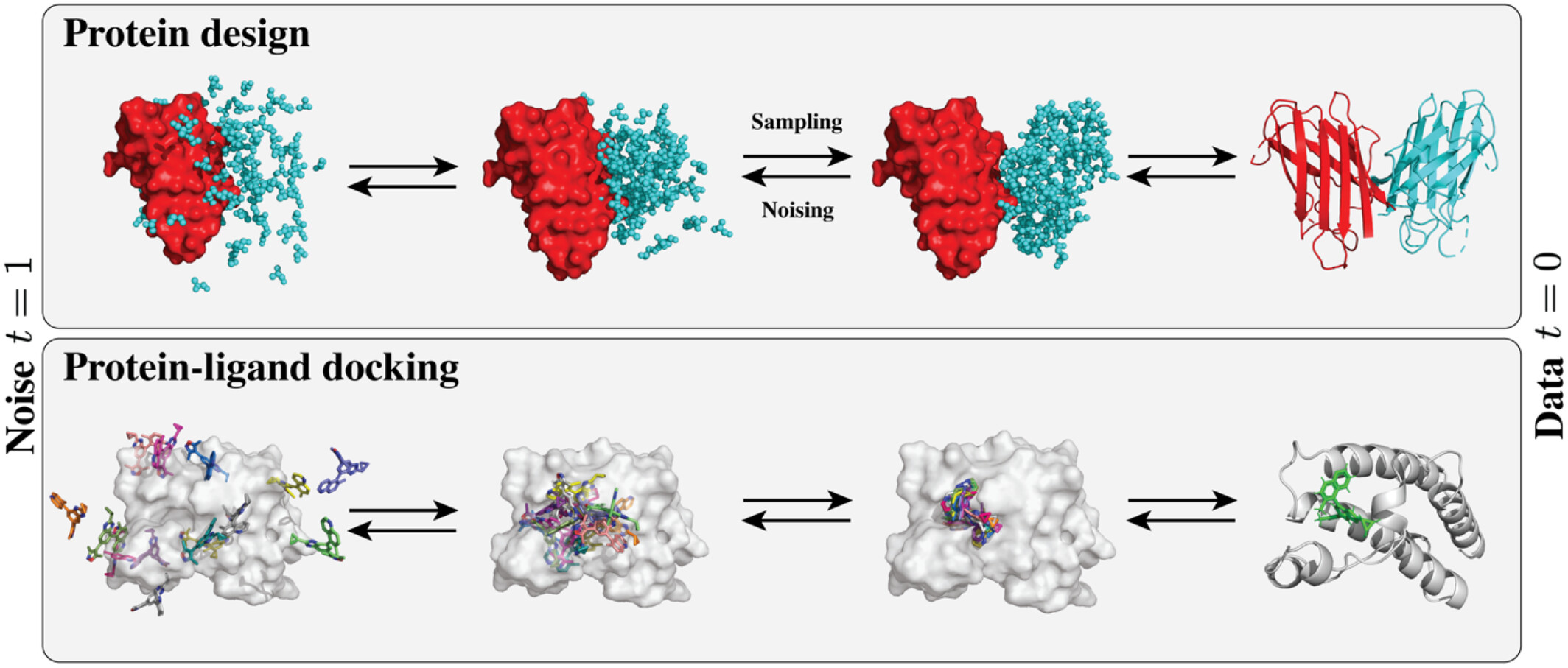}
    \caption{Protein diffusion models illustration adopted from \cite{yim2024diffusion} with CC-BY-NC 4.0. The top plot shows diffusion process for protein backbone generation, the bottom plot shows the diffusion process for protein ligand docking.}
    \label{fig:diffusion_protein}
\end{figure}

The other crucial aspect of successful computational protein design is sequence optimization also known as fixed-backbone sequence design or the inverse folding problem. The goal of this task is to find the most suitable sequence with a given backbone template such that it will fold and perform desired function in its intended environments\cite{dahiyat1997novo,xiong2014protein}. This approach is crucial for both redesigning natural proteins and selecting sequences for \textit{de novo} backbones\cite{huang2016coming,o2018spin2,polizzi2020defined}. We will describe two main class of methods that
are most commonly used:\\
\textbf{Energy Function-Based Methods:}These methods, developed over the past three decades, are based on well-understood physical principles and can provide insights into the physical basis of protein stability and function\cite{dahiyat1997novo,xiong2014protein,mackenzie2016tertiary}. They are often computationally efficient for small to medium-sized proteins. However, they have shown relatively low success rates in experimental validation and high sensitivity to target structures\cite{fleishman2011computational,schreier2009computational}. These methods may struggle to capture complex, long-range interactions and are limited by the accuracy of the underlying energy functions. Despite these limitations, they have been the mainstay of computational protein design for many years, with examples including Rosetta Design \cite{leaverchapter}, Proteus \cite{simonson2013computational}, ABACUS \cite{xiong2014protein}, and TERM\cite{mackenzie2016tertiary}.\\
\textbf{Deep Learning Methods:}More recent approaches use deep learning for sequence design, demonstrating superior performance in both computational tests and wet experiments\cite{ingraham2019generative,dauparas2022robust,li2014direct,anand2022protein}. These methods can capture complex, non-linear relationships in protein structure and sequence, and have the ability to learn from large datasets of known protein structures. They are potentially more robust to variations in target structures. However, deep learning methods are often difficult to interpret the basis of their predictions. Additionally, they may struggle with novel protein folds not well-represented in training data\cite{li2014direct,mcpartlon2022deep,gao2022pifold}. Besides, many application specific models are developed for sequence design aimed to enhance performance in domain specific tasks such as antibody design\cite{hoie2024antifold,dreyer2023inverse}. In this study, we will utilize our structure generative model and the iterative design framework in combination with a wide range of inverse folding models to improve the efficiency and robustness of computational protein design pipelines.\\

Structure-based \textit{de novo} functional protein design, which involves creating proteins distinct from those exist in the nature based on biophysical principles, has achieved notable successes including the design of novel protein folds\cite{huang2016novo} and the creation of \textit{de novo} enzymes\cite{rothlisberger2008kemp,yeh2023novo}. Protein redesign and optimization, which involves modifying existing proteins for enhanced stability, altered specificity, or new functions, has also seen substantial advancements \cite{wijma2014computationally}. Although significant effort and progress has been made towards computationally designing novel structure and sequence to achieve new functional proteins, current successes in protein design often require extensive manual input from both biochemistry and computational experts, limiting the accessibility of advanced protein design tools to the broader scientific community and hindering their ability to achieve diverse research goals. This thesis will explore and discuss novel methods for developing a more accessible computational framework, aiming to allow easier access of advanced protein design tools and enable their efficient use for \textit{de novo} protein design across the life science and computational communities.

\section{Machine learning}
\begin{center}
    \textbf{Convolutional Neural Network}\\
\end{center}
Convolutional Neural Networks (CNNs) are a type of deep learning architecture initially developed for computer vision and image processing applications \cite{lecun1989backpropagation} by adaptively learning spatial hierarchies of features from input data with specialized layers, namely convolutional layers and pooling layers designed to exploit the 2D structure of image data \cite{lecun2015deep,fukushima1980neocognitron}. CNNs have since become the backbone of numerous state-of-the-art models for image classification, object detection, as well as semantic segmentation\cite{krizhevsky2012imagenet}.

The convolutional layer performs a convolution operation on the input data with learnable kernels. For a 2D input $x$ and a filter $w$ of size $m \times n$, the convolution operation can be expressed as:
\begin{equation}
    (x * w)_{i,j} = \sum_{u=0}^{m-1} \sum_{v=0}^{n-1} x_{(i + u, j + v)} \cdot w_{(u, v)}
\end{equation}
Where $i$ and $j$ are the spatial indices of the output feature map. This operation is typically followed by a non-linear activation function, such as the Rectified Linear Unit (ReLU) \cite{glorot2011deep}:
\begin{equation}
   f(x) = \max(0, x) 
\end{equation}

Pooling layers, often inserted between successive convolutional layers, serve to reduce the spatial dimensions of the feature maps\cite{scherer2010evaluation} The most common pooling operation is max pooling, which can be defined as:
\begin{equation}
    y_{ij} = \max_{(m,n) \in R_{ij}} x_{mn}
\end{equation}

where $R_{ij}$ is a local neighborhood around position $(i,j)$.
The combination of these operations allows CNNs to learn increasingly abstract representations of the input data as information flows through the network\cite{zeiler2014visualizing}.

Besides computer vision and image processing, convolutional neural networks have also been widely successful in the domain of computational biology. In genomic sequence analysis, CNNs are powerful tools for analyzing DNA and RNA protein binding sites and profiling for non-coding variant effects \cite{zhou2015predicting,lai2022annotating}. In protein structure prediction, RaptorX pioneered the use of CNN in protein structure prediction and AphFold later showed drastically improved structure modeling accuracy in CASP13 \cite{wang2017accurate,senior2020improved}. In protein-protein interaction prediction, CNNs are used to model inter-protein interaction with sequence data.\cite{hashemifar2018predicting}.\\
\begin{center}
    \textbf{Geometric Graph Neural Network}\\
\end{center}
Graph Neural Networks (GNNs) have emerged as a powerful tool for learning on graph-structured data, making them particularly relevant for problems in computational biology and chemistry, including protein design\cite{battaglia2018relational,isert2023structure}. Unlike conventional neural networks that operate on fixed-size inputs, GNNs can process data of arbitrary size and structure, making them well-suited for three dimensional structural data.
\begin{figure}[h]
    \centering
    \includegraphics[width=0.6\linewidth]{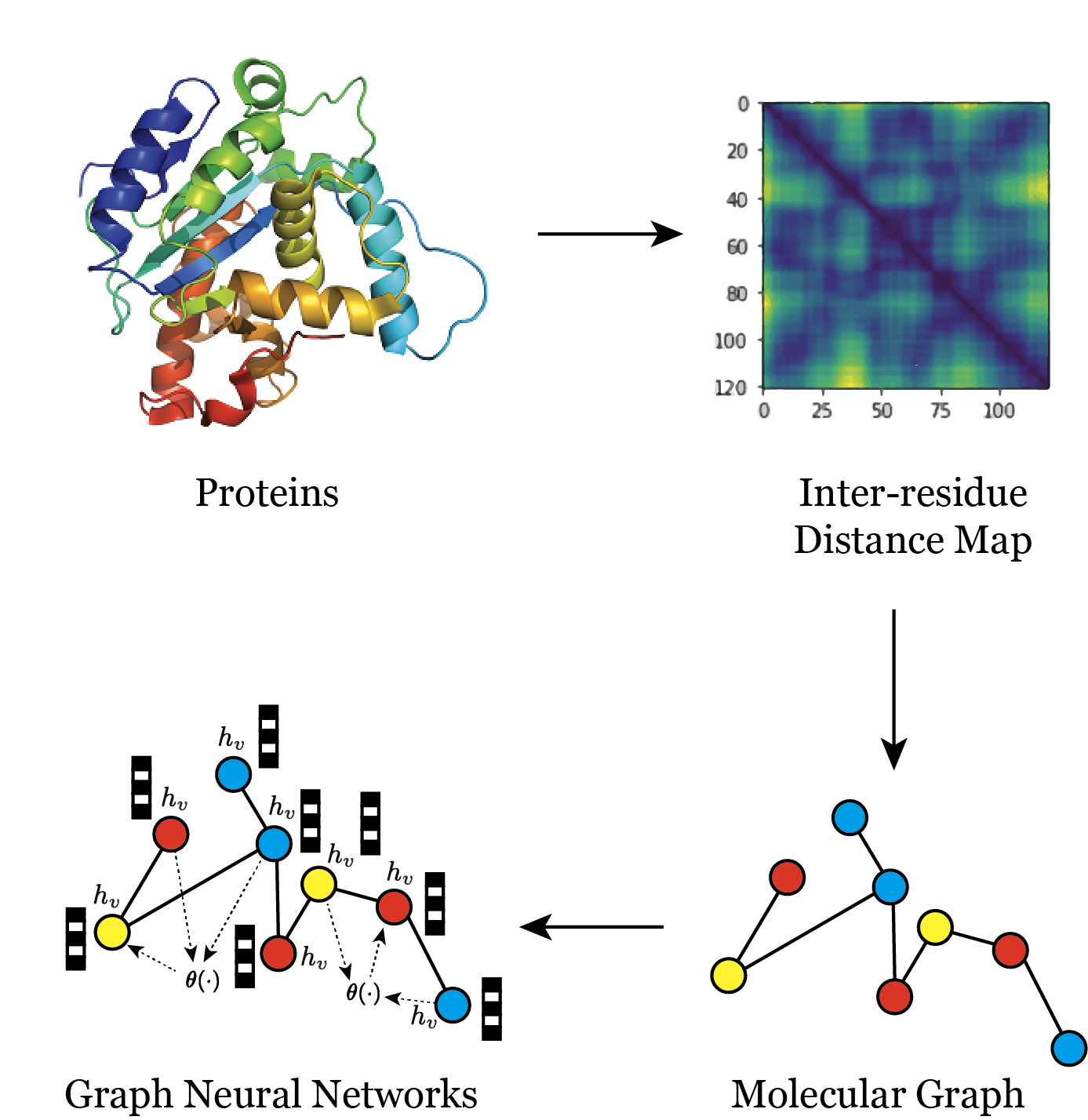}
    \caption{Illustration of graphical representation of protein structures GNNs.}
    \label{GNN-illustration}
\end{figure}

GNNs operate on graph data structures $G = (V, E)$, where $V$ represents a set of nodes and $E$ represents the edges connecting these nodes\cite{zhou2020graph}. In the context of proteins, nodes might represent amino acid residues and their atomic structures, while edges could represent chemical bonds or spatial proximity\cite{gainza2020deciphering}. GNNs work by iteratively updating node representations based on the features of neighboring nodes and edges. This process, often referred to as message passing, allows the network to capture both local and global structural information. The general form of this update can be expressed as:
\begin{equation}
\begin{split}
    a_v^{(k)} &= \text{AGGREGATE}^{(k)}({h_u^{(k-1)} : u \in N(v)})\\
    h_v^{(k)} &= \text{UPDATE}^{(k)}(h_v^{(k-1)}, a_v^{(k)})
\end{split}
\end{equation}
where $h_v^{(k)}$ is the feature vector of node $v$ at the $k$-th iteration, $N(v)$ is the set of neighbors of $v$, and $\text{AGGREGATE}$ and $\text{UPDATE}$ are learnable functions.Several variants of GNNs have been developed, each with unique properties. These include Graph Convolutional Networks (GCNs)\cite{kipf2016semi}, Graph Attention Networks (GATs)\cite{velivckovic2017graph}, and Message Passing Neural Networks (MPNNs)\cite{gilmer2017neural}. For instance,
in the case of Graph convolution networks(GCNs) with simple neighbor aggregation:
\begin{equation}
    h_v^{(k)} = \sigma\left(W^{(k)} \sum_{u \in N(v)} \frac{1}{|N(v)|} h_u^{(k-1)} + B^{(k)}h_v^{(k-1)}\right) 
\end{equation}
where $W^{(k)}$ and $B^{(k)}$ are learnable weight matrices and $\sigma$ is a non-linear activation function.

In the context of protein design, nodes typically represent amino acids or atoms, while edges represent chemical bonds or spatial proximity. The node features $x_v$ can include amino acid properties or molecular embedding, while edge features $e_{vu}$ could represent distances or bond types.
\begin{center}
    \textbf{Graph Transformer}\\
\end{center}
Graph transformer merges the structural inductive bias of Graph Neural Networks (GNNs) with the powerful attention mechanisms of Transformer models\cite{vaswani2017attention}. This combination allows for more expressive and flexible representations of graph-structured data, making them particularly suited for complex tasks in protein design and analysis especially for direct structural generation. The Attention architecture introduced in \cite{vaswani2017attention} revolutionized sequence modeling with the self-attention mechanism:
\begin{equation}
    \text{Attention}(Q, K, V) = \text{softmax}\left(\frac{QK^T}{\sqrt{d_k}}\right)V 
\end{equation}

where $Q$, $K$, and $V$ are query, key, and value matrices, respectively, and $d_k$ is the dimension of the key vectors. To integrate the attention mechanism in graphical settings between node $i,j$ \cite{shi2020masked}:
\begin{equation}
\begin{split}
    q^l_{(c,i)} &= W^l_{(c,q)}h^l_{i} + b^l_{(c,q)} \\
    k^l_{(c,j)} &= W^l_{(c,k)}h^l_{j} + b^l_{(c,k)} \\
    e_{(c,ij)} &= W^l_{c,e}e_{(i,j)} + b^l_{c,e} \\
    \alpha^l_{(c,ij)} &= \frac{<q^l_{(c,i)}, k^l_{(c,j)} + e_{(c,ij)}>}{\sum_{u \in N(i)}<q^l_{(c,i)}, k^l_{(c,u)} + e_{(c,iu)}>}
\end{split}
\end{equation}

Where $<q,k> = exp(\frac{q^Tk}{\sqrt{d}})$ and $d$ is the size of each head, For the $c-th$ head attention, the node features $h^l_{i}$ and $h^l_{j}$ are first transformed into the query vector $q^l_{(c,i)} \in \mathbb{R}^d$ and key vector $k^l_{(c,j)} \in \mathbb{R}^d$ respectively with the trainable weights $W^l_{(c,q)},W^l_{(c,k)}, b^l_{(c,q)},b^l_{(c,k)}$. Then the edge features $e_{ij}$ will be encoded and add to the key vector in each layer. After computing the multi-head attention, message aggregation was performed:
\begin{equation}
    \begin{split}
        v_{(c,j)} &= W^l_{(c,v)}h^l_{j} + b^l_{(c,v)}\\
        \hat{h}^{l+1} &= \Vert_{c=1}^{C}\sum_{j\in N(i)} \alpha^l_{(c,ij)}[v^l_{c,j} + e_{(c,ij)}]
    \end{split}
\end{equation}

Where $\Vert$ concatenate for $C$ attention heads. The node feature $h_j$ is transformed into $v_{(c,j)} \in \mathbb{R}^d$. Together with residual connection, and non-linear transformation, the node feature update will be 
\begin{equation}
    \begin{split}
        r^{l}_i &= W_r^lh_i^l+b_r^l\\
        \beta_i^l &= Sigmoid(W_g^l[\hat{h}^{l+1};r^{l}_i;\hat{h}^{l+1}-r^{l}_i])\\
        h^{l+1} &= ReLU(LayerNorm((1-\beta_i^l)\hat{h}^{l+1} + \beta_i^lr_i^l))
    \end{split}
\end{equation}

Graph Transformers represent a powerful tool for learning on protein structures, combining the structural inductive bias of graphs with the expressive power of attention mechanisms. We will use the above formulation of graph transformer in our structure generative model for direct coordinate generation. 
\begin{center}
    \textbf{Variational Autoencoder}\\
\end{center}
Variational methods have become increasingly important in machine learning and statistical inference. They provide powerful tools for approximating complex probability distributions and learning latent representations of data. This section will provide the background for variational autoencoder used as a framework to model the protein structures.  

Variational autoencoders(VAEs) also known as auto-encoding variational bayes, introduced by \cite{kingma2013auto} are a type of deep generative model leverages both the autoencoder architecture and variational inference to learn a compressed latent representation of the data. In many Bayesian models, computing the exact posterior distribution is computationally infeasible. Variational Bayes addresses this by approximating the true posterior with a simpler distribution, typically from a tractable family. The goal is to find the member of this family that is closest to the true posterior, where "closest" is measured by the Kullback-Leibler (KL) divergence. Let's consider the following simple latent variable graphical model:
\begin{figure}[h]
    \centering
    \includegraphics[width=0.3\linewidth]{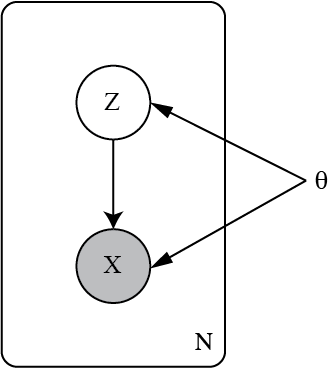}
    \caption{Probabilistic graphical model for Variational Autoencoder}
    \label{fig:enter-label}
\end{figure}

 Where $\theta$ is a set of deterministic parameters; $x$ is our observation which can either be discrete or continuous, $z$ is a continuous latent variable. We can write the joint distribution of the graphical model as $p(x,z;\theta) = p_{\theta}(z)p_{\theta}(x|z)$, where $p_{\theta}(z)$ is the prior of the latent variable and $p_{\theta}(x|z)$ is the likelihood function of our observation. 
 
 For inference on latnet variable models, one common challenge is such that the posterior distribution of the latent variable $p_{\theta}(z|x)$ is often computationally intractable because it requires integration over all possible value of $z$.\\
 \begin{equation}
    p_{\theta}(z|x) = \frac{p_{\theta}(x|z)p_{\theta}(z)}{\underbrace{\int_{z}p_{\theta}(x|z)p_{\theta}(z)dz}_{intractable}}
\end{equation}

Variational inference approaches this challenge by imposing a tractable variational distribution $q_{\phi}(z|x)$ in place if the true posterior and minimize the Kullback–Leibler divergence between the target posterior and the variational distribution.\\
\begin{equation}
\begin{split}
    KL(q_{\phi}(z|x)||p_{\theta}(z|x)) & = \int_{z} q_{\phi}(z|x) \log(\frac{q_{\phi}(z|x)}{p_{\theta}(z|x)})dz\\
    & = E_{z \sim q_{\phi}(z|x)}[\log(q_{\phi}(z|x)) - \log(p_{\theta}(z|x))]\\
    & = E_{z \sim q_{\phi}(z|x)}[\log(q_{\phi}(z|x)) - \log(\frac{p_{\theta}(x|z)p_{\theta}(z)}{p_{\theta}(x)})]\\
    & = E_{z \sim q_{\phi}(z|x)}[\log(q_{\phi}(z|x))] - (\log(p_{\theta}(x,z) - \log (p_{\theta}(x))]\\
\end{split}
\end{equation}

After rearranging the terms, we can write the log likelihood of $x$ as:\\
\begin{equation}
    \begin{split}
        \log(p_{\theta}(x)) &= E_{q_{\phi}}[\log(p_{\theta}(x,z))] - E_{q_{\phi}}[\log(q_{\phi}(z|x))] + KL(q_{\phi}(z|x)||p_{\phi}(z|x))\\
        ELBO &= \log(p_{\theta}(x)) - \underbrace{KL(q_{\phi}(z|x)||p_{\theta}(z|x))}_{\geq 0}
    \end{split}
\end{equation}

KL divergence is non-negative according to the Jensen's inequality. The log likelihood(evidence) is then lower-bounded by $E_{q_{\phi}}[\log(p_{\theta}(x,z))] - E_{q_{\phi}}[\log(q_{\phi}(z|x))]$, which is referred to as the Evidence Lower Bound(ELBO). We can estimate the gradient of the ELBO with a naive Monte Carlo estimator, however, it is often challenging due to its high variance\cite{kingma2013auto}. Another important observation is that the evidence is a constant and the KL divergence is non-negative, maximizing the ELBO can be seen as simultaneously minimizing the KL divergence between the variational distribution and the true posterior and maximizing the marginalized data likelihood. We can then rewrite the ELBO as\\
\begin{equation}
    \begin{split}
        ELBO &= E_{q_{\phi}}[\log(p_{\theta}(x,z))] - E_{q_{\phi}}[\log(q_{\phi}(z|x))]\\
        & = E_{q_{\phi}}[\log(p_{\theta}(x|z)p(z))] - E_{q_{\phi}}[\log(q_{\phi}(z|x))]\\
        &= E_{q_{\phi}}[\log(p_{\theta}(x|z)) + \log(p_{\theta}(z)) - \log(q_{\phi}(z|x))]\\
        & = E_{q_{\phi}}[\log(p_{\theta}(x|z))] - KL(q_{\phi}(z|x) || p_{\theta}(z))
    \end{split}
\end{equation}

Here, $q_{\phi}(z|x), p_{\theta}(x|z)$ are not defined specifically. Instead of assuming specific distributions, we can take advantage of the fact that we can generate any distribution by mapping a normal distribution with sufficiently sophisticated function\cite{doersch2016tutorial}. DNNs are used here for their ability to approximate complicated functions and its differentiability, therefore, we can optimize it with stochastic gradient descent and back propagation. For $q_{\phi}(z|x)$ we define two neural encoder $Enc_{\phi}^{\mu}(x), Enc_{\phi}^{\sigma^2}(z|x)$ that takes input $x\in \mathcal{R}^{n}$ into two vectors $\mu, \sigma^2 \in \mathcal{R}^k$ that encodes the mean and variance of the Gaussian distribution of our latent variable $z$; this model is also called as a \textit{recognition model}. For $p_{\theta}(x|z)$ we define a neuro decoder $Dec_{\theta}(z)$ that takes a latent vector $z\in \mathcal{R}^k$ and output a reconstructed input $\hat{x} \in \mathcal{R}^n$, this is also called as a \textit{generative model}.

\begin{equation}
    \begin{split}
        q_{\phi}(z|x) &= \mathcal{N}(Enc_{\phi}^\mu (x), diag(Enc_{\phi}^{\sigma^2}(x)))\\
        \hat{x} &= Dec_{\theta}(z) \quad z \sim q_{\phi}(z|x)
    \end{split}
\end{equation}
The optimization objective of VAE can then be written as
\begin{equation}
    \begin{split}
        l(x, \theta, \phi) = \underbrace{- E_{q_{\phi}(z|x)}[\log(p_{\theta}(x|z))]}_{\textit{Reconstruction loss}} + \underbrace{KL(q_{\phi}(z|x) || p_{\theta}(z))}_{\textit{Regularizer}}
    \end{split}
\end{equation}
We use VAE as a crucial component to build the deep generative models for protein structures and downstream protein design tasks. 
\chapter{End-to-end Deep Structure Generative Model for Protein Design and Optimization}
\section{Motivation}
Computationally designing protein systems has long been a challenging problem due to the complexity of protein structures and interactions among various components within the system. Successful \textit{in silico} protein design demands both accurate modeling of the system of interest and effective structure and sequence generation to accommodate the desired functional properties. Though promising progress has been made towards more accurate design methods \cite{dauparas2022robust,yeh2023novo,trippe2022diffusion,jumper2021highly}, the gap between computational design and experimental realization remains significant. \\

A protein design project often commences with a predefined biochemical objective, such as enhancing enzyme activity, devising binders for specific target molecules, or creating interaction systems involving multiple protein entities. The first step involves the identification of an initial structural template, which can be achieved through two primary methods: a database search aimed at locating existing proteins sharing a similar biochemical function or the \textit{de novo} approach achieved through \textit{in silico} backbone generation targeting specific functional objectives. Following the initial structure template determination, the subsequent phase involves assembling the appropriate amino acid sequence given the predetermined backbone template, also known as inverse protein folding. This process entails the search of amino acid sequences that possess a high likelihood of folding into the designated template structure. Inverse protein folding is often accomplished using deep learning models or energy-based sequence optimization methods. The final phase involves further \textit{in silico} selection pipelines that sift through the generated candidates to identify the most promising ones for downstream experiments. \cite{kortemme2024novo,yang2021bottom,yeh2023novo}\\

There are a couple of main drawbacks to the above design paradigm. 1) Structure templates identified in the first design phase are always assumed to be static, disregarding the inherent dynamics of proteins. Therefore, restricting the inverse folding phase to only a snapshot of the protein system results in a constrained exploration of the potential sequence design landscape. 2) A successfully designed protein system must both be viable in the desired cellular environment and meet functional objectives. Frequently, a single iteration through the outlined design paradigm proves inadequate to accomplish these goals.

To address the abovementioned shortcomings, this thesis project will attempt to develop a structural generative model for variational structure sampling. By doing so, expanding the sampling space available to inverse folding models, a revolutionized adaptive computational design paradigm is proposed. This paradigm integrates structural generative models and advanced system simulation for robust and efficient \textit{in silico} design optimization. Moreover, promising preliminary results suggest that our structural generative model has the potential to serve as a pre-training framework for protein structure, comparable to the protein language models tailored for protein sequences, as discussed earlier. Part of this thesis will be dedicated to the exploration of upscaling the structural generative model for structure embedding and its application to various downstream tasks, including protein function annotation and inverse protein folding.

\section{Abstract}
Designing protein with desirable structure and functional properties is the pinnacle of computational protein design with unlimited potentials in the scientific community from therapeutic development to combating the global climate crisis. However, designing protein macromolecules at scale remains challenging due to hard-to-realize structures and low sequence design success rate. Recently, many generative models are proposed for protein design but they come with many limitations. Here, we present a VAE-based universal protein structure generative model that can model proteins in a large fold space and generate high-quality realistic 3-dimensional protein structures. We illustrate how our model can enable robust and efficient protein design pipelines with generated conformational decoys that bridge the gap in designing structure conforming sequences. Specifically, sequences generated from our design pipeline outperform native fixed backbone design in 856 out of the 1,016 tested targets(84.3\%) through AF2 validation.  We also demonstrate our model's design capability and structural pre-training potential by structurally inpainting the complementarity-determining regions(CDRs) in a set of monoclonal antibodies and achieving superior performance compared to existing methods. 

\section{Introduction}
Computational protein design has been of great interest to the scientific community for decades. Designing protein macromolecules with specific function and structure is a highly sought after technique with broad application to therapeutics, biosensors, and enzyme engineering \cite{griss2014bioluminescent,yang2021bottom,lu2022machine,zhang2022thermodynamically}. However, despite years of effort and advancements, computational protein design still remains a very challenging problem. 

Traditionally, the protein design pipeline is often regarded as a two step process where the practitioner first determines the protein backbone structure accommodating the specified structural and biochemical properties, and then designs the amino acid sequence with the given backbone structure. In this regime, the most successful applications rely heavily on template-based fragment sampling and domain-expert specified topologies for backbone determination and energy minimization based sequence design \cite{courbet2022computational,cao2022design}. While this approach is widely adopted, there are apparent drawbacks. For example, the resulting backbone structure from the first step may not be optimal or designable and the designed sequence in the second step may not readily conform to the desired backbone template. As a result, the success rate for computational protein design remains relatively low \cite{huang2016coming,anishchenko2021novo}.\\

In recent years, progress in machine learning and deep learning research has contributed to significant advances for protein modeling such as mutation effect estimation \cite{frazer2021disease,meier2021language,riesselman2018deep}, protein function prediction \cite{lai2022accurate,gligorijevic2021structure}, and structure prediction \cite{jumper2021highly,baek2021accurate,wang2017accurate}.These advancements have also played a part in aspects of the computational protein design problem. For fixed backbone sequence design, a series of deep learning methods have emerged to improve conventional energy based approaches \cite{alford2017rosetta,khatib2011algorithm} by directly incorporating structural information using SE(3)-equivaraint frameworks \cite{mcpartlon2022deep,jing2020learning,hsu2022learning}. An array of recent works have studied the use of generative models for structure generation\cite{lin2021deep,anand2019fully,anand2018generative}, however, these methods often generate topological constraints and rely on downstream tools for 3-dimensional structure determination. One of the major difficulties for direct coordinate generative modelling is  properly accounting for the rotation and translation equivariance in the target conformation.\\

In the Chapter, we present a versatile VAE-based deep structural generative model that seeks to bridge the gap for robust computational protein design. Our method makes contributions to three fundamental aspects of protein design: First, we directly model protein structure in the 3-dimensional coordinate space which avoids downstream coordinate recovery in constraints based model. Second, Our method is universal such that it can model proteins of arbitrary size and thus exposes our model to the whole fold space while previous generative models are restricted to protein with certain size and can only be trained on a small subset of all available folds for a given model. Third, We address the translation and rotation equivariance in both the input space and the use of a locally aligned coordinate loss proposed by \cite{jumper2021highly}. It is important to note that structure entries in database such as the Protein Data Bank(PDB) \cite{berman2000protein} often represent a single sample from its conformational landscape.  By conditioning on a given backbone structure, our model is able to generate conformational decoys from the latent space. Combined with fixed-backbone sequence design models and accurate protein structure prediction tools, our model can enable efficient \textit{in silico} design screening.\\ 


To evaluate our models from multiple aspects. First, we demonstrate our model's ability to generate high-quality, realistic protein structure ensembles by comparing the generated three-dimensional coordinates to experimentally determined structures. This comparison allows us to assess the accuracy and realism of our model's output. Second, we showcase improved efficiency and success rates in conventional design pipelines by producing sequences that recapitulate backbone templates derived from our model. This evaluation highlights the practical applicability of our approach in enhancing existing protein design methodologies. Lastly, we corroborate our model's design capabilities through the inpainting of backbone coordinates in the complementarity-determining regions (CDRs) of monoclonal antibodies. In this task, we achieve state-of-the-art results, exemplifying our model's versatility as a structure generator. These evaluations collectively validate our model's effectiveness in generating, improving, and manipulating protein structures, highlighting its potential as a powerful tool in computational protein design.

\section{Literature Review}
\subsection{Deep Generative models}. 
Generative adversarial networks(GAN) \cite{goodfellow2020generative} and auto-encoding variational Bayes (VAE)\cite{kingma2013auto} are powerful generative frameworks that are used from image synthesis\cite{brock2018large,razavi2019generating} to language modeling\cite{bowman2015generating}. \cite{riesselman2018deep,shin2021protein,frazer2021disease} also used VAE for protein sequence design and variant effect prediction. Most of the aforementioned structure generative models also used VAE or GAN as their generative framework. Recently, Diffusion generative models and flow-based models are emerged in popularity specifically for tasks in protein structure prediction and generation\cite{yim2024diffusion,watson2023novo,ingraham2023illuminating,ho2020denoising}.
We build our structural model based on the VAE framework where we first encode the invariant protein structure representations(See Methods) into the latent space then a decoder is used to generate the corresponding three-dimensional coordinates from the latent space. This approach allows our model to generate flexible protein conformations conditioned on the input backbone constraints. Trained on masked input constraints, our model can easily be adopted for backbone inpainting for structural design.\\
\subsection{Protein structure generative models}. 
Existing generative methods for protein structure design focus on generating invariant topological constraints such as inter-residue contacts and distance maps which are then converted into three dimensional structures via downstream coordinate recovery tools such as AMDD\cite{boyd2011distributed} or pre-trained structure predictors\cite{yang2020improved,baek2021accurate}. While generative models can produce protein structures through 1D and 2D constraints, they often require a second step to recover the Cartesian coordinates. This recovery process can be challenging, especially for 1D representations where small errors in backbone torsion can propagate and lead to unrealistic structures. For 2D representations, methods have been developed to convert distance matrices into 3D structures, but these can be sensitive to systematic noise produced by neural networks.
Though these methods have garnered some success, there is no a priori guarantee that the generated constraints are geometrically viable. Consequently, the resulting three dimensional structures are often of low quality or biochemically infeasible. Moreover, it is generally not possible to perform conditioned structure generation on these models\cite{lin2021deep,anand2019fully,anand2018generative,karimi2020novo}. On the other hand, methods that focuses on generating structure ensembles are often fold specific and also relying on generating topological constraints\cite{mansoor2024protein,janson2023direct}.
While \cite{eguchi2022ig} proposed the first direct coordinate structure generative model, its application is limited to proteins of a fixed length, and is trained only to recover inter-atom distance and torsion maps. In contrast, our model, addresses rotation and translation equivariance in both the input and output space by distilling invariant representations of protein geometry and by using a equivariant locally aligned coordinate loss function to perform gradient optimization directly on the coordinate space. In this way, our model can directly and flexibly model the three-dimensional structure and generate conformational snsembles. For diffusion based models such as \cite{watson2023novo,ingraham2023illuminating,ho2020denoising}, though highly flexible, they are often limited by intensive compute requirement and the quality consistency of the generated structures.
\subsection{Fixed backbone sequence design}
For sequence optimization or often known as the inverse folding problem or fixed backbone sequence design, Traditionally, this problem was addressed by optimizing energy functions through Markov chain Monte Carlo (MCMC) methods, combining physical and statistical potentials \cite{shapovalov2011smoothed}. However, these conventional methods often resulted in limited sequence diversity and struggled with designing multi-body interactions crucial for protein function\cite{boyken2016novo,maguire2018rapid}.\\
DL-based sequence design algorithms have emerged to address these limitations. These methods utilize various representations of protein structures, including graphs \cite{strokach2020fast,ingraham2019generative,dauparas2022robust}, 2D matrices \cite{chen2019improve,norn2021protein}, torsional angles \cite{o2018spin2}, and voxelized volumes \cite{qi2020densecpd,anand2022protein,zhang2020prodconn}, to generate sequence probability profiles or full sequences. Some approaches frame the problem as a constraint satisfaction problem \cite{strokach2020fast}, while others use MCMC\cite{anand2022protein} or autoregressive models\cite{ingraham2019generative} to generate sequences. DL methods have shown promise in introducing more diverse sequences and capturing multi-body interactions more effectively than conventional approaches. However, it is important to note that while these methods show great potential, their generalizability and reliability have not been extensively validated through experimental testing\cite{anand2022protein,strokach2020fast,dauparas2022robust}. As the field progresses, DL approaches may offer a more flexible and powerful toolkit for protein design, potentially leading to more diverse and functionally optimized sequences.\\
Although our model does not directly design the sequence alongside the conformation decoys, we demonstrate how our generated conformational ensembles can be utilized and address some of the problems mentioned above and significantly improve the design capability of inverse folding models.\\

\section{Methods}
\begin{figure}[H]
    \centering
    \includegraphics[width=1\linewidth]{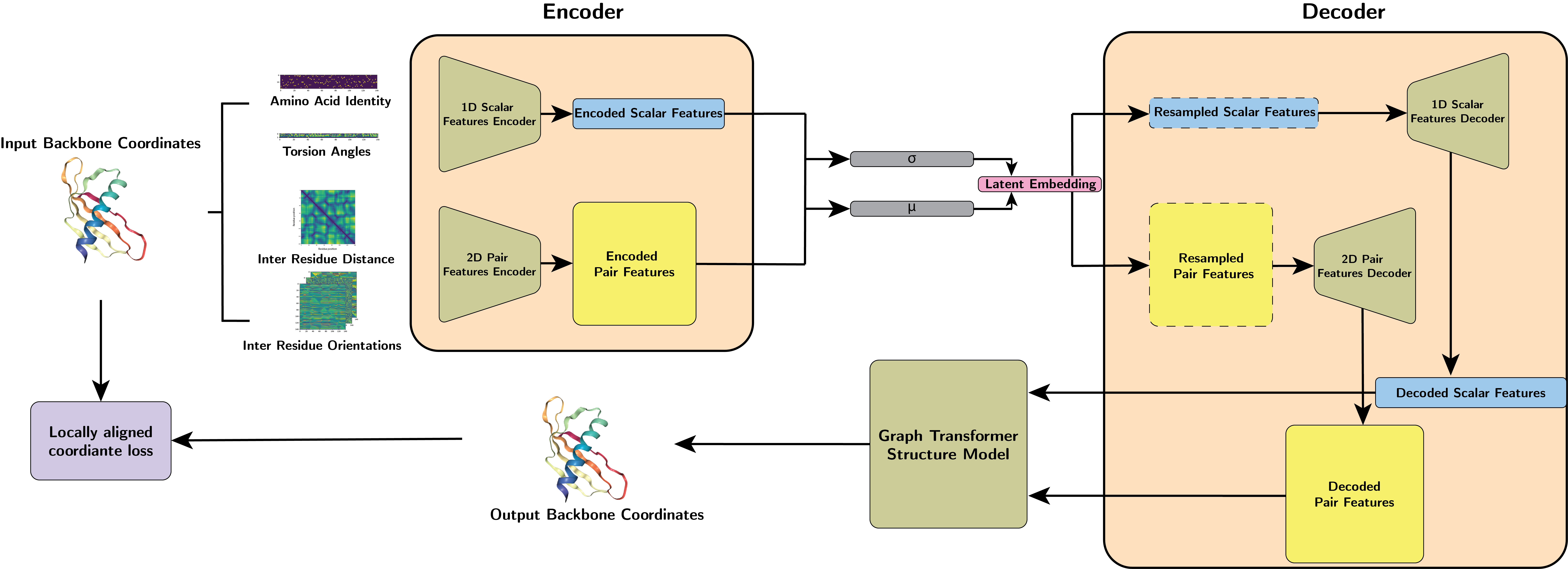}
    \caption{Model Overview. Structures are first distilled into pairwise (inter-residue distance \& orientation) and scalar (torsion angles \& amino acid sequence) features, Then our model separately encodes the pairwise and scalar features into the latent space. The latent representation is then sampled and decoded by the pair and scalar feature decoder before being fed into a graph-transformer based structure module for coordinate generation. A locally aligned loss is then computed between the reconstructed  coordinates and the native input structure.}
    \label{fig:C1_Fig1}
\end{figure}

In this section we illustrate the overall model design as shown in Figure \ref{fig:C1_Fig1}. To prepare for input features, each protein structure is distilled into invariant pairwise representations of inter-residue distance and orientations as described in \cite{yang2020improved} and scalar representations of amino acid sequence and backbone torsion angles. This input is then fed through an encoder network which produces a latent representation of each residue. These representations are reassembled and passed to a decoder module which reconstructs the backbone coordinates. \\
\subsection{Protein representations}
We represent a protein as a complete molecular graph $\mathcal{G} = \big(\mathcal{V},\mathcal{E}\big)$ where $\mathcal{V}$ consists of scalar residue features $v_{i}$ and $\mathcal{E}$ consists of pairwise features $e_{ij}$ between residues $i$ and $j$.\\
\textbf{Scalar Features}\\
The nodes $v_i$ of our input graph correspond to protein residues $ i \in \{1...n\}$. The input scalar feature $\mathcal{F}_{i}^{scalar}$ associated with residue $i$ consists of amino acid identity and dihedral angle encodings:
\begin{equation}
\mathcal{F}_{i}^{scalar}= \{\boldsymbol{f}_{AA}(s_{i}),\boldsymbol{f}_{dihedral}(\boldsymbol{\theta_{i}})\}
\label{eq:residue_feats}
\end{equation}
The first, $\boldsymbol{f}_{AA}\left(s\right) \in \mathbb{R}^{20}$, is a one-hot encoding of the amino acid type $s$ using 20 bins for each naturally occurring amino acid. The second, $\boldsymbol{f}_{dihedral}(\boldsymbol{\theta}) \in \mathbb{R}^6$, is an encoding of the three dihedral angels with Fourier features $\{\sin, \cos \} \circ \{\phi_i, \psi_i, \omega_i \}$, where $\phi_i, \psi_i, \omega_i$ are dihedral angles computed from the coordinates from $C_{i-1}, N_i, C_{\alpha i}, C_i, N_{i+1}$ atoms.\\
\textbf{Pairwise Features}\\
for a given pair of residues $i$ and $j,$ we define the edge $e_{ij}$ features as

\begin{equation}
\mathcal{F}_{ij}^{pair}=\{\boldsymbol{f}_{dist}^{(v_i,v_j)}(\vec{X}^{C_\alpha}_j,\vec{X}^{C_\alpha}_i),\boldsymbol{f}_{ori}(\boldsymbol{\theta_{ij}})\})\label{eq:pair_struct_feats}
\end{equation}
The first encoding, $\boldsymbol{f}_{dist}^{(v_i,v_j)}(\vec{X}^{C_\alpha}_j,\vec{X}^{C_\alpha}_i) \in \mathbb{R}^{16}$ is the distance encoding that embeds the inter-residue distance $d_{C_\alpha} = \Vert \vec{X}^{C_\alpha}_j - \vec{X}^{C_\alpha}_i \Vert_2$ with 16 Gaussian radial basis functions with centers evenly spaced in $[0,20] \angstrom$ as described in \cite{jing2020learning}. $\boldsymbol{F}_{ori}(\theta) \in \mathbb{R}^3$ is the encoding of the angle $\theta$ performed in the same manner as the backbone dihedral encoding for residue features. The input angles $\theta_{ij}\in\left\{ \phi_{ij},\psi_{ij},\omega_{ij}\right\} $ are pairwise inter-residue orientations defined in \cite{yang2020improved}. To produce pairwise orientation information, we impute a unit vector in the direction $C\beta_{i}-C\alpha_{i}$ before computing the respective angles. The imputed vector is calculated as in \cite{jing2020learning} using
$$
 \sqrt{\frac{1}{3}}\left\langle \textbf{n}\times\textbf{c}\right\rangle -\sqrt{\frac{2}{3}}\left\langle \textbf{n}+\textbf{c}\right\rangle    
$$

where $\left\langle x\right\rangle =x/\Vert x\Vert_{2}$, $\textbf{n}=N_{i}-C\alpha_{i},$ and $\textbf{c}=C_{i}-C\alpha_{i}$.\\
\textbf{Masking Schemes for structure inpainting and pre-training}\label{sec:masking}\\
To help facilitate our model's ability to inpaint protein structural regions, we employ a masking scheme for a given protein denoted as $Mask(n)$. We implemented three types of masks:\\
\textbf{linear} mask where a random residue is selected uniformly at random with $p = 1/L$ where $L$ is the length of the respective protein and the mask will then span 10 residues around the selected one.\\
\textbf{spatial} mask where a random residue is selected uniformly at random with $p = 1/L$ and all residues within $12 \angstrom$ are masked.\\
\textbf{random} mask where each residue will be masked at the rate of $p \sim Uniform(0,0.5)$ where $p$ is sampled for each individual residue.\\

For each masked residue, we use zero masks in both the scalar and pairwise features such that $\mathcal{F}_i^{scalar} = \mathsf{0}$ and $\mathcal{F}_{i,:}^{pair} = \mathcal{F}_{:,i}^{pair} = \textbf{0} \quad \forall i \in Mask(n)$. In the hybrid masking training strategy, 50\% of the input proteins are mask with one of the aforementioned masks with equal probability. We also analyzed the effect of various masking probability and strategies, see the Appendix for more details.

\subsection{Model Architecture}
\textbf{Feature encoders \& decoder}\\
The CoordVAE model uses different architectures for its encoder and decoder networks. The encoder network consists of a 1D feature module for the scalar input and a 2D feature module for the pairwise input using dilated convolution network architecture \cite{yu2017dilated}.
\begin{equation}
    \mathcal{H}^{1D}_{l+1} = \text{LeakyReLU}(\text{1D-InstanceNorm}(\text{1D-Conv}(\mathcal{H}_l^{1D},\mathcal{K}_{l}^{1D})))
\end{equation}
\begin{equation}
    \mathcal{H}^{2D}_{l+1} = \text{LeakyReLU}(\text{2D-InstanceNorm}(\text{2D-Conv}(\mathcal{H}_l^{2D},\mathcal{K}_{l}^{2D})))
\end{equation}

Where $\mathcal{H}_l^{1D}$ and $\mathcal{H}_l^{2D}$ denotes the scalar and pairwise representation at layer $l$ respectively and $\mathcal{K}_l^{1D}$ and $\mathcal{K}_l^{2D}$ denotes the kernel size for the scalar and and pairwise convolution. Each convolutional block applies a leakyReLU nonlinearity with negative slope parameter set to 0.2, and instance normalization \cite{ulyanov2016instance}.

The decoder network applies a mirrored architecture to the latent scalar and pairwise representations. Finally, the decoded scalar and pairwise features are processed jointly with a graph transformer described in \cite{shi2020masked} for coordinate generation.
\begin{equation}
    \mathcal{H}_{out}^{2D}, \mathcal{H}_{out}^{1D} = \text{GraphTransformer}(\mathcal{H}_{dec}^{1D}, \mathcal{H}_{dec}^{2D})
\end{equation}

Where $\mathcal{H}_{dec}^{1D}$ and $\mathcal{H}_{dec}^{2D}$ are the scalar and pairwise output of the decoder network respectively. For Coordinate generation, we process the output of the graph transformer module with a dense projection.
\begin{equation}
    \vec{X_i} = \text{Linear}(\mathcal{H}_{out}^{1D})_i
\end{equation}

Where $\vec{X_i}$ are the backbone coordinates of residue $i$ and the projection is implemented with as a dense layer without bias.
For CNN baseline models, the graph transformer module is removed and coordinates are directly projected form the 1D feature decoder output $\mathcal{H}_{dec}^{1D}$.\\
\textbf{VAE}\\
To implement a VAE framework within our encoder-decoder architecture, we combines the encoder output by a horizontally average-pooling(HAP) the pairwise output feature and concatenate it with the scalar output before applying the mean and variance projection networks.\\
\begin{equation}
    \mathcal{H}_{latent} = \text{HAP}(\mathcal{H}_{enc}^{2D}) \Vert \mathcal{H}_{enc}^{1D}
\end{equation}
Where $\mathcal{H}_{enc}^{1D}$ and $\mathcal{H}_{enc}^{2D}$ are the scalar and pairwise of the encoder output respectively and $\Vert$ denotes concatenation. To produce the mean and variance of the latent variable we use two separate projections on $\mathcal{H}_{latent}$.
\begin{align}
    \vec{\mu}_i &= \text{Linear}^{\mu}(\mathcal{H}_{latent})_i\\
    \vec{\sigma}^2_i & = \text{Linear}^{\sigma^2}(\mathcal{H}_{latent})_i\\
    \mathcal{Z}_i & \sim \mathcal{N}(\vec{\mu}_i, \vec{\sigma}^2_i)
\end{align}
To produce the input scalar and pairwise features to the decoder networks, we take the outer product of the sampled latent representations.
\begin{align}
    \mathcal{Z}^{1D} &= Linear(Z)\\
    \mathcal{Z}^{2D} &= \mathcal{Z}^{1D} \otimes \mathcal{Z}^{1D}
\end{align}
Where $\mathcal{Z}^{1D}$ and $\mathcal{Z}^{2D}$ are the input to the scalar and pairwise feature decoder respectively.
For the architecture, we used [32, 63, 128, 256] number of channels in both the 1D and 2D feature encoder and the inverse for 1D and 2D feature decoder. we used 64 as the latent dimension in our experiments. Models are optimized using Adam with learning rate of $1e^{-3}$
\textbf{Loss function and objectives}\\
Let $\vec{X}_i$ denote the backbone coordinates of residue $v_i$. We predict backbone coordinates of $\{ N_i, C_{\alpha i}, C_i, O_i\}$  for each residue $i$. After predicting coordinates $\{ N_i, C_{\alpha i}, C_i, O_i\}$, we follow \cite{jumper2021highly}, and define the local $C\alpha$ - frame of residue $i$ as the rigid transformation $T_i=(R_i,\vec{t}_i)\in SE(3)$ such that\\
\begin{align} \label{eq:rigid_frames}
\centering
&T_i \circ \vec{C_\alpha}_i  = (0,0,0,0) \\
&T_i \circ \vec{N_i}  = (\Vert \vec{C\alpha_i} - \vec{N_i}\Vert_2,0,0,0) \\
&T_i \circ \vec{C_i}  = (0,\Vert \vec{C\alpha_i} - \vec{C_i}\Vert_2,0,0) \\
&T_i \circ \vec{O_i}  = (0,0,\Vert \vec{C\alpha_i} - \vec{O_i}\Vert_2,0)
\end{align}

where $T \circ \vec{x} \triangleq R \vec{x} + \vec{t}$. Assuming linear independence between the displacement vectors $\vec{C\alpha}_i - \vec{N_i}$ and $\vec{C\alpha_i} - \vec{C_i}$, this transformation is unique and well defined. With a single local frame defined from each residue's predicted coordinates, we are able to apply per-residue frame aligned point error (pFAPE) loss against the native coordinates and local frames as 
\begin{equation}
\text{pFAPE}\left(T_{i},T_{i}^{\ast},\text{\ensuremath{\{\vec{X_j}\},\{\vec{X_j}^{\ast}\};\theta}}\right)
=\frac{1}{\theta}\sum_{j}\min\left(\left\Vert \left(T_{i}\right)^{-1}\circ \vec{X}_j-\left(T_{i}^{\ast}\right)^{-1}\circ \vec{X}_{j}^{*} \right\Vert _{2},\theta\right)\label{eq:pFAPE}
\end{equation}
where $\theta = 10$ is a threshold determining when the loss value should be clamped, and an asterisk is used to differentiate between native and predicted frames and coordinates. The pFAPE loss is averaged over all frames $T_i$ and all atom types $X$ to produce the final loss $\mathcal{L}_{FAPE}$.
For the VAE loss, in addition to the FAPE  reconstruction loss, we also compute the Kullback–Leibler divergence loss
\begin{equation}
    \mathcal{L}_{total} = \mathcal{L}_{FAPE} + \beta D_{KL}(Z||p(Z))
\end{equation}
Where $D_{KL}$ is the KL Divergence between the latent variable distribution and the prior multivariate Normal distribution.\\
\textbf{Fixed backbone sequence design}
We use the pre-trained fixed backbone design model of \citep{dauparas2022robust} to generate sequences from backbone structures. For each test target, we generate 500 conformational decoys and sequences are deranged for each of the decoys independently.\\
\textbf{Structrue prediction oracles}
We use the AF2\citep{jumper2021highly} implemented by ColabFold\citep{mirdita2022colabfold} as well as ESMFold\cite{lin2023evolutionary} for structure prediction, for all structures, we use single sequence without MSA and set the number of recycles to three for all sequences. We make structure prediction to each sequence designed from the conformational decoy library and select sequence based on the prediction results.\\

\subsection{Data}
\textbf{CATH4.2}\\
We obtained the the CATH4.2 data from \cite{ingraham2019generative} which contains 19,752 structures and structurally split with into train/validation/test sets by CATH fold annotations. To ensure sequence and structural independence between the train/test structures, the 40\% non-redundant set is used and split is done on the topology/fold level.\\
\textbf{Antibody Structures}\\
For antibody structures, we used the structural antibody Database (SAbDab)\cite{dunbar2014sabdab} obtained from \cite{jin2021iterative} which contains 1266, 1564, 2325 structures for CDR-H1, CDR-H2, and CDR-H3 respectively after filtering and splitting, there is no more than 40\% sequence identity in the inpainted regions for each set of structures between the train/test structures. Please refer to \cite{jin2021iterative} for further details.

\section{Results}
In this section, We first demonstrate our model's ability to generate realistic backbones structures through comparing with a set of experimentally determined and structurally independent structures. Then we show how the conformational decoys generated by our model outperforms native structure backbones for designing structure conforming amino acid sequences. Lastly, we tested out model's structure design capability by testing CDR inpainting on a set of monoclonal antibodies.\\
\subsection{End-to-End structure reconstruction \& generation}
In this section, we will demonstrate our model's ability to reconstruct high-quality protein 3-dimensional structures from invariant protein representations. The most crucial capabilities for structure generative models is to construct topologically feasible protein conformations such that the generated backbone structures are viable candidates for downstream design pipelines. For this propose, we presented structural similarity metrics including the local distance difference test (lddt) score\citep{mariani2013lddt}, TM-score\citep{zhang2004scoring}, root mean square deviation (RMSD), and the $L_1$ norm of inter-residue distance between the reconstructed structures and the native input structures to measure how well our models reconstruct the input structures.

\begin{figure}[H]
    \centering
    \includegraphics[width=0.9\linewidth]{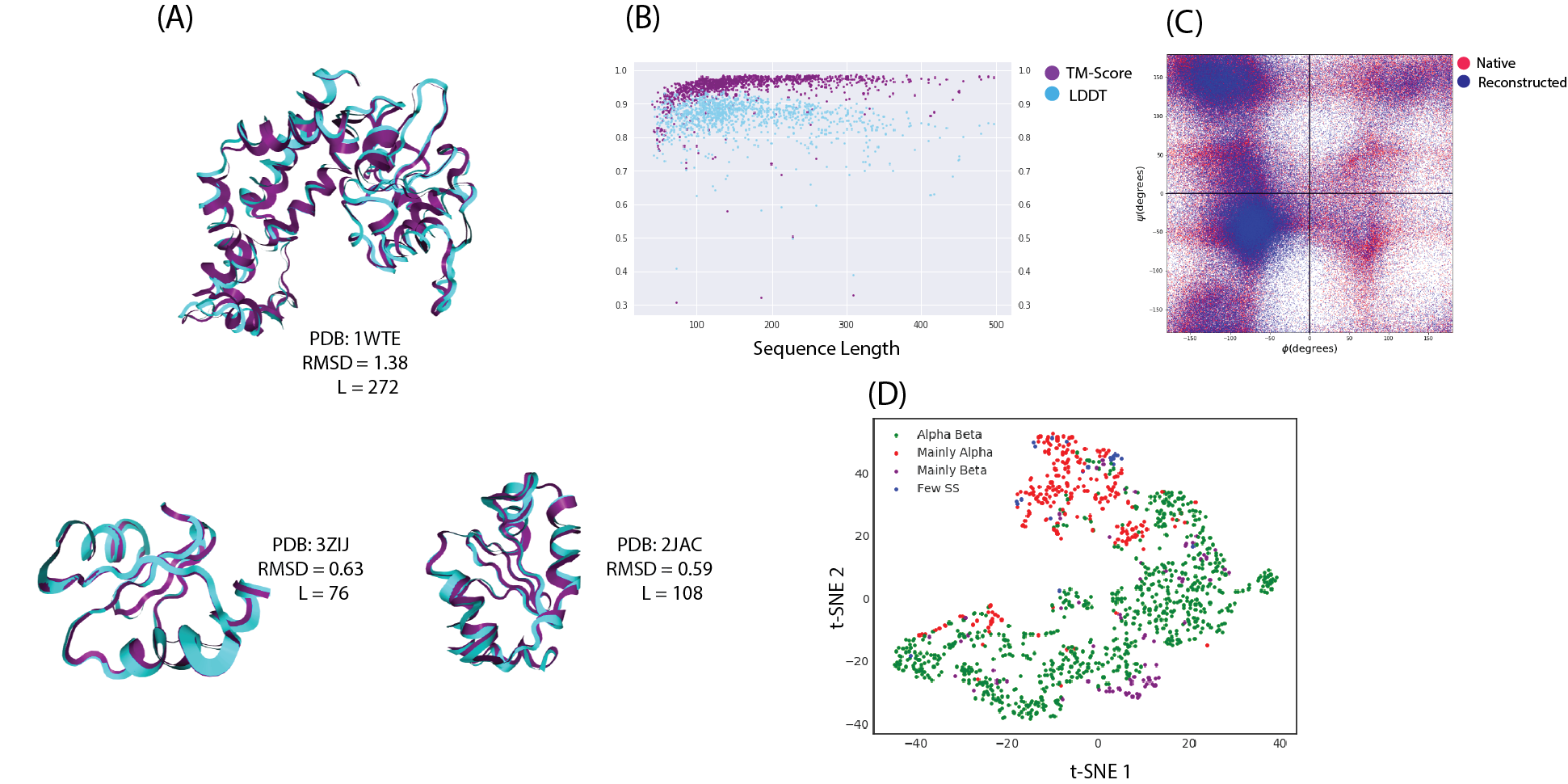}
    \caption{(A) Examples of reconstructed backbone structures of different folds and sizes. 1WTE (top), 3ZIJ (bottom left), 2JAC (bottom right). (B) LDDT (blue) and TM-score (purple) against native structure vs. the length of the test target. (C) Ramachandran plot of the native (red) and reconstructed(blue) structures. (D) t-SNE embedding of the latent space for the test targets colored by their CATH class annotation.}
    \label{fig:C1_Fig2}
\end{figure}

Our model is able to consistently construct high-quality structure coordinates of the target proteins up to $L=500$ residues as shown in Figure \ref{fig:C1_Fig2}(A \& B). To further inspect the geometric feasibility of the reconstructed proteins, we examine the torsion angle distributions and compared it to the native structures as shown in Figure \ref{fig:C1_Fig2}(C) where we notice the torsion angles from the generated structures are mostly in the feasible regions and matches up closely to the native torsion angels. To illustrate how our model's learned latent representation of the input protein contains useful information in its fold space, we plot the t-SNE embedded components colored by its CATH class annotations in Figure \ref{fig:C1_Fig2}(D). Structures that share the same CATH fold class are clustered in the latent space accordingly.

In summary, our model demonstrates remarkable consistency in generating high-quality structural decoys across diverse fold classes and protein sizes. The reconstruction accuracy of protein structures not included in the training sets establishes a lower bound for our model's performance, consistently achieving backbone RMSD values around 1Å. This level of accuracy, combined with the model's broad applicability, underscores its robustness and flexibility compared to prior methods. 

\subsection{Ablation study on input features}
To understand how different input features impact model performance, we compared our model with different features ablated in table \ref{tab:table1}. When both inter-residue distance and orientation are used along with the amino acid sequence, our model achieved average RMSD of 1.122 and TM-score of 0.941 at experimentally comparable resolution. We notice model performance degrades when we remove the inter-residue orientation as input features. This suggests our model relies on spatial orientation information to accurately construct the structure coordinates. It is worth noting that even tough our model achieves the best performance with amino acid sequence input, our model has comparable result when operate under sequence free mode for backbone-only decoy generation. 
\begin{table}[h]
\centering
\begin{tabular}{ccccc}
\hline
Model                    & lddt↑ & TM↑   & RMSD↓ & Distance L1↓ \\ \hline
CoordVAE CNN  & 0.472 & 0.420 & 9.049 & 2.666        \\
CoordVAE w/o orientation & 0.604 & 0.649 & 4.429 & 1.413        \\
CoodVAE w/o seq          & 0.788 & 0.905 & 1.489 & 0.699        \\ \hline
CoordVAE &  \textbf{0.841} & \textbf{0.941} & \textbf{1.122} & \textbf{0.502} \\ \hline
\end{tabular}
\caption{Average structural similarity metrics evaluated on the test targets over different models}
\label{tab:table1}
\end{table}

In addition, we also compared the short, medium, long range contact accuracy where the predicted contacts are derived from the inter-residue distance of the generated 3-dimensional coordinates as shown in Table 2.2\\
\begin{table}[h]
\centering
\begin{tabular}{cccc}
\hline
Model                    & Contact(S)↑ & Contact(M)↑   & Contact(L)↓ \\ \hline
CNN Baseline             & 0.321 & 0.3.5 & 0.193        \\
CoordVAE w/o orientation & 0.500 & 0.436 & 0.378         \\
CoodVAE w/o seq          & 0.875 & 0.833 & 0.785         \\
CoordVAE & \textbf{0.885} & \textbf{0.857} & \textbf{0.819}\\ \hline
\end{tabular}
\caption{Average short(S), medium(M), long(L) contact accuracy evaluated on the 1,120 test targets over different models}
\label{tab:table2}
\end{table}
We benchmark the contact accuracy across different models, evaluated on 1,120 test targets. The results demonstrate a clear progression in performance from the CNN Baseline to the full CoordVAE model. The CNN Baseline shows the lowest accuracy, with values of 0.321, 0.305, and 0.193 for short, medium, and long-range contacts respectively. The CoordVAE without orientation significantly improves upon this, achieving accuracies of 0.500, 0.436, and 0.378. Further improvement is seen with the CoordVAE without sequence information, which reaches high accuracies of 0.875, 0.833, and 0.785. The full CoordVAE model, incorporating both orientation and sequence information, demonstrates the best performance across all contact ranges, with the highest accuracies of 0.885, 0.857, and 0.819 for short, medium, and long-range contacts respectively. This consistent improvement across all contact ranges, particularly for the challenging long-range contacts, suggests that the full CoordVAE model effectively captures complex structural relationships in proteins despite no loss function was imposed on the contact map. The results highlight the importance of both orientation and sequence information in accurately predicting protein contacts.\\

\subsection{Conformational decoy sampling for robust protein sequence design}
\begin{figure}[H]
    \centering
    \includegraphics[width=0.9\linewidth]{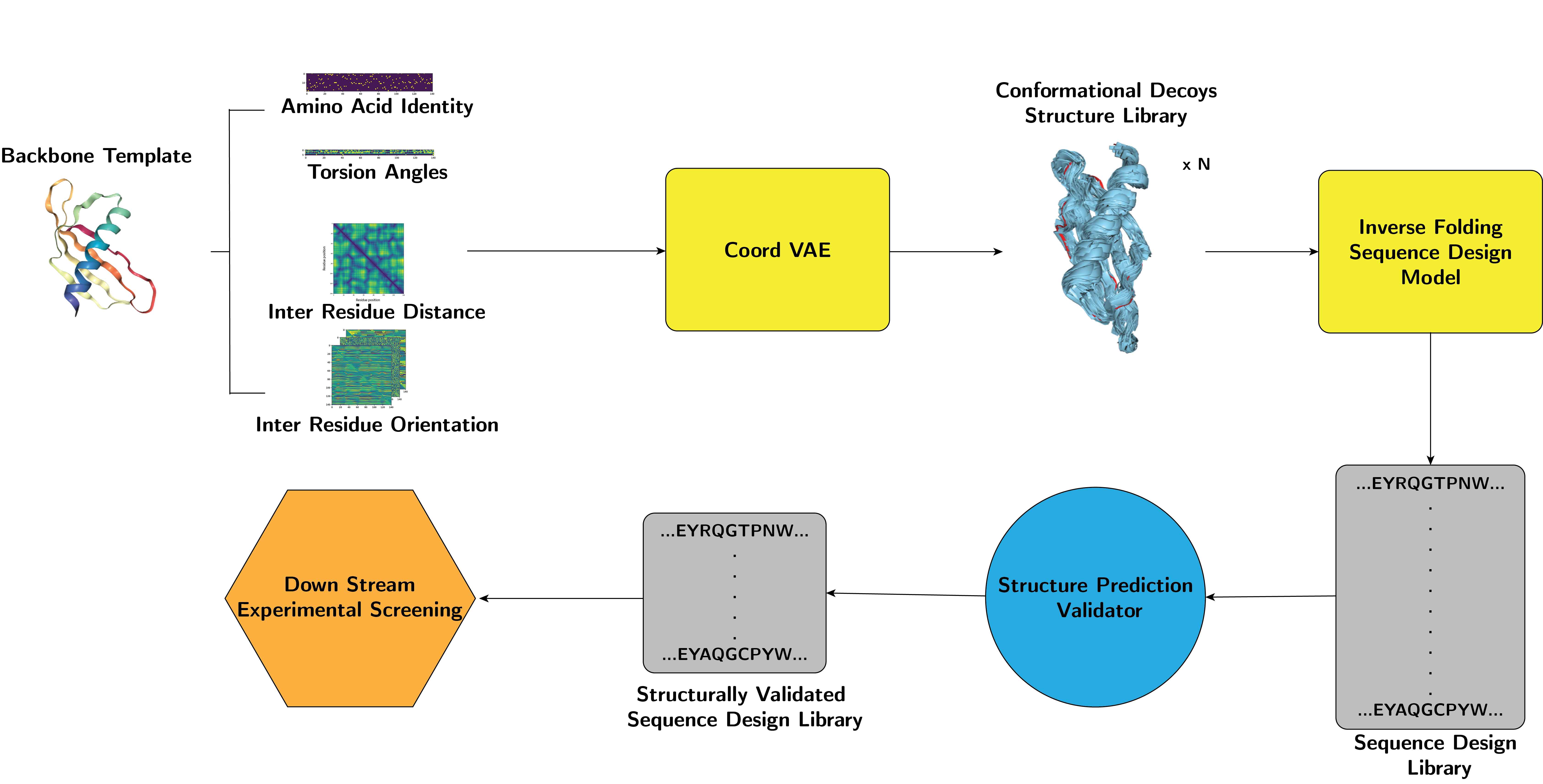}
    \caption{Outline of robust protein design procedure using conformational decoy library. Starting with a primary backbone structure target, our structure generative model embeds it into a latent space and decodes into conformational decoys to form a structure library. Using a fixed backbone sequence design program on the backbone structure library, one can obtain a preliminary sequence library. To filter and structurally validate the primary sequence library, a structure prediction oracle is used and the validated sequence library is ready for further downstream tasks.}
    \label{fig:C1_Fig3}
\end{figure}

In this section, we describe how our structural generative model can be used for robust \textit{de novo} protein design. The design pipeline using conformational decoys is outlined in Figure \ref{fig:C1_Fig3}. A primary design target backbone structure is provided as the design template, such a template can be generated by topology design program such as TopoBuilder\citep{sesterhenn2020novo} or other template based methods. Our structure generative model(CoordVAE) will embed the input backbone into a latent representation then a set of conformational decoys is generated by the structure decoder to form a library. A sequence library is then produced by running a fixed backbone inverse folding program on the decoy library. Each sequence will then be folded and validated by a structure prediction oracle such as AlphaFold\citep{jumper2021highly}, and those which pass the structure validation criteria will be used to form a structurally validated sequence library for further filtering and downstream experimental screening. Our structure model can generate thousands of decoys per minute on a single GPU for fast and efficient library generation.

To elucidate how sequences designed from the conformational decoys compare to sequence designed with native one shot backbone templates, we designed an experiment such that we designed sequences from both a backbone decoy library generated from our structural model and single shot experimentally determined backbones from PDB and we computationally predict the structure of sequences designed from both set and see which set can produce more computationally confident sequences.

\begin{figure}[H]
    \centering
    \includegraphics[width=0.99\linewidth]{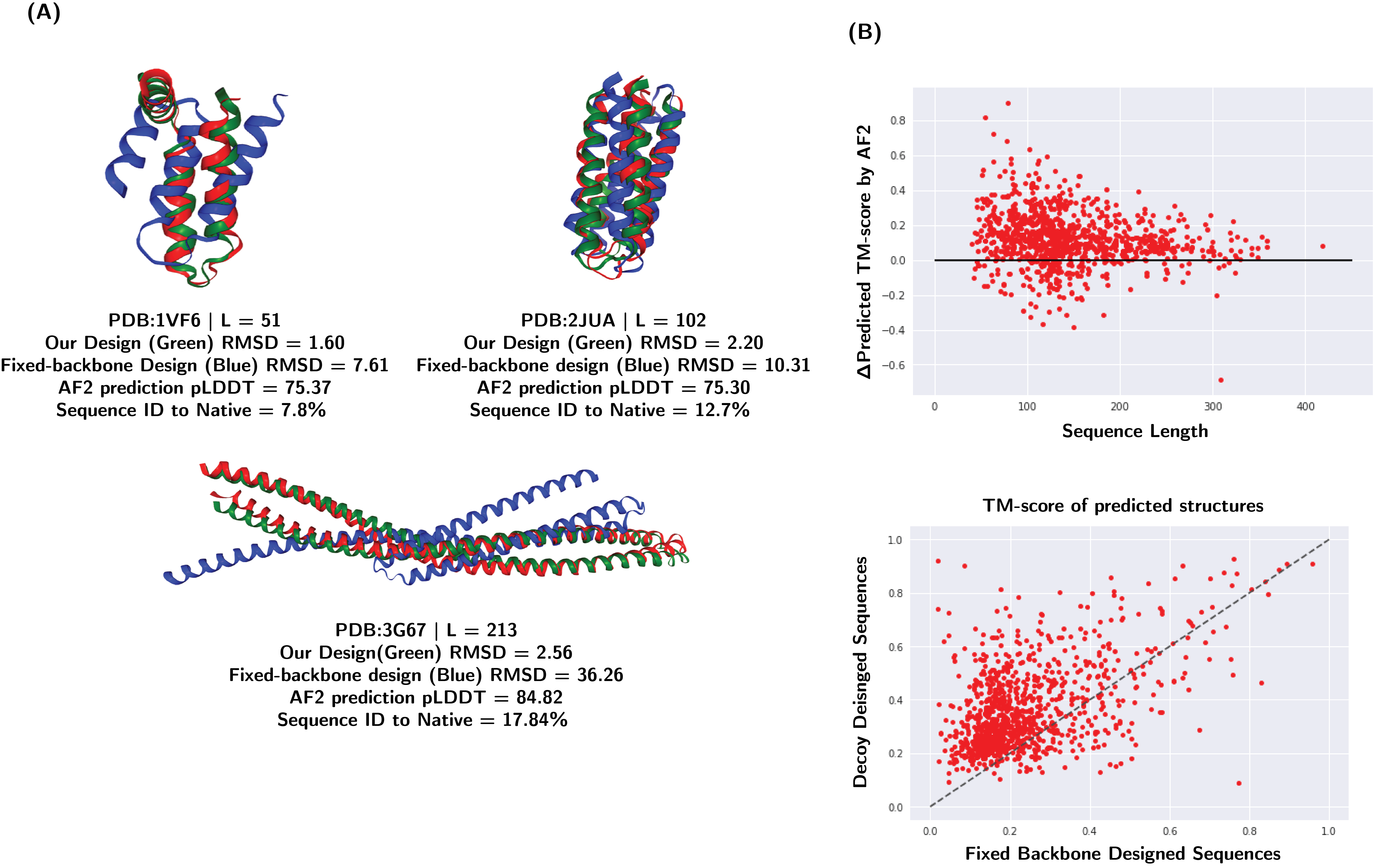}
    \caption{Examples of protein designed from conformational decoys. \textbf{(A)} AF2 folded sequence designed from conformational decoys. Overlay of AF2 predicted structure of decoy designed sequence \& native backbone designed sequence (green \& blue), native backbone(red) for PDB:1VF6 (top left), PDB:2JUA (top right), PDB:3G67 (bottom). \textbf{(B)} Scatter plots of TM-scores between the best folded conformational decoy designed sequences and sequences designed from the native backbones(bottom). $\Delta$ TM-score between decoy designed sequences and fixed backbone designed sequences versus the target size.(top)}
    \label{fig:C1_Fig4}
\end{figure}
We provide examples in Figure \ref{fig:C1_Fig4}(A) where our designed sequence folded more successfully than the native sequence when MSA and template information is not available. Particularly, as shown for PDB:3G67, the native sequence is folded with RMSD = 36.26 while our best decoy designed sequence folded with RMSD = 2.56 with high confidence. We observed that our model can consistently improve designed quality regardless of the size of the target backbone.  In Figure \ref{fig:C1_Fig4} (B), we deployed the design pipeline to a set of 211 backbone targets and plotted the TM-score of the predicted structure compared to native sequence(bottom), and sequence designed from the native backbone structure(top). Sequences designed from our pipeline outperform sequences designed from fixed native backbones in 856 out of the 1,016 tested targets(84\%).

 \begin{figure}[H]
    \centering
    \includegraphics[width=0.99\linewidth]{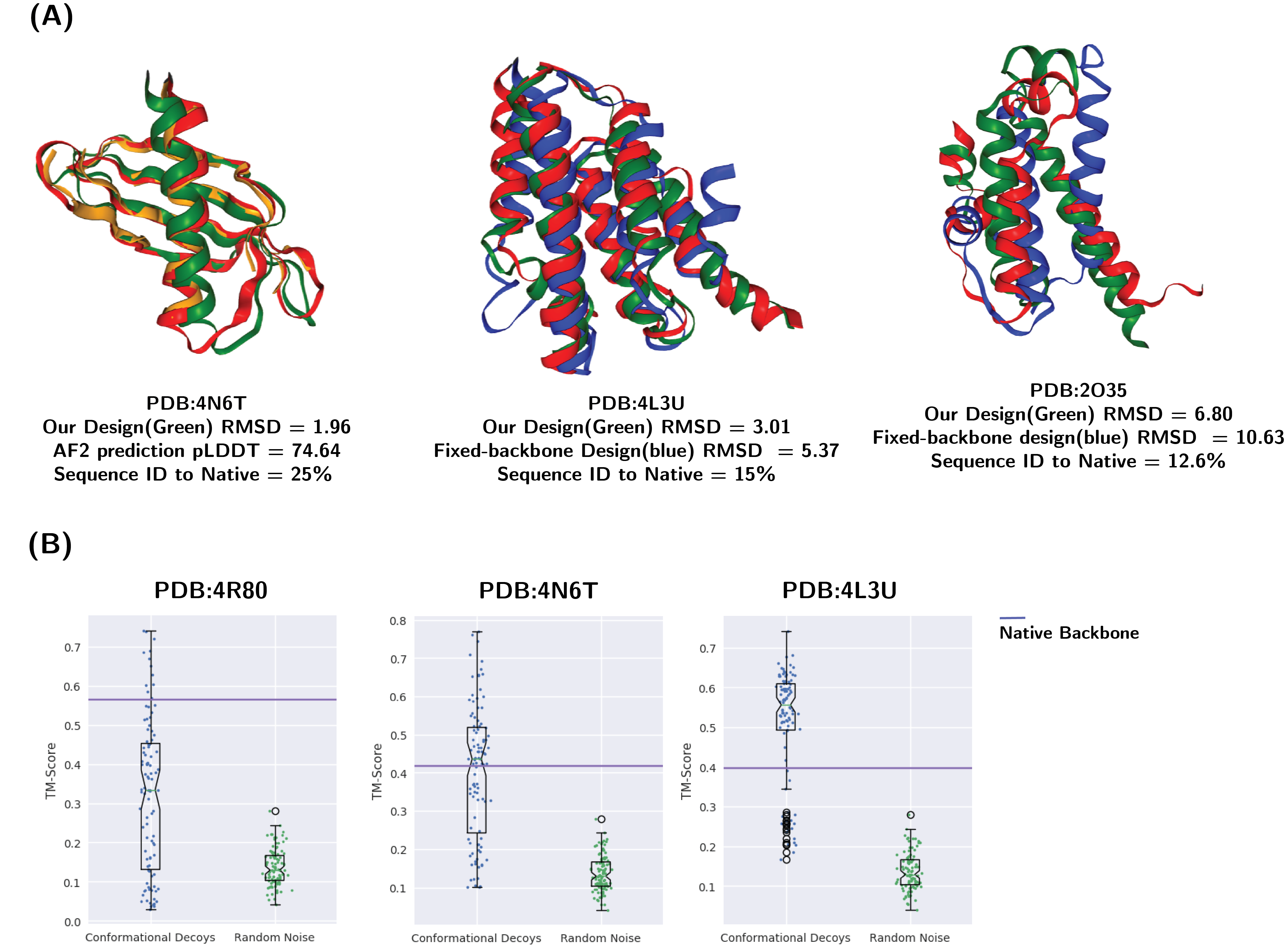}
    \caption{\textbf{(A)} Examples of AF2 predicted designed sequences and native sequences. \textbf{(B)} TM-score distributions of decoy designed sequences vs. noisy backbone designed sequences.}
    \label{fig:C1_Fig5}
\end{figure}
To demonstrate our approach can simultaneously drive sequence diversity and design confidence, we showed more examples of conformer based sequence design at Fig \ref{fig:C1_Fig5}(A). To confirm that our structure generative model does not simply generate noisy versions of the input structure, we compared with sequences designed from backbone coordinates with added random Gaussian noise. As shown in Figure \ref{fig:C1_Fig5}(B), the results for conformatioal decoys are clearly favorable to those of noisy backbones. 
\begin{figure}[H]
    \centering
    \includegraphics[width=0.99\linewidth]{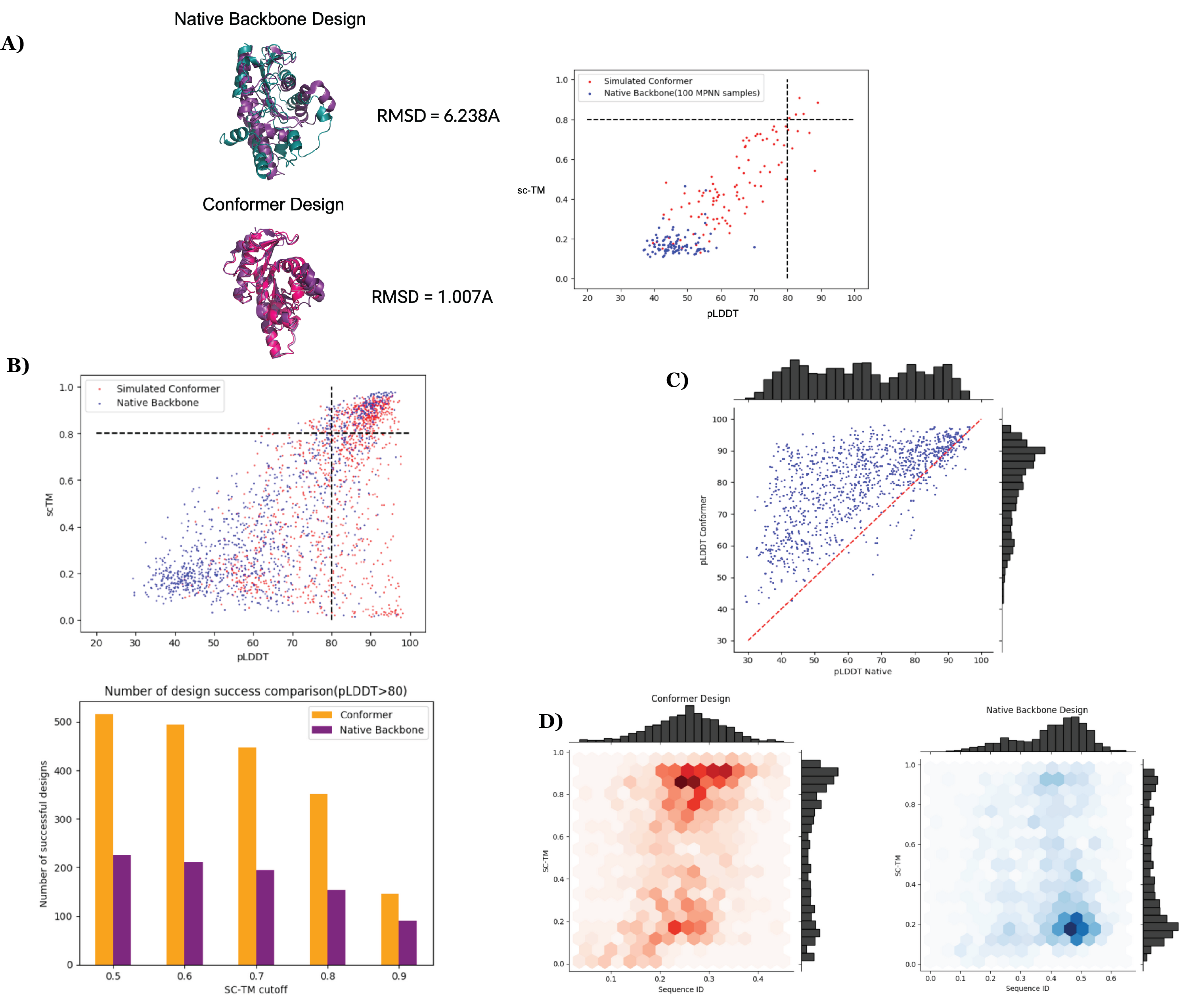}
    \caption{A) Example of backbone ensemble sampling vs. sequence design sampling of PDB:1JDI. B) Self-consistent TM score comparison between backbone ensemble sampling and single shot native backbone design(Top). Design success rate across different sc-TM threshold comparison(Bottom). C) pLDDT of designed sequence comparison between conformer sampling vs. single shot native backbone. D) sc-TM distribution vs. sequence identity distribution of the conformer based sequence design(Left). sc-TM distribution vs. sequence identity distribution of the native backbone sequence design(Right)}
    \label{fig:C1_Fig6}
\end{figure}

To further demonstrate our deep structure generative model presents an edge in providing diverse backbone ensembles that can overcome limitation in inverse folding methods relying on single-shot native backbones. Fig. \ref{fig:C1_Fig6}(A) illustrates an example where proteinMPNN temperature-based sequence sampling with a native backbone failed to generate structure-conforming sequences, while our structure ensemble approach yielded many successful designs. The advantages of our method is further evidenced in Fig. \ref{fig:C1_Fig6}(B), which shows the distribution of sc-TM scores versus pLDDT across all tested targets. Notably, our conformer-based designs (red points) achieved a higher concentration of successful designs in the upper-right quadrant compared to single-shot native backbone templates (blue points). The bar chart below quantifies this advantage, showing that conformer-based designs consistently outperform native backbone designs across various sc-TM cutoffs, with approximately twice the success rate in many cases. Fig. \ref{fig:C1_Fig6}(C) further supports this conclusion by showing that our conformer-based designs (blue points) consistently achieve higher pLDDT scores compared to native backbone designs across the majority of tested folds.\\

A particularly important finding is illustrated in Fig. \ref{fig:C1_Fig6}(D), which reveals the relationship between sequence diversity and design success. Contrary to the conventional wisdom that suggests a trade-off between sequence diversity relative to native proteins and design success rate, our method demonstrates the ability to simultaneously optimize for both. The conformer-based approach (left heatmap) generates much more diverse sequences while increasing design success, as evidenced by the higher density of points in the upper-left region compared to the native backbone design (right heatmap). This is a significant improvement, as it allows for broader exploration of the sequence space without compromising designability.\\
These results underscore the power of our deep structure generative model in enhancing protein design capabilities. By providing diverse backbone ensembles, our approach not only improves the success rate of inverse folding but also expands the accessible sequence space, potentially leading to the discovery of novel protein designs with unique properties and functions.\\
In summary, we show that conformational decoy ensembles generated by our model can be used to improve protein design pipeline by providing a structure library for downstream fixed-backbone sequence design applications therefore significantly increase the available sampling space. Our experiments demonstrate that structure-conforming sequences can be reliably designed from the conformational decoys compared to single backbone targets. With this approach, we present a new path towards robust and efficient protein design.\\

\subsection{Unconditional structure inpainting for antibody design}
Monoclonal antibodies are important targets for therapeutics development and considerable effort has been dedicated by the community towards computational antibody design\citep{pinto2020cross,alford2017rosetta,eguchi2022ig,adolf2018rosettaantibodydesign}. While previous methods mostly focused on sequence design, we adopt our structure generative models to inpaint the complementarity-determining regions (CDRs) for structure based antibody design. CDR grafting is a a fundamental technique in antibody engineering which transplants the CDR region from one antibody to another. However, it comes with significant challenges that can impact the success and efficiency of the process due to the delicate nature of antibody binding characteristics. In this section we show how our model can be used for self-consistent structure design with a hybrid mask scheme and provide a direct comparison with existing structure design methods.

To inpaint protein structures, we employed three types of masks; linear, spatial and random and a hybrid masking strategy(see details in the Method section) that impose masks in the input scalar and pair representation and the completed structure will be recovered by the structure generative model. To further test the ability of our model to distill meaningful representation from a larger fold space, we pre-traiend our model on the CATH4.2 dataset and fine-tune the model with a set of monoclonal antibody structures.

\begin{table}[H]
\centering
\resizebox{\textwidth}{!}{
\begin{tabular}{cccccccc}
\hline
 &
   &
  \multicolumn{2}{c}{CDR-H1} &
  \multicolumn{2}{c}{CDR-H2} &
  \multicolumn{2}{c}{CDR-H3} \\ \hline
\multicolumn{1}{c|}{Model} &
  \multicolumn{1}{c|}{Training Data} &
  lddt↑ &
  \multicolumn{1}{c|}{RMSD↓} &
  lddt↑ &
  \multicolumn{1}{c|}{RMSD↓} &
  lddt↑ &
  RMSD↓ \\ \hline
\multicolumn{1}{c|}{CoordVAE(10AA linear)} &
  \multicolumn{1}{c|}{} &
  0.587 &
  \multicolumn{1}{c|}{2.847} &
  0.588 &
  \multicolumn{1}{c|}{2.864} &
  0.498 &
  4.427 \\
\multicolumn{1}{c|}{CoordVAE(spatial)} &
  \multicolumn{1}{c|}{} &
  0.515 &
  \multicolumn{1}{c|}{3.944} &
  0.514 &
  \multicolumn{1}{c|}{3.914} &
  0.487 &
  4.350 \\
\multicolumn{1}{c|}{CoordVAE(Random)} &
  \multicolumn{1}{c|}{} &
  0.447 &
  \multicolumn{1}{c|}{4.216} &
  0.534 &
  \multicolumn{1}{c|}{4.152} &
  0.450 &
  5.185 \\
\multicolumn{1}{c|}{CoordVAE(mixed mask)} &
  \multicolumn{1}{c|}{\multirow{-4}{*}{CATH4.2}} &
  0.591 &
  \multicolumn{1}{c|}{2.903} &
  0.575 &
  \multicolumn{1}{c|}{3.223} &
  0.530 &
  3.889 \\ \hline
\multicolumn{1}{c|}{CoordVAE} &
  \multicolumn{1}{c|}{} &
  0.81 &
  \multicolumn{1}{c|}{1.55} &
  0.872 &
  \multicolumn{1}{c|}{1.00} &
  0.809 &
  1.55 \\
\multicolumn{1}{c|}{\textbf{CoordVAE(Transfer)}} &
  \multicolumn{1}{c|}{} &
  \textbf{0.88} &
  \multicolumn{1}{c|}{\textbf{0.81}} &
   \textbf{0.90} &
  \multicolumn{1}{c|}{ \textbf{0.85}} &
   \textbf{0.823} &
   \textbf{1.35} \\
\multicolumn{1}{c|}{AR-GNN} &
  \multicolumn{1}{c|}{} &
  N/A &
  \multicolumn{1}{c|}{2.97} &
  N/A &
  \multicolumn{1}{c|}{2.27} &
  N/A &
  3.63 \\
\multicolumn{1}{c|}{RefineGNN} &
  \multicolumn{1}{c|}{\multirow{-4}{*}{SABDAB}} &
  N/A &
  \multicolumn{1}{c|}{1.18} &
  N/A &
  \multicolumn{1}{c|}{0.87} &
  N/A &
  2.50 \\ \hline
\end{tabular}}
\caption{Structure inpainting performance on the test monoclonal antibody dataset in CDR-H1(left), CDR-H2(middle), CDR-H3(right) across different models.}
\label{tab:table2}
\end{table}
\begin{figure}[H]
    \centering
    \includegraphics[width=1\linewidth]{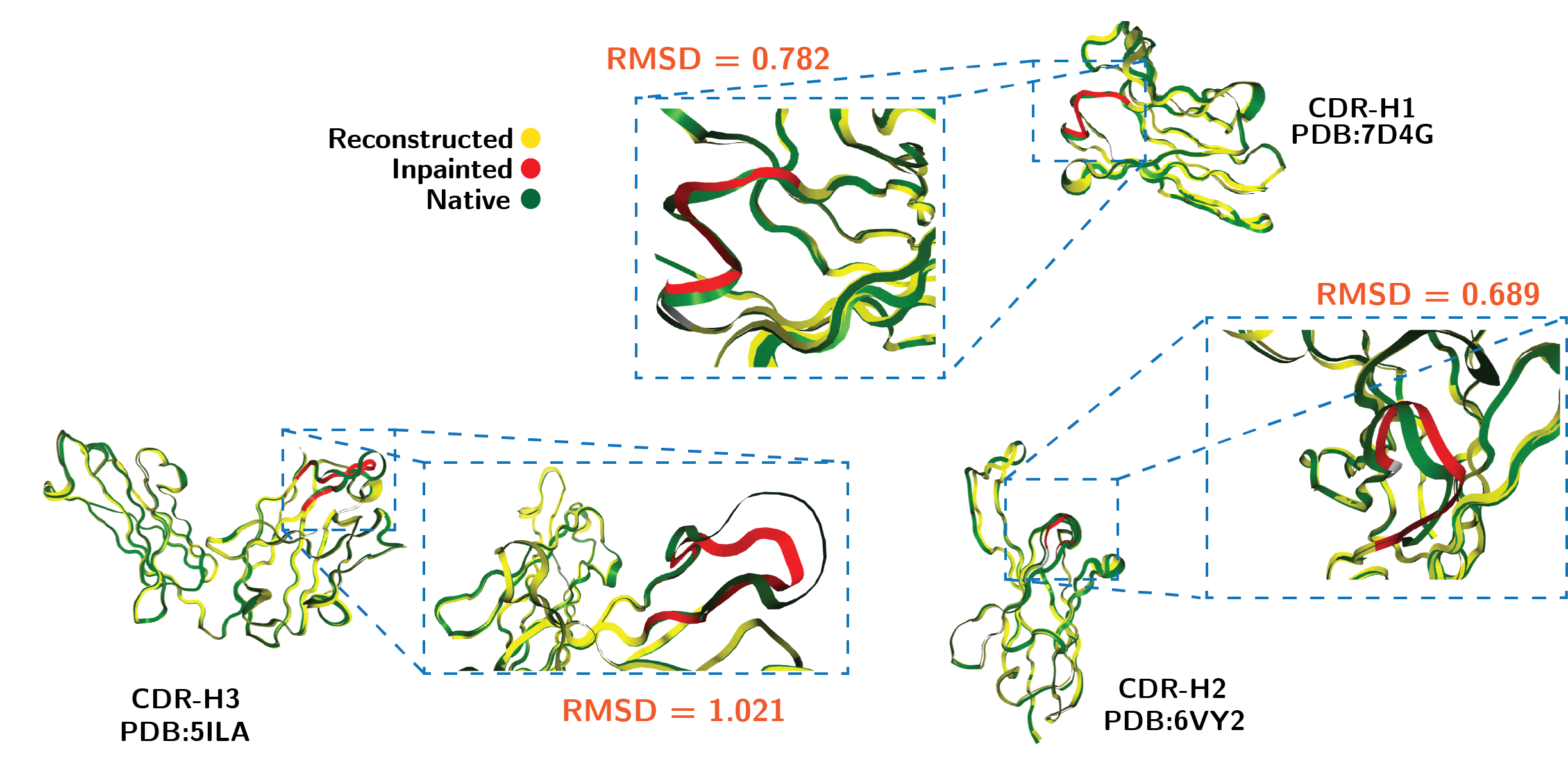}
    \caption{Examples of antibody structure inpainting. overlay of the CDR-H3 region of PDB:5ILA (bottom left) native structure(green), unmasked region (yellow) and masked inpainting region (red) reconstruction with RMSD = 1.021. overlay of the CDR-H2 region of PDB:6VY2 (bottom right) native structure(green), unmasked region (yellow) and masked inpainting region (red) reconstruction with RMSD = 0.689. overlay of the CDR-H1 region of PDB:7D4G (top) native structure(green), unmasked region (yellow) and masked inpainting region (red) reconstruction with RMSD = 0.782}
    \label{fig:C1_Fig7}
\end{figure}

We first tested our model's inpainting performance on three CDRs(H1, H2, H3) with models trained on the CATH4.2 dataset without fine-tuning on antibody structures. Our best model trained on hybrid masking scheme achieved RMSD of 2.903, 3.223, and 3.889 on CDR-H1, CDR-H2, and CDR-H3 respectively as shown on table \ref{tab:table2}. Next, we test models trained on antibody structure with masked CDRs, without CATH4.2 pre-training, our model achieved RMSD of 1.55, 1.00, 1.55 respectively. With model pre-trained on the CATH4.2 dataset and fine-tuned on antibody structure with masked CDRs, our model achieved RMSD of 0.81, 0.85, 1.35 compare to state-of-the-art method RefineGNN\citep{adolf2018rosettaantibodydesign} with RMSD of 1.18, 0.87, 2.50 respectively. To avoid information leakage, we filtered the CATH4.2 dataset with the antibody test sets for redundancy in pre-training.

To summarize, our structure generative model can be adopted to perform structure inpainting by properly masking the input features. Through our experiment, we show the hybrid masking scheme can improve design performance and our model can adequately inpaint antibody CDRs. With pre-training on large structure databases, our model outperforms state-of-the-art antibody structure design models and we observe that our model has the largest improvement on longer inpainting regions(CDR-H3). In this chapter, we only performed unconditional CDR structure generation in this study with our structure generative model. We will further explore conditional CDR structure generation along with sequence design experiments in next chapter.\\ 

\section{Discussion}
This chapter of the thesis presents a novel structural generative model, CoordVAE, for protein design that addresses key limitations in existing computational approaches. The model demonstrates high-quality reconstruction of protein backbone structures, generates diverse conformational ensembles, and enables robust sequence design through a decoy sampling approach. Additionally, it shows promising capabilities in structure inpainting for antibody design. These results have several important implications for the field of computational protein design and suggest exciting directions for future work.

One of the key innovations of CoordVAE is its ability to directly model protein structures in three-dimensional coordinate space while addressing rotational and translational equivariance. This allows the model to generate realistic protein conformations without relying on intermediate topological constraints or downstream coordinate recovery steps. The high-quality reconstructions achieved across a range of protein sizes and folds, as evidenced by results shown above, demonstrate the model's ability to capture complex structural relationships. The clustering of latent representations by CATH fold classes further indicates that the model has learned meaningful structural embeddings.

The conformational decoy sampling approach enabled by CoordVAE represents a significant advance for fixed-backbone sequence design. By generating ensembles of backbone conformations, this method expands the search space for compatible sequences beyond what is possible with a single fixed template. The improved performance of sequences designed from decoy ensembles compared to native backbones, as validated by structure prediction, suggests this approach can lead to more designable and stable protein sequences. This could have major implications for \textit{de novo} protein design efforts, potentially increasing success rates and expanding the range of achievable functions.

The model's success in antibody CDR inpainting, particularly when pre-trained on a diverse protein dataset, highlights its potential for structure-based antibody engineering. The ability to generate plausible CDR conformations while maintaining the overall antibody framework could streamline antibody design processes and potentially lead to improved therapeutic candidates. The superior performance compared to existing methods, especially for longer CDR regions, is particularly promising.

Several aspects of the model design contributed to its strong performance. The use of both distance and orientation information in the input features proved crucial for accurate coordinate reconstruction. The hybrid masking scheme during training likely enhanced the model's ability to handle incomplete structural information, as evidenced by its success in inpainting tasks. The effectiveness of transfer learning from a diverse protein dataset to antibody-specific tasks suggests the model captures generalizable principles of protein structure.

While the results are promising, there are limitations and areas for future investigation. The current model focuses on backbone structure generation and does not directly address side chain packing or non-protein components like ligands or cofactors. Extending the model to incorporate these elements could further improve its utility for real-world design challenges. Additionally, while the conformational decoy approach shows clear benefits, further work is needed to understand the optimal sampling strategies and to develop methods for efficiently filtering and selecting the most promising decoys.

In conclusion, this work represents a step forward in computational protein design, introducing a versatile structural generative model that addresses key limitations of existing approaches. The demonstrated capabilities in structure reconstruction, conformational decoy sampling, and structure inpainting open up new possibilities for robust and efficient protein design. As the field continues to advance, integrating such generative models with other computational and experimental techniques promises to accelerate the development of novel proteins for a wide range of applications in biotechnology and medicine.

In the following chapters, I will employ the structural generative model developed and explore \textit{de novo} protein design and protein optimization through a iterative framework and applying the model to challenging real-world protein design problems and validating designs experimentally.  Extending the model to handle multi-chain protein complexes and further antibody design. However, other potential directions for future work includes integrating the structural generative model more tightly with sequence design algorithms, potentially in an end-to-end differentiable framework and exploring the use of the learned structural embeddings for other tasks such as function prediction or protein-protein interaction modeling.

\chapter{Adaptive \textit{de novo} Protein Design via iterative sequence structure co-optimization}
\section{Motivation}
In the first chapter, we developed a deep structure generative model for protein structure generation. In this chapter we will explore how we can embed our model in a iterative design framework for robust and efficient \textit{de novo} protein design. 

Computational protein design has emerged as a powerful tool for creating novel proteins with targeted functional attributes, offering immense potential in fields ranging from therapeutics to industrial enzymes\cite{huang2016coming,kortemme2024novo,yeh2023novo}. However, despite significant progress, the gap between computationally designed proteins and those that perform successfully in experimental settings remains substantial\cite{gainza2018computational}. This challenge stems from the complex nature of protein folding and function, which involves a vast design space and intricate relationships between sequence, structure, and activity.

Current approaches to \textit{de novo} protein design pipeline often rely on one-shot design with generated backbone templates\cite{yeh2023novo,watson2023novo} which constraints the design space and therefore relatively low success rate. The complexity of protein design often requires satisfying multiple, sometimes competing objectives simultaneously, such as stability, solubility, and specific functional properties\cite{rocklin2017global}. These multi-objective optimization goals are difficult to meet in a single design cycle, motivating the need for an iterative approach.

In this chapter of the thesis, we will outline the proposed iterative algorithm inspired by ideas from directed evolution\cite{arnold2018directed} and evolutionary design algorithms such as genetic algorithm and evolutionary programming\cite{sivanandam2008genetic}, detail its integration with our structure generative model, and discuss strategies for incorporating various forms of feedback into the design process. We will also present case studies demonstrating the effectiveness of this adaptive approach in addressing challenging protein design problems, with the goal of advancing the field towards more reliable and efficient computational protein engineering.
\section{Introduction}
Protein design and engineering have become an indispensable tool in biotechnology, enabling the development of novel enzymes, therapeutics, and biomaterials with enhanced or completely new functions. One of the most successful approaches in this field has been directed evolution, a method that mimics natural evolution in a laboratory setting to optimize protein properties\cite{arnold2018directed}. Directed evolution typically involves iterative cycles of genetic diversity generation followed by screening or selection for desired traits. This approach has led to numerous breakthroughs, including the development of enzymes with improved catalytic efficiency, stability, and even novel functions\cite{packer2015methods}.

Despite its successes, directed evolution faces several challenges. The method is often labor-intensive and time-consuming, requiring the screening of vast libraries of mutants and often the fitness measure is technically sophisticated. Moreover, the random nature of mutations means that fitness advancing variants are rare, and multiple rounds of evolution are typically necessary to achieve significant improvements \cite{arnold2018directed}. The method also struggles with navigating complex fitness landscapes, where beneficial mutations may be separated by fitness valleys that are difficult to traverse through random mutagenesis alone\cite{poelwijk2007empirical}.

To address these limitations, researchers attempted to leverage machine learning techniques to accelerate directed evolution. This approach uses computational models to guide the evolution process, potentially reducing the number of variants that need to be experimentally tested and accelerating the discovery of improved proteins\cite{yang2019machine}. Machine learning algorithms can learn from previous experimental data to predict which mutations are likely to be beneficial, thereby focusing the search on more promising regions of the fitness landscape\cite{mazurenko2019machine}. 

While machine learning-assisted directed evolution has shown promise, recent years have seen a surge in interest in purely computational protein design methods, particularly those based on deep generative models\cite{watson2023novo,yeh2023novo,wang2022scaffolding}. These approaches aim to design proteins \textit{de novo} or to predict beneficial mutations without the need for iterative experimental testing.

However, these methods also face challenges. The predictions made by deep learning models can sometimes be difficult to interpret and often unreliable compared to experimental standard. where it's not always clear why certain designs are predicted to be successful and the generated backbones have limited designability. Additionally, while these models can generate numerous designs \textit{in silico}, experimental validation is still necessary to confirm their function, and the success rate of purely computational designs remains lower than that of evolutionarily refined proteins\cite{huang2016coming}.

Moreever, recent years have witnessed remarkable advancements in deep learning-based approaches for protein structure prediction and molecular simulation. The release of AlphaFold2 by DeepMind in 2020 marked a watershed moment in protein structure prediction, achieving unprecedented accuracy in the Critical Assessment of protein Structure Prediction (CASP) competition\cite{jumper2021highly}. This was quickly followed by other powerful models such as RoseTTAFold\cite{baek2021accurate} and ESMFold\cite{lin2023evolutionary}. These breakthroughs have dramatically expanded our ability to model protein structures \textit{in silico}. In the field of small molecule docking, deep learning methods have also made significant strides. Models like AtomNet\cite{wallach2015atomnet} and DeepDock\cite{liao2019deepdock} have demonstrated improved accuracy and speed compared to traditional docking algorithms. More recently, end-to-end differentiable docking models like EquiBind\cite{stark2022equibind}  and DiffDock\cite{corso2022diffdock} have emerged, offering the potential for gradient-based optimization of binding poses. However, while these methods generally performs better in known ligands and computationally predicted protein structures, it still struggles to extend to novel ligand and protein pairs and high resolution local atomic interactions\cite{buttenschoen2024posebusters}. In practice, We found energy based models such as AutoDock Vina\cite{eberhardt2021autodock} generates more physically feasible docking poses. 

In an effort to harness the full potential of our structure generative model for practical \textit{de novo} protein design and to enhance the robustness and efficiency of current protein engineering pipelines, we have developed an innovative iterative design framework. This framework adaptively optimizes multiple \textit{in silico} fitness objectives within a flexible and modular pipeline. The versatility and power of our proposed framework are demonstrated through its ability to:
\begin{itemize}
    \item Design \textit{de novo} structures that are distinct from those found in current structural databases, thereby expanding the protein fold space.
    \item Optimize existing functional proteins with subsequent experimental validation, bridging the gap between computational predictions and real-world performance.
    \item Engineer scaffolds for functional proteins, resulting in the creation of \textit{de novo} functional proteins with tailored properties.
    \item Structure based conditional \textit{de novo} CDR design for antibody engineering.
\end{itemize}

\section{Literature Review}
In this Chapter, we will challenge our iterative design framework that utilizes previously developed structure generative model to an array of real-world design tasks such as unconditional protein structure generation, small-molecule binding . In this section, I will briefly describe those tasks and provide the context for our experiments.\\ \\
\subsection{Structure based protein design}
\begin{figure}[H]
    \centering
    \includegraphics[width=0.6\linewidth]{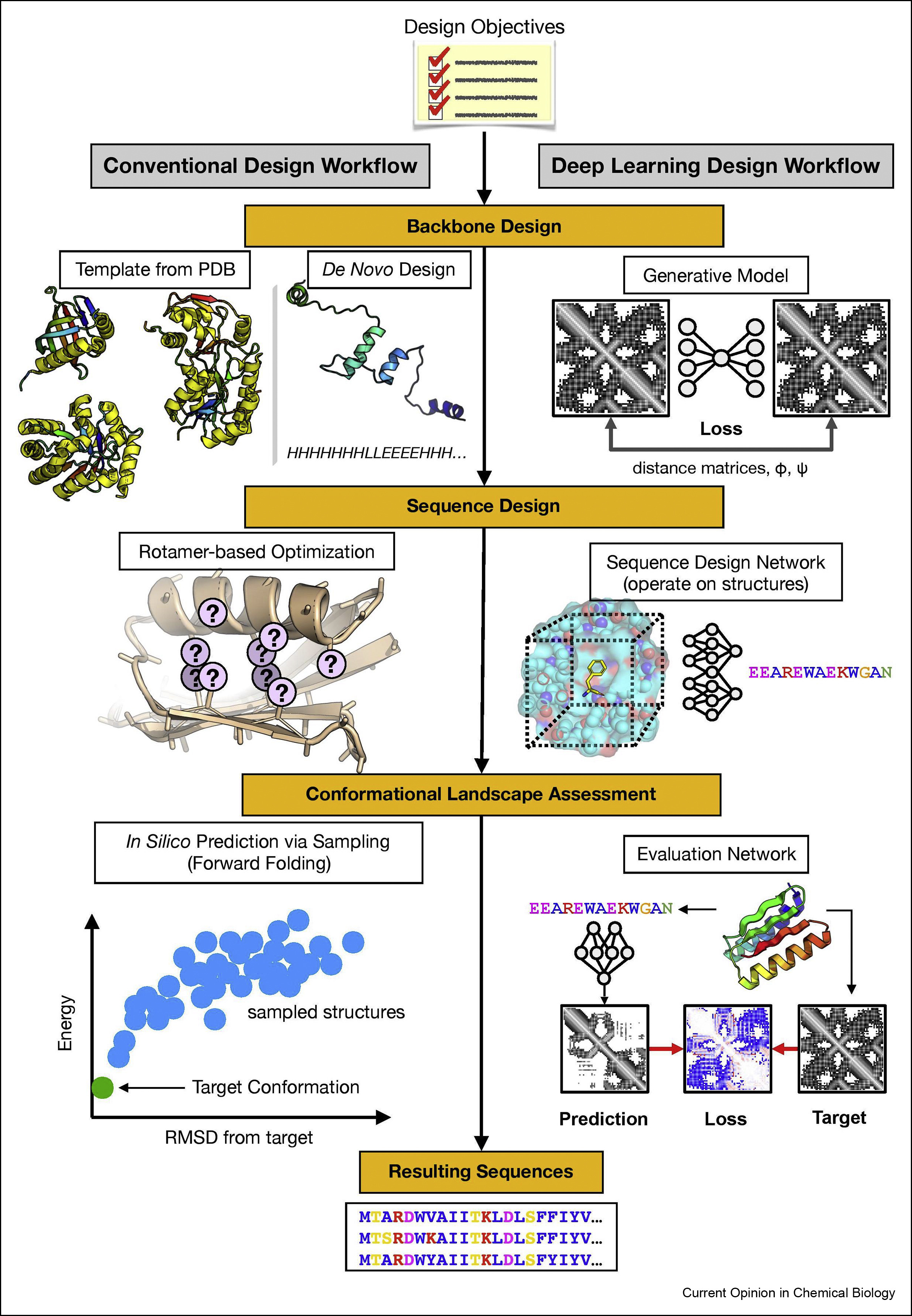}
    \caption{Structure-based protein design workflow illustration adopted from \cite{ovchinnikov2021structure} CC-BY 4.0. Comparison between conventional structure-based protein design workflow vs. DL-based design workflow. For structure generation, the conventional approaches leverages existing structure fragments or functional motifs in the structure database and along with experts' knowledge to build the structure blueprint for subsequent design steps while DL-based approaches uses neural networks trained on the vast structure database that can generate structure templates. For sequence optimization, conventional approaches uses physics based method with energy minimization, DL-based methods use structure conditioned machine learning models to predict the amino acid sequences. For design scoring, conventional methods use energy based simulation such as molecular dynamics simulation or Rosetta energy to select viable candidates. DL-based workflow use \textit{in silico} structure and property prediction models to evaluate the fitness of design candidates.}
    \label{fig:structure-based-design}
\end{figure}

Structure-based protein design is a fundamental approach to engineering proteins with specific functions, leveraging the intricate relationship between a protein's three-dimensional structure and its functions. Proteins achieve their functions through the precise spatial organization of amino acids, such as the hydrogen-bonded networks in enzyme active sites that create the ideal chemical environment for catalysis. Understanding this sequence-structure-function relationship is crucial for designing proteins with desired properties\cite{huang2016coming}. Conventional structure-based design pipeline can often be regarded as a three step process Fig.\ref{fig:structure-based-design}, the first step is to build the structure blueprint of the protein with desired function and design objectives. The second step is to generate the amino acid sequences optimized for the structure templates that can stably fold and perform the desired function in its intended environments. The last step will be to score the designs to come up with the final design candidates for further in-depth validation and analysis. There have been many success by employing this design principle to design functional proteins in a variety of applications such as enzyme design\cite{kaplan2004novo,kiss2013computational}, protein binder design \cite{vazquez2024novo,watson2023novo,gainza2023novo,cao2022design}, and biosensors\cite{quijano2021novo}.\\ \\
Traditionally, protein design has relied on building 3D structural models to satisfy functional constraints derived from design objectives, using accurate energy models to guide atomic movements in simulated systems\cite{huang2016coming}. However, the advent of deep learning (DL) algorithms has offered unprecedented opportunities with data-driven methods. These DL approaches offer the potential to enhance the design process by capturing complex patterns in protein structures and sequences more effectively than conventional methods. For more background on methods for \textit{de novo} structure generation and sequence design please refer to the Background section.\\ \\
\subsection{Unconditional structure generation}
Generating novel protein structures without relying on specific constraints or predetermined templates has gained attention in the filed of \textit{de novo} protein design as deep generative models are being more and more popular as structure modeling tools. This approach aims to explore and expand the known protein fold space, potentially leading to the discovery of new structural motifs and functional proteins that may not exist in nature\cite{huang2016coming}. Traditionally, protein design relied on conventional backbone-design methods, which broke down structure design into hierarchical components of topology and syntax\cite{koga2012principles,chualexander2022novo}. These methods, while interpretable, were limited in their ability to explore the full space of designable sequences. The advent of deep learning has dramatically changed this landscape, offering new ways to manipulate protein structures in response to functional constraints\cite{winnifrith2024generative}.\\ \\
Deep generative modeling has emerged as a powerful strategy for efficient sampling from high-dimensional distributions of protein structures \cite{goodfellow2020generative,kingma2013auto,ho2020denoising}. Particularly noteworthy is the rise of diffusion-based generative models, which have shown remarkable success in protein design \cite{anand2022protein,watson2023novo,lee2023score}. These models benefit from an iterative generation mechanism that aligns well with the hierarchical nature of protein structure, breaking down the structure-generation problem into high-level tertiary organization, followed by local secondary structure, and finally chemical detail. The stochastic nature of theses generative model allows sampling structures that are previously not being observed in the nature and can potentially harbor novel tology and function. Methods such us GAN, Ginie, and FoldingDiff\cite{anand2019fully,lin2023generating,wu2024protein} all showed promising results in generating novel protein folds while \cite{anishchenko2021novo,watson2023novo} experimentally validated numerous novel structures generated by their methods.\\ \\
\subsection{Computational enzyme and small-molecule binder design}
Computational enzyme and small-molecule binder design have made significant success in recent years, leveraging advances in computational methods, deep learning, and experimental techniques. This field aims to create novel proteins with specific catalytic or binding functions from first principles, offering a complementary approach to traditional protein engineering methods. The general process of computational enzyme design typically involves designing a 'theozyme' - an idealized active-site model that includes a quantum mechanically calculated transition state and key functional groups from amino acid side chains required for transition state stabilization. This theozyme is then docked into structurally characterized proteins to identify suitable scaffolds, followed by redesigning residues in and around the active site to optimize interactions\cite{kiss2013computational,hilvert2013design}.\\
\begin{figure}[h]
    \centering
    \includegraphics[width=0.7\linewidth]{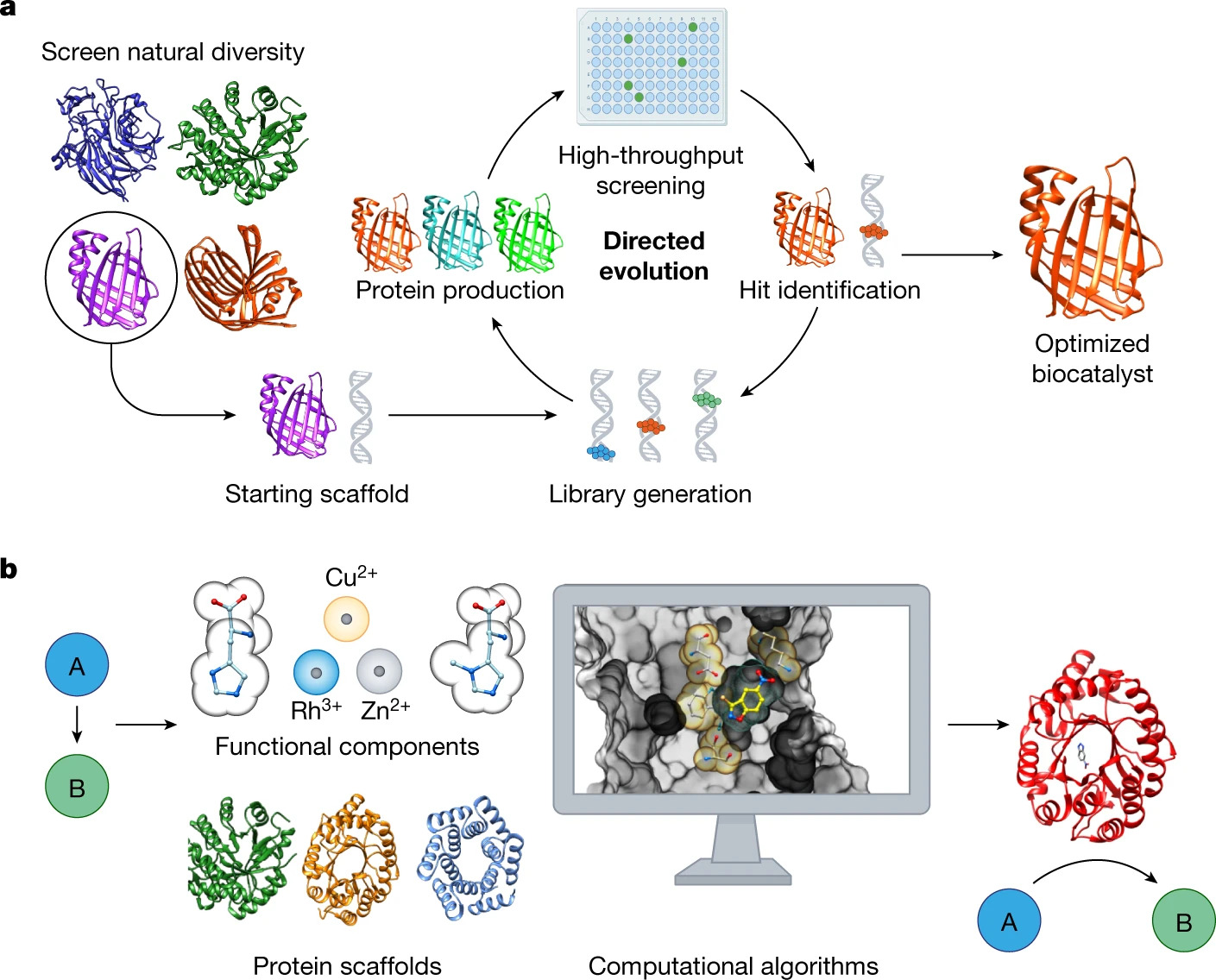}
    \caption{Enzyme engineering workflow illustration adopted from \cite{lovelock2022road}. a) Conventional enzyme engineering workflow. Natural protein scaffold with desired structure and function is picked and fitness is optimized via directed evolution. b) Computational \textit{de novo} enzyme design which starts by selecting or building suitable protein scaffold from scratch via generative models or simulation based filtering, then \textit{in silico} design and scoring scheme is employed to produce design candidates for downstream validation.}
    \label{fig:computational-enzyme-design}
\end{figure}
This approach has led to the successful design of protein catalysts for various model transformations, including the Kemp elimination \cite{rothlisberger2008kemp}, retro-aldol reactions \cite{jiang2008novo}, and Diels-Alder reactions \cite{siegel2010computational}. While initial designs often show low activity, they can be significantly improved through directed evolution \cite{yeh2023novo,crawshaw2022engineering}. Notably, \cite{yeh2023novo} used a generative model\cite{anishchenko2021novo} to generate the scaffold with family-wide hallucination, distinct from others which use existing scaffolds. Although the initial success rate of functional design is quite low, this example paved a promising path to emerging \textit{de novo} enzyme design. Despite these advances, challenges remain. Designing highly active enzymes with efficiencies comparable to natural systems is still difficult, and expanding the range of chemistries achievable with \textit{de novo} enzymes remains a key goal \cite{crawshaw2022engineering}. To address this, researchers are exploring hybrid design strategies that combine the strengths of deep learning with fundamental biophysical understanding. These approaches aim to leverage the pattern recognition capabilities of machine learning while incorporating known principles of enzyme catalysis and protein structure.\\ \\
\subsection{Protein-protein binder design and motif scaffolding}
Functional-motif scaffolding is a critical aspect of computational \textit{de novo} protein design, with applications ranging from designing enzyme active sites to creating high-affinity binders. Recent advancements in deep learning methods, have significantly improved our ability to scaffold protein structural motifs that carry out binding and catalytic functions. \\ \\
One common challenge of protein engineering is to optimize large and unstable proteins in its natural scaffolds and therefore difficult to apply powerful techniques such as directed evolution on theses proteins. Interestingly, in many cases, the critical functional elements, such as active sites or binding interfaces, comprise only a small portion of the protein's overall structure. A promising strategy to overcome these limitations involves scaffolding these essential functional motifs into smaller, more stable structures. By transplanting active sites or key functional elements into compact, robust scaffolds, we can potentially create proteins that maintain their function and activity while gaining improved stability, solubility, and expressibility. Most importantly, this approach can increase their availability for experimental optimization. Methods such as \cite{wang2022scaffolding,watson2023novo,trippe2022diffusion} showed promising \textit{in silico} results and \cite{watson2023novo} experimentally validated some of their designs.\\ \\
\subsection{Structure based antibody design}
Structure based antibody design represents the grand challenge in therapeutic protein engineering. This approach aims to overcome the limitations of traditional antibody development techniques, which are often costly, laborious, and may not always produce antibodies that bind to the desired epitope on an antigen\cite{bradbury2011beyond,lu2020development}. By leveraging computational tools and structural information, structure-based antibody design offers the potential to rapidly create binders for specific targets, whether for combating new diseases or facilitating research.\\ \\
\begin{figure}[h]
    \centering
    \includegraphics[width=0.99\linewidth]{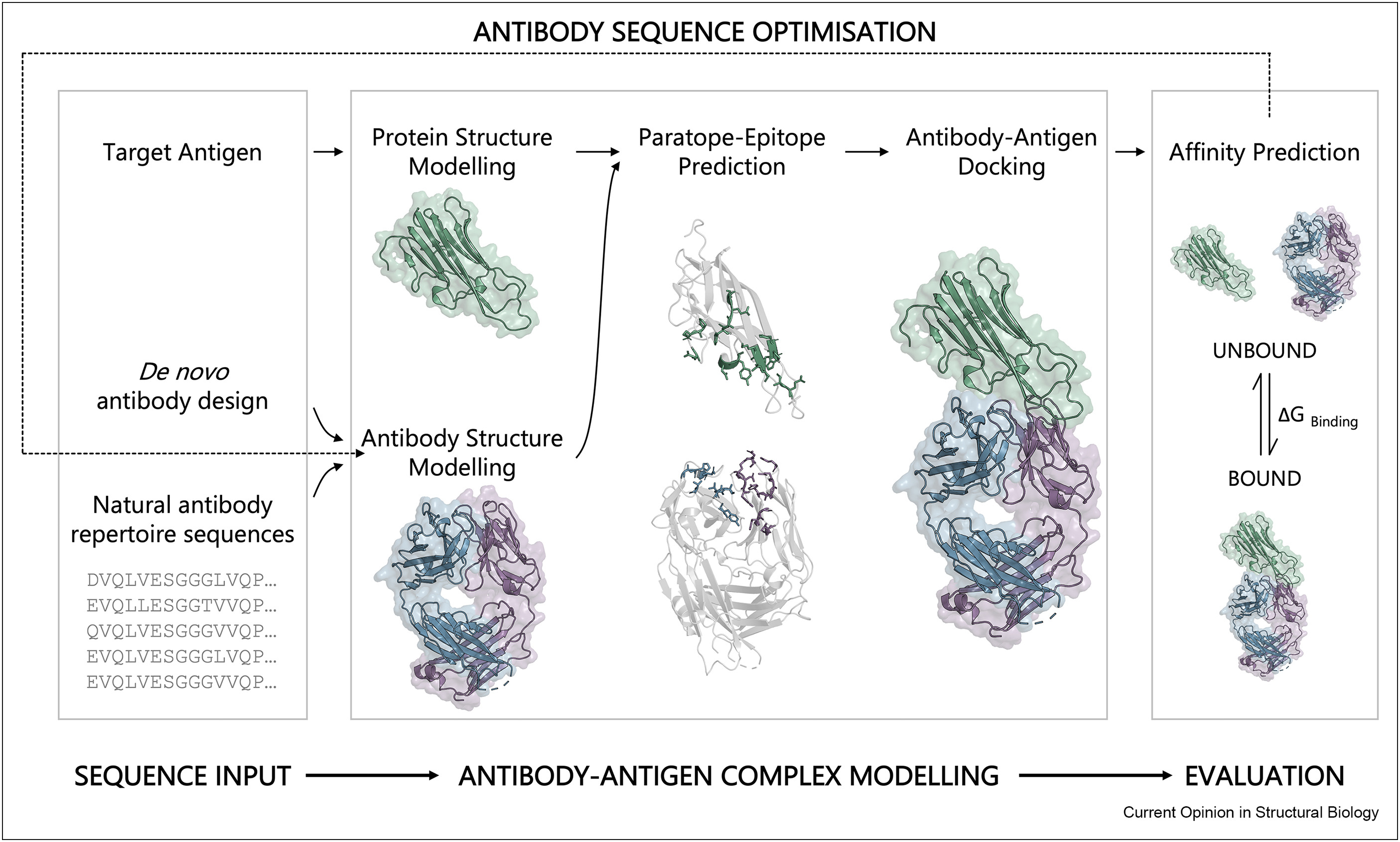}
    \caption{Structure based antibody design workflow illustration adopted from \cite{hummer2022advances} CC-BY 4.0. Workflow of \textit{in silico} structure based antibody design, the pipeline starts with a antibody framework of choice and the target antigen. A generative model or CDR minding can be used to initialize the seed AB candidates. Computational structure modeling tools and antibody docking tools will then be used to evaluate the designed complex \textit{in silico}. A scoring function or binding affinity prediction tool will then be applied to filter the design candidates. The best candidates with a given metric will be presented as the resulting design. This process can be iterative and the best candidate can be fed back into the pipeline for further optimization and design.}
    \label{fig:structure_based_ab_design}
\end{figure}
Recent breakthroughs in computational structure prediction, particularly through deep learning methods, have ushered in a new era for structure-based antibody design\cite{baek2021accurate,jumper2021highly}. With the increasing availability of accurate protein structures, including those of antibodies and antigens, it is now possible to perform large-scale structural antibody virtual screening. This approach mirrors the successful strategies employed in small molecule drug development\cite{gorgulla2020open} and opens up new possibilities for antibody engineering.\\ \\
Recent advancements in the field have focused on improving various aspects of the structure-based design pipeline. Deep learning-based methods specifically developed for antibody structure prediction, such as DeepAb\cite{ruffolo2023fast} and ABlooper\cite{abanades2022ablooper}, have shown promising results in accurately predicting the structure of complementarity determining regions (CDRs), particularly the challenging CDR-H3 loop. These methods are not only more accurate but also substantially faster than general structure prediction tools, enabling rapid generation of large numbers of antibody structures. Progress has also been made in paratope and epitope prediction, which are crucial steps in assessing binding potential. Methods like PECAN\cite{pittala2020learning}, EPMP\cite{del2021neural}, and PInet\cite{dai2021protein} have demonstrated improved accuracy in predicting these binding interfaces, with PInet achieving state-of-the-art performance in epitope prediction. These advancements contribute to more effective antibody design by helping to identify the key residues involved in antigen binding. While challenges remain, particularly in modeling antibody-antigen complexes\cite{evans2021protein,akdel2022structural}, these developments are paving the way for more effective virtual screening of antibodies against desired antigen targets.\\\\
On the other hand, antibody optimized sequence design models\cite{hoie2024antifold} that builds up on general inverse folding models\cite{hsu2022learning} have emerged to address the challenge of CDR sequence design with promising \textit{in silico} but lacking experimental validation. While current methods may not yet produce optimal binders directly, they provide a foundation for virtual screening and subsequent \textit{in silico} affinity maturation. The integration of machine learning models to predict the effects of mutations on binding affinity\cite{akbar2022silico} further enhances the potential for computational optimization of antibody-antigen interactions.
\section{Methods}
\subsection{Overall Approach}
In this section, we present a comprehensive overview of our iterative design framework, which integrates a structure generative model with multiple \textit{in silico} structural and functional assessment oracles for design optimization. This versatile pipeline can be tailored to accommodate various design objectives, ensuring its applicability across a wide range of protein design tasks.

The design process initiates with an input structure template, the selection of which is flexible and task-dependent. For instance, when optimizing the structural stability of an existing functional protein, a complete structure template along with active sites can be provided. In scenarios requiring unconditional structure generation, a randomized amino acid sequence may serve as the starting point. For partial structure scaffolding, the structure of the functional motif forms the initial template. Once a template is selected, it will be processed where it is distilled into one-dimensional and two-dimensional topological features. These features then serve as input for our structure generative model. Subsequently, the model generates a diverse library of structural decoys, which form the template pool for downstream sequence design.

Following structural generation, a fixed-backbone sequence design model\cite{dauparas2022robust} is employed to create a library of sequences compatible with the backbone library. These designed sequences are then evaluated using a state-of-the-art structure prediction oracle, such as AlphaFold2\cite{jumper2021highly} or RoseTTAFold\cite{baek2021accurate}. Sequences are filtered based on prediction confidence scores, with only those passing this \textit{in silico} structure validation filter progressing to the next stage.

For designs where specific protein functions are desired, the filtered sequences undergo customized functional oracles such as computational docking and function prediction\cite{eberhardt2021autodock}. This step allows for the evaluation of the designed proteins' potential to perform the intended function. Candidates that successfully pass both the structural and functional filters are then ranked, with the top designs selected as input templates for the subsequent iteration. This iterative process continues until either the specified design objectives are achieved or a predetermined number of iterations is reached. At this point, the pipeline halts and outputs the best candidates from the accumulated design pool.

By integrating structure generation, sequence design, and multi-faceted evaluation within an iterative framework, our approach offers a powerful and flexible tool for protein engineering. It allows for the exploration of vast design spaces while maintaining a focus on both structural integrity and functional requirements. This methodology represents a significant step forward in our ability to design novel proteins with tailored properties and functions.

\begin{figure}
    \centering
    \includegraphics[width=1\linewidth]{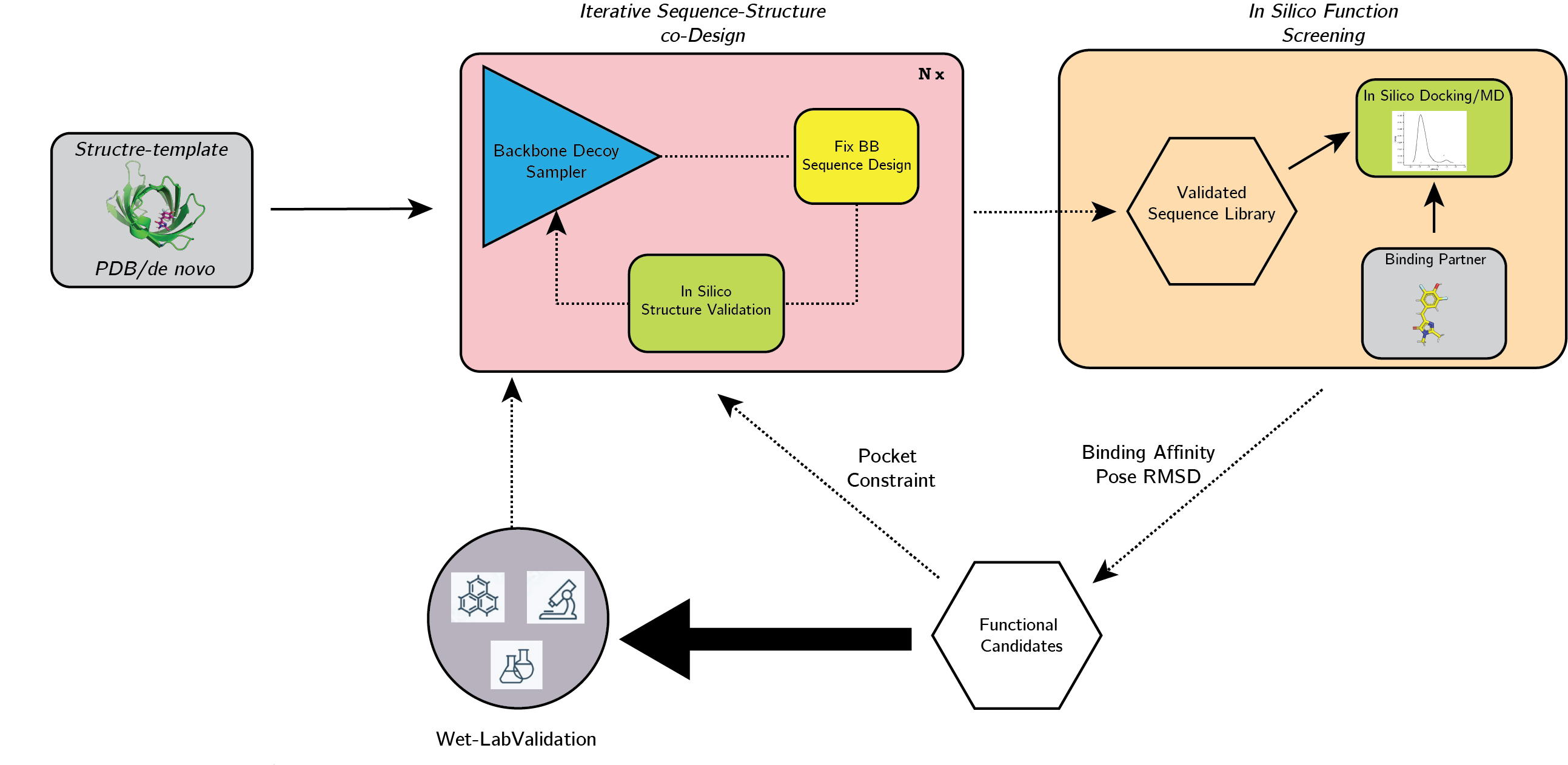}
    \caption{Overview of iterative design pipeline for adaptive protein design and optimization. The pipeline can be dissected into three modules, the design module which performs structure-sequence co-optimization on the input structure templates. The resulting structurally validated library is then fed into the scoring module which can include various \textit{in silico} simulation and fitness prediction tools. The resulting functionally validated candidates are then selected as templates for next design iteration or further experimental validation. }
    \label{fig:enter-label}
\end{figure}

\subsection{Deep structural generative models}
 For designs presented in this chapter, we use the generative model illustrated in Figure \ref{fig:C1_Fig1}. To prepare for input features, each protein structure is distilled into invariant pairwise representations of inter-residue distance and orientations as described in \cite{yang2020improved} and scalar representations of amino acid sequence and backbone torsion angles. This input is then fed through an encoder network which produces a latent representation of each residue. These representations are reassembled and passed to a decoder module which reconstructs the backbone coordinates. For more detail, see the Methods section in Chapter 2.
\subsection{in silico validation and sequence design}
\textbf{Computational stricture prediction}:For \textit{in silico} structure validation, we employed two protein structure prediction methods: AlphaFold2\cite{jumper2021highly} and ESMFold\cite{lin2023evolutionary}. To avoid evolutionary bias in \textit{de novo} structure generation, we used AlphaFold2's single sequence mode, which predicts structures based solely on amino acid sequences without relying on multiple sequence alignments or templates. For optimizing design throughput and partial structure scaffolding where evolutionary bias is preferred, we leveraged ESMFold, which offers faster prediction times and stable motif modeling. The choice between these tools was flexible, adaptable to specific design requirements.\\
\textbf{computational docking}:For computational molecular docking, we employed AutoDock Vina\cite{eberhardt2021autodock}, a widely-used open-source program for protein-ligand docking. AutoDock Vina utilizes a sophisticated scoring function and efficient optimization algorithm to predict the binding modes of small molecules to protein targets.
The docking simulations were performed using default parameters, with a search space centered on the predicted binding site of the target protein. The grid box dimensions were set to encompass the entire binding pocket, allowing for comprehensive sampling of possible ligand conformations and positions. Multiple independent docking runs were conducted for each ligand-protein pair to ensure thorough exploration of the conformational space and to assess the consistency of predicted binding modes.\\
\textbf{Fixed backbone sequence design}: For fixed backbone sequence design, we employed ProteinMPNN\cite{dauparas2022robust}, a deep learning-based method for protein sequence design. We primarily utilized the default settings for sequence generation, which have been optimized for a wide range of design tasks. Temperature-based sampling was applied selectively, depending on the specific requirements of each design task. In cases of functional protein design, we implemented constraints to fix the active sites, thereby preserving the critical functional elements of the target protein. For partial structure scaffolding tasks, we extended this approach by fixing not only the functional motif sequence but also its local structure. This strategy ensured the preservation of both the motif structure and the intended function of the designed proteins. These targeted constraints allowed us to explore sequence space effectively while maintaining essential functional and structural features, thus optimizing our protein design process for specific applications. For antibody design, we also employed AntiFold\cite{hoie2024antifold} for enhanced antibody specific design performance.\\
\textbf{Structure DB search}: To assess the novelty of the \textit{de novo} generated structures, we used Foldseek\cite{van2024fast} to perform structure search with the PDB\cite{berman2000protein} and the CATH\cite{sillitoe2021cath} databases using structure alignment.
\subsection{Experimental Validation Methods}
\textbf{Plasmid and strain construction}\\
Genes were synthesized and inserted between the BamH I and Not I sites of pET-28b vector by GenecefeBiol (Jiangsu, China), resulting in the plasmids pET-28b. The plasmids were then transformed into E. coli BL21 (DE3) competent cells (TransGen Biotech, Beijing, China) for expression, 6 × His-tag was added to the N-terminus for purification purposes.\\
\textbf{Protein expression and purification}\\
The recombinant E. coli BL21 (DE3) cells were cultured in LB medium containing 50 $\mu$g/mL kanamycin at 37 °C with shaking (200 rpm) to OD600 ~ 0.8. Protein expression was induced by adding isopropyl $\beta$-D-1-thiogalactopyranoside (IPTG) to a final concentration of 0.5 mM, and cell growth was continued overnight at 25 °C. The cells were harvested by centrifugation (10,000 × g, 10 min, 4 °C).\\
For purification, the harvested cells were resuspended in buffer A (50 mM Tris, 500 mM NaCl, pH 8.0) and sonicated on ice to lyse the cells using a Scientz JY92-IIN sonicator (Ningbo, China). As the proteins were expressed with 6 × histidine-tag in their N-terminus, they were then purified using a HiTrap™ Chelating HP column (Cytiva, MA, USA) and equilibrated with buffer A. Unbound proteins were eluted from the column with buffer W1 (50 mM Tris, 500 mM NaCl, 20 mM imidazole, pH 8.0) and buffer W2 (50 mM Tris, 500 mM NaCl, 50 mM imidazole, pH 8.0), respectively. Then, the proteins was eluted from the column with buffer B (50 mM Tris, 500 mM NaCl, 250 mM imidazole, pH 8.0). The purified proteins were further desalted using a Desalting column (Cytiva, MA, USA) with buffer C (20 mM Tris, 150 mM NaCl, pH 8.0).\\
The protein concentration was measured using the Bradford assay (Thermo Fisher, MA, USA), and the purity was determined using SDS-PAGE.\\
\textbf{Fluorescence binding assay}\\
Protein-activated DFHBI fluorescence signals were measured in 96-well plate format (Corning 3650) on a SpectraMax M2 plate reader (Molecular Devices, CA, USA) with $\lambda$ex = 467 nm, cutoff = 495 nm, and $\lambda$em = 495 to 595 nm. Binding reactions were performed at 200 µl total volume in buffer C, containing 25 µM proteins and 25 $\mu$M DFHBI. DFHBI (Sigma) was suspended in DMSO as instructed to make 100 mM stock.\\
\textbf{Differential scanning fluorimetry}\\
The desugbed protein solutions (~10 mg/mL) were diluted to 0.2 mg/mL in their respective buffered solutions (PBS at pH 7.4). Protein Thermal Shift Dye (4461146, Applied Biosciences), initially provided at a concentration of 1000x, was diluted to a concentration of 8x using Milli-Q water. The antibody solutions (12.5 $\mu$L in each well) were dispensed into 96-well white PCR plates (04729692001, Roche) in triplicate, 2.5 $\mu$L of the dye solution was added per well, and the solution was mixed by pipetting up and down ten times. The plates were then sealed with foil (04729757001, Roche Diagnostics).Thermal melts were performed using a LightCycler 480 real-time PCR instrument (Roche Diagnostics). The fluorescence (Ex: 558 nm, Em: 610 nm) was measured as the plate was heated from 25 to 99 °C. Many (>50) acquisitions were collected per 1 °C, and the heating rate was ~0.6 °C /min. The apparent melting temperatures (Tm) of the GFP mutants were determined by analyzing the first derivative of the fluorescence with respect to temperature. This involved fitting a second order polynomial to the major peak and solving for the temperature at which the maximum occurred.

\subsection{Iterative optimization algorithms}
Genetic algorithms (GAs) represent a powerful class of optimization techniques inspired by the principles of natural selection and evolution\cite{holland1992adaptation}. These algorithms simulate the process of natural evolution, including inheritance, mutation, selection, and crossover, to solve complex optimization problems. In the context of computational protein design, GAs have emerged as a valuable tool for exploring the sequence and structural space of proteins that dates back to the 90s\cite{jones1994novo,desjarlais1995novo,hellinga1994optimal} for task such as optimizing side-chain conformations for protein core packing and functional site design.

One of the key advantages of genetic algorithms in protein design is their ability to efficiently sample large, complex search spaces\cite{voigt2001computational}. This is particularly useful in protein design problems, where the number of possible sequences grows exponentially with protein length. GAs can navigate this vast space by maintaining a population of candidate solutions and evolving them over multiple generations, often leading to innovative and non-obvious design solutions. 

For unconditional structure generation we used the following design algorithm

\begin{algorithm}[H]
\caption{Unconditional Structure Generation}
\begin{algorithmic}[1]
\Procedure{UnconditionalStructGen}{$Seq_{init}, T, N_{decoy}, N_{template}$}
    \State $Population \gets []$
    \State $StructLib_{init} \gets$ StructurePredictor($Seq_{init}$)
    \State $StructLib_{0} \gets StructLib_{init}$
    \For{$t \gets 1$ to $T$}
        \State $Templates_{t} \gets$ RankPTM($StructLib_{t-1}, N_{template}$)
        \State $DecoyLib_{t} \gets$ CoordVAE($Templates_{t}, N_decoy$)
        \State $SeqLib_{t} \gets$ FixedBBDesign($DecoyLib_{t}$)
        \State $StructLib_{t} \gets$ StructurePredictor($SeqLib_{t}$)
        \State $Population \cup StructLib_{t}$
    \EndFor
    \State \textbf{return} RankPopulation($Population$)
\EndProcedure
\end{algorithmic}
\end{algorithm}
 To begin, a randomly sampled sequence pool $Seq_{init}$ of length $L$ is provided. For each iteration, we predetermine the number of templates we select for each design iteration and the number of decoys to generate for each template. Then we set the number of iteration to be $T$. In the design pipeline, Alphafold2 with single sequence mode was used as StructurePrediction and ProteinMPNN was used as the FixedBBDesign algorithm with 1 sequence generated for each decoy. RankPTM is a function that rank a set of prediction structure by their pTM scores.  

For functional protein optimization we used the following design algorithm

\begin{algorithm}[H]
\caption{Functional Protein Optimization}
\begin{algorithmic}[1]
\Procedure{FunctionalProtOpt}{$Tempate_{init}, T, N_{decoy}, N_{template}, FuncSites$}
    \State $Population \gets []$
    \State $DecoyLib_{0} \gets$ CoordVAE($Template_{init}$)
    \State $SeqLib_{0} \gets$ FixedBBDesign($DecoyLib_{0}, FuncSites$)
    \State $StructLib_{0} \gets$ StructurePredictor($SeqLib_{0}$)
    \For{$t \gets 1$ to $T$}
        \State $Templates_{t} \gets$ RankFuncRMSD($Template_{init}, StructLib_{t-1},FuncSites,N_{template}$)
        \State $DecoyLib_{t} \gets$ CoordVAE($Templates_{t}, N_decoy$)
        \State $SeqLib_{t} \gets$ FixedBBDesign($DecoyLib_{t}, FuncSites$)
        \State $StructLib_{t} \gets$ StructurePredictor($SeqLib_{t}$)
        \State $Population \cup StructLib_{t}$
    \EndFor
    \State \textbf{return} RankPopulation($Population$)
\EndProcedure
\end{algorithmic}
\end{algorithm}

We initiate our optimization pipeline with a functional protein template $Template_{init}$ of interest as well as the protein functional cites, for each iteration, we predetermine the number of templates we select for each design iteration and the number of decoys to generate for each template. Then we set the number of iteration to be $T$. In the design pipeline, Alphafold2 with single sequence mode is used as StructurePrediction and ProteinMPNN is used as the FixedBBDesign algorithm with 1 sequence generated for each decoy and the functional cites were fixed in this step. RankFuncRMSD is a function that rank a set of prediction structure by their the RMSD w.r.t the input template.

For motif grounded protein scaffolding we used the following design algorithm

\begin{algorithm}[H]
\caption{Motif Grounded Protein Scaffolding}
\begin{algorithmic}[1]
\Procedure{MotifGroundedScaffold}{$Template_{init}, T, N_{decoy}, N_{template}, MotifSites$}
    \State $Population \gets []$
    \State $SeqLib_{0} \gets$ MotifFill($Template_{init}, MotifSites, N_{decoy}$)
    \State $StructLib_{0} \gets$ StructurePredictor($SeqLib_{0}$)
    \For{$t \gets 1$ to $T$}
        \State $Templates_{t} \gets$ RankMotifRMSD($Template_{init}, StructLib_{t-1},MotifSites,N_{template}$)
        \State $DecoyLib_{t} \gets$ CoordVAE($Templates_{t}, N_decoy$)
        \State $SeqLib_{t} \gets$ FixedBBDesign($DecoyLib_{t}, MotifSites$)
        \State $StructLib_{t} \gets$ StructurePredictor($SeqLib_{t}$)
        \State $Population \cup StructLib_{t}$
    \EndFor
    \State \textbf{return} RankPopulation($Population$)
\EndProcedure
\end{algorithmic}
\end{algorithm}

We initiate our design pipeline with a structure motif $Template_{init}$ of interest as well as the relative infilling among the segments of the motif, for each iteration, we predetermine the number of templates we select for each design iteration and the number of decoys to generate for each template. Then we set the number of iteration to be $T$. In the design pipeline, ESMFold is used as StructurePredictor and ProteinMPNN is used as the FixedBBDesign algorithm with 1 sequence generated for each decoy and the motif sequence were fixed in this step. RankMotifRMSD is a function that rank a set of prediction structure by their the RMSD w.r.t the input motif.
\subsection{Data}
\textbf{DIPS for complex structures}\\
We used the DIPS(Database of Interacting Protein Structures) dataset presented by \cite{townshend2019end} to fine tune our structure generative model for binding surface generation. We capped the size of the complex structure to 500 amino acid to account for memory usage. Then we computed the binding surface with a 10A radius inter-chain contact, that is, any residue that is within 10A to any residue of the other chain. During training, the binding interface of one component of the protein complex is randomly masked.\\
\textbf{Antibody Structures}\\
For antibody structures, we used the structural antibody Database (SAbDab) obtained from \cite{jin2021iterative} which contains 1266, 1564, 2325 structures for CDR-H1, CDR-H2, and CDR-H3 respectively after filtering and splitting, there is no more than 40\% sequence identity in the inpainted regions for each set of structures between the train/test structures. Please refer to \cite{jin2021iterative} for further details.\\
\section{Results}
Our iterative design framework demonstrates high versatility and efficacy across a spectrum of protein engineering challenges. In this section, we present the results of three distinct applications, each showcasing a different aspect of the framework's capabilities. First, we test our model's ability to design \textit{de novo} protein folds by unconditional structure generation, where our approach successfully generates novel protein folds with high \textit{in silico} folding confidence. This not only expands the known protein structure space but also demonstrates the potential of our framework to explore beyond naturally occurring protein architectures. Second, we present a \textit{de novo} design of a small molecule-activated fluorescent protein. This case study illustrates the framework's ability to engineer proteins with complex, environmentally responsive functions. We provide both \textit{in silico} and experimental validation, offering a comprehensive view of the design process and its outcomes. Lastly, we demonstrate the framework's ability to perform structure-based protein engineering through a motif-grounded scaffolding of PD1 (Programmed Cell Death Protein 1). This application highlights the potential of our approach in the field of therapeutic protein design, showcasing how functional motifs can be integrated into novel structural contexts. Together, these results underscore the power and flexibility of our iterative design framework, spanning from fundamental advances in protein structure exploration to the creation of functional proteins with potential real-world applications.
\subsection{Unconditional Structure Generation}
The ability to generate novel protein structures is a fundamental challenge in protein engineering, with implications ranging from understanding protein evolution to designing new functional proteins. Traditional approaches to protein design have often been limited by the known protein fold space, typically relying on modifications of existing structures \cite{huang2016coming}. Our iterative design framework, however, aims to push beyond these boundaries by enabling the generation of entirely new protein folds for a wide range of sizes that do not naturally occur.\\
\begin{figure}[h]
    \centering
    \includegraphics[width=0.99\linewidth]{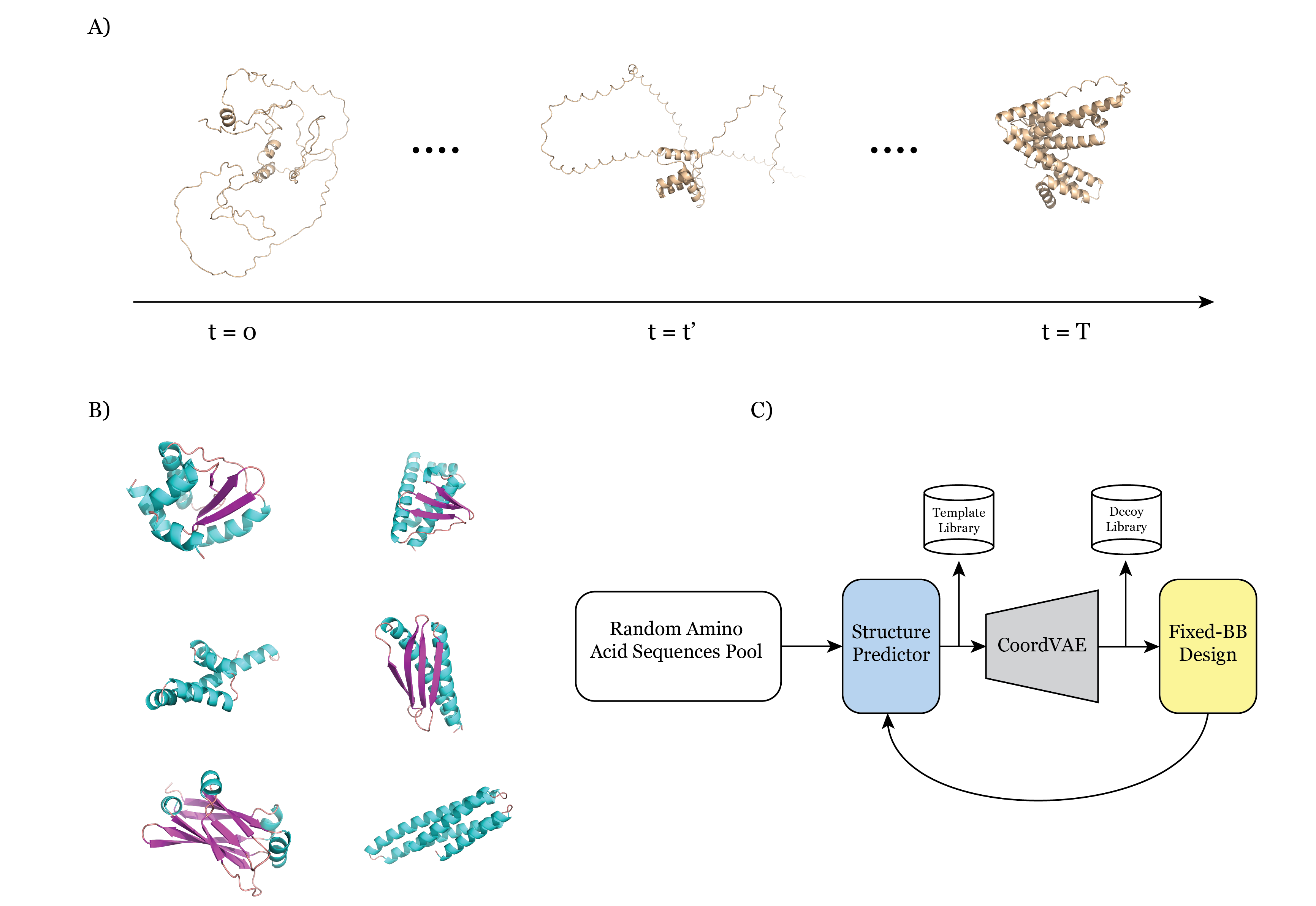}
    \caption{\textbf{Unconditional Generation of monomeric structures.} A) Illustration of iterative structure evolution by design iterations. t=0 indicates random initialization of amino acid sequences, t=T indicates design convergence. B) Example of generated \textit{de novo} structures, helical structures are colored cyan and beta strands are colored purple. C) Conceptual graph of the iterative design framework.}
    \label{fig:C2_R1}
\end{figure}
In \ref{fig:C2_R1}.A we show an example of structure evolution through our iterative design algorithm, which we found in practice, structures converges within 10 iterations with 5-10 templates chosen at each round of design. Also, we found it advantageous to use the predicted Template Modeling (pTM) score as the primary fitness criterion, rather than relying solely on pLDDT. This strategy helped us avoid a bias towards long helical structures, which computational structure predictors tends to favor for higher pLDDT scores. By incorporating pTM, we were able to generate a more diverse set of structures with varied secondary structure compositions. As illustrated in Fig. \ref{fig:C2_R1}.B, our approach successfully produced designs spanning a wide range of secondary structure elements, including $\alpha$-helical, $\beta$-sheet, and mixed $\alpha/\beta$ topologies. This diversity in secondary structure composition demonstrates the versatility of our framework and its ability to explore a broad spectrum of protein folds, rather than being limited to particular folds.
\begin{figure}[h]
    \centering
    \includegraphics[width=0.99\linewidth]{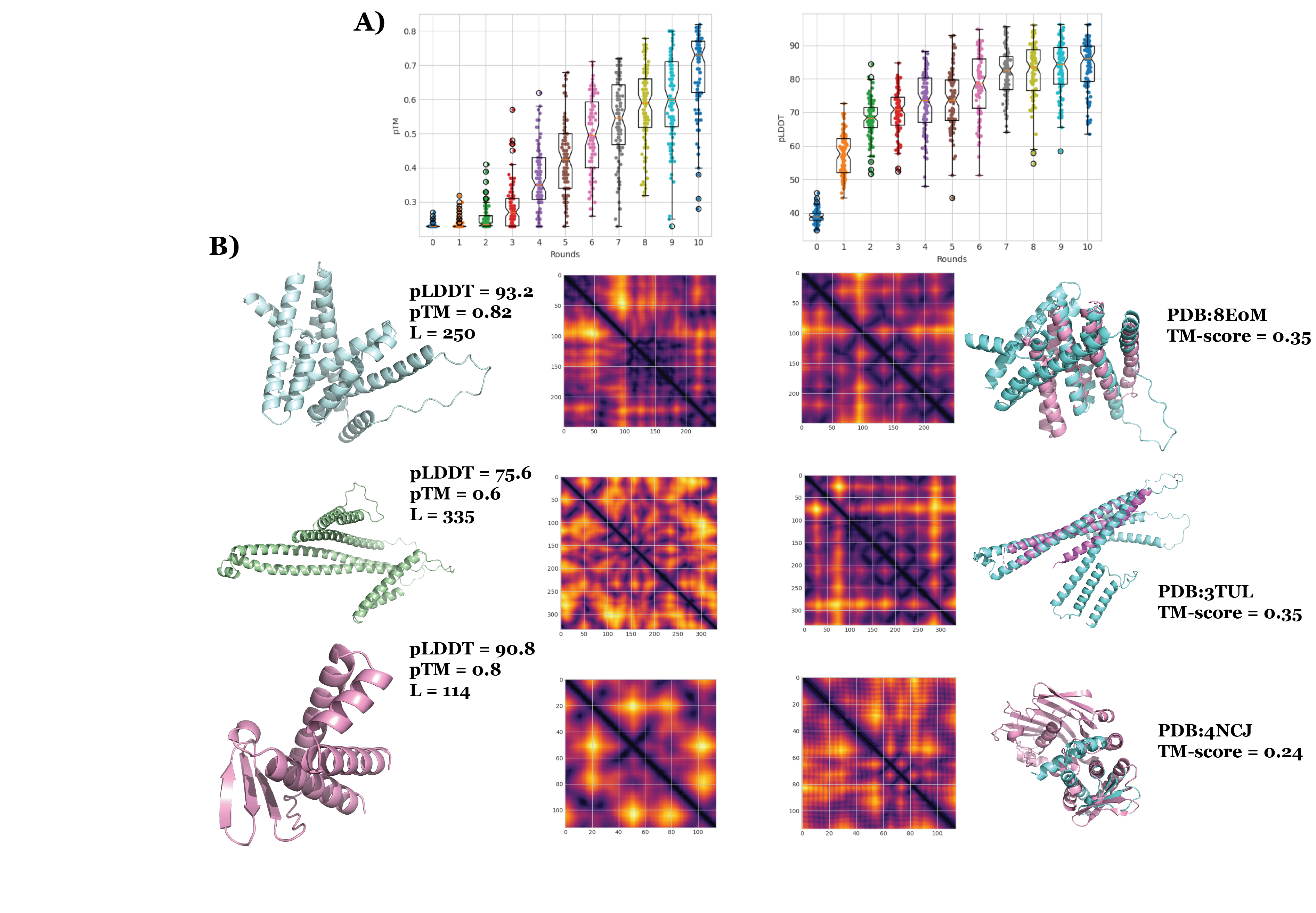}
    \caption{\textbf{Unconditional Generation of novel protein structures.} A) Example pLddT and pTM evolution with design iterations. C) Column 1,  \textit{de novo} generated protein folds. Column 2, initial inter-residue distance map of randomly initialized sequences. Column 3, inter-residue distance map for the \textit{de novo} generated protein folds. Column 4 overlay of \textit{de novo} generated protein folds with its closest hit from the PDB. In this case, we found designs from our iterative design framework to exhibit both high \textit{in silico} folding viability and  structural novelty. }
    \label{fig:C2_R2}
\end{figure}
We began by assessing our framework's ability to generate protein sequences that can be folded \textit{in silico} with high confidence. This is a crucial first step, as it demonstrates the framework's capacity to produce designs that are likely to adopt stable, well-defined structures\cite{anishchenko2021novo}. We generated a diverse set of protein sequences using our adaptive algorithm, iteratively refining them based on \textit{in silico} folding predictions.
To evaluate the quality of our designs, we employed state-of-the-art protein structure prediction tools, including AlphaFold2 \cite{jumper2021highly} and ESMFold\cite{lin2023evolutionary}. We considered a design successful \textit{in silico} if it achieved a high predicted Local Distance Difference Test (pLDDT) score, which indicates the confidence of the structure prediction [6].  Our framework was able to generate a substantial number of sequences with high pLDDT scores($\geq 90$), indicating very high confidence in their predicted structures\ref{fig:C2_R2}.A. 

Next, we evaluated the novelty of the generated structures. To assess this, we compared our designs against known protein structures in the Protein Data Bank (PDB) using both global and local structural alignment methods \cite{van2024fast}. We found that a significant proportion of our high-confidence designs exhibited structural features that were distinct from any known protein fold. This novelty is demonstrated in Fig. \ref{fig:C2_R2}B, which showcases three representative designs of varying lengths and their closest structural matches in the PDB. For the first design shown in Fig. \ref{fig:C2_R2}B, the closest structural homolog in the PDB (PDB ID: 8EOM) has a TM-score of only 0.35, indicating substantial structural divergence from known folds. We also demonstrated the ability to generate novel folds for longer proteins. The second design in Fig. \ref{fig:C2_R2}B, a protein with 335 amino acids, found its closest structural match in the PDB (PDB ID: 3TUL) with a TM-score of 0.35. We observed that our method could produce diverse and novel folds for shorter proteins as well. The third design in Fig. \ref{fig:C2_R2}B, despite its smaller size, exhibited remarkable novelty. Its closest structural match in the PDB (PDB ID: 4NCJ) had a TM-score of only 0.24. These results collectively demonstrate the capability of our iterative design framework to generate protein structures that are substantially different from any known folds across a range of protein sizes.

Our iterative design framework demonstrated success in generating novel protein structures with high \textit{in silico} folding confidence. We produced a diverse array of protein designs spanning various sizes and secondary structure compositions. Notably, many of these designs exhibited significant structural novelty, with low TM-scores to any existing experimental structures when compared to their closest matches in the Protein Data Bank. This ability to generate stable, novel protein folds across different protein sizes  potentially pave new paths for exploring protein structure-function relationships beyond the confines of naturally evolved proteins. In contrast to previous methods such as network hallucination\cite{anishchenko2021novo} and diffusion models\cite{wu2024protein,lin2023generating}, our model is much more compute efficient and flexible. One can use other fitness criterion such as topology preference and the globularity. However, there are several limitations and considerations to keep in mind. The results are based \textit{in silico} on  predictions, and experimental validation of these structures is crucial to confirm their stability and folding in real-world conditions. While the designs show novelty compared to known structures, their functional potential remains to be explored. The reliance on computational structure prediction tools like AlphaFold2 and ESMFold means that the method's success is partially dependent on the accuracy and limitations of these prediction tools. The approach may still have biases or limitations in the types of folds it can generate, which may not be immediately apparent from the \textit{in silico} results. Future work should focus on experimental validation of these designs, exploration of their functional potential, and further refinement of the design algorithm to address any biases or limitations
\subsection{De novo design of DFHBI activated fluorescent protein}
Computational design of functional proteins represents a frontier in protein engineering, offering the potential to create tailored molecular tools for a wide range of applications in biotechnology and medicine\cite{huang2016coming}. This approach not only allows for the optimization of existing protein functions but also enables the creation of entirely new functionalities not found in nature. Despite many recent advances in the field, there is still a gap between the computationally generated design and their experimental success. In this section we demonstrate how to utilize our iterative design framework for \textit{de novo} functional protein design with $100\%$ experimental success rate and improved thermal stability and production yield. 

Fluorescent proteins, have been a prime target for computational design due to their immense utility in biological imaging and their relatively well-understood structure-function relationships\cite{pedelacq2006engineering}. While naturally occurring fluorescent proteins like GFP have been widely used, there is a growing demand for proteins with tailored properties, such as specific activation mechanisms or spectral characteristics \cite{rodriguez2017growing}. Small molecule-activated fluorescent proteins are especially valuable as they allow for temporal control of fluorescence, enabling more precise experimental manipulations \cite{dou2018novo}.

\begin{figure}[h]
    \centering
    \includegraphics[width=0.99\linewidth]{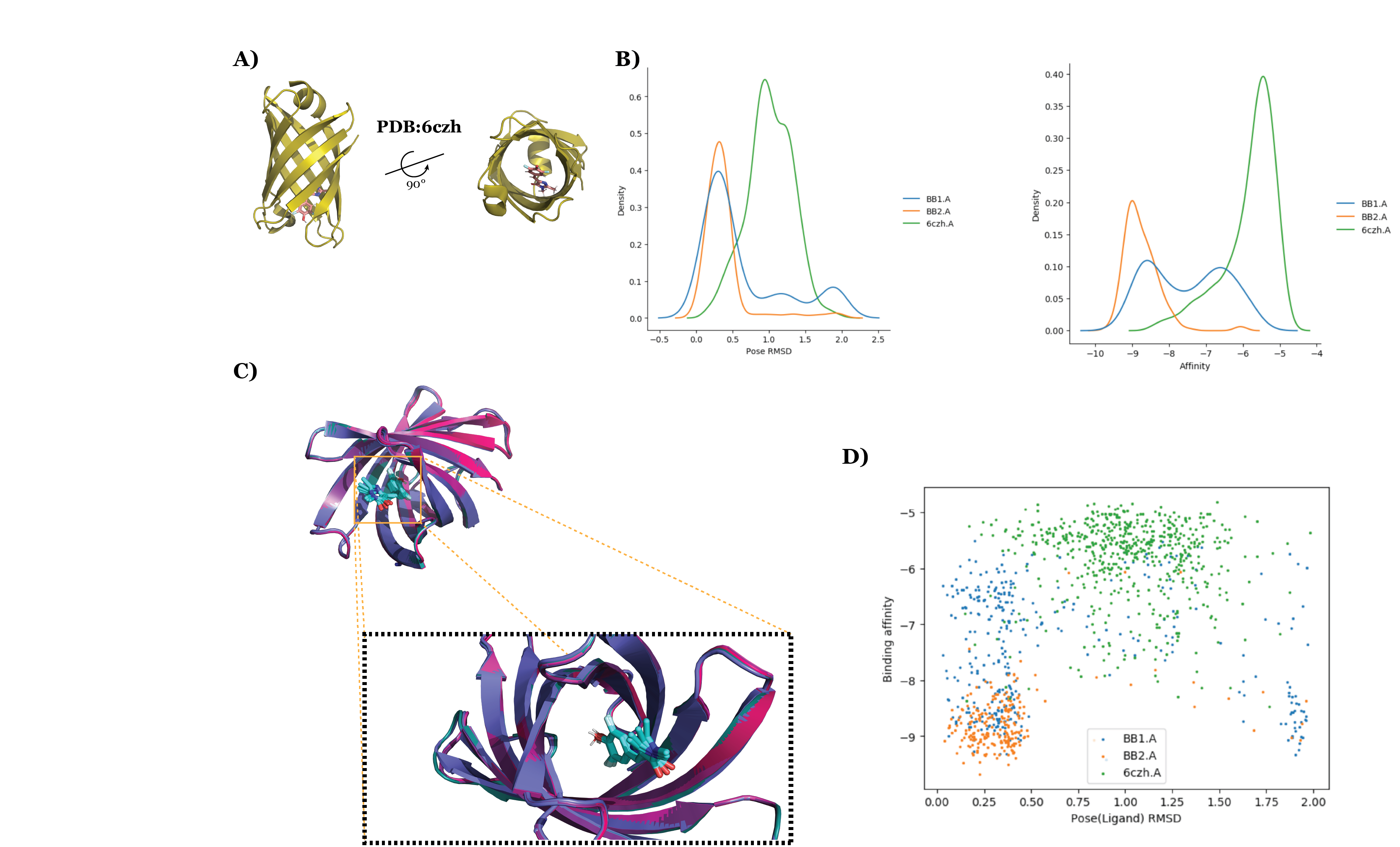}
    \caption{\textbf{De novo design of DFHBI-activated fluorescent $\beta$-barrels} A) PDB:6CZH previously designed DFHBI-activated $\beta$-barrel. B) Distribution of ligand pose RMSD(Left) of the \textit{de novo} designed $\beta$-barrels with starting template 6czh.A and the first design batch BB1.A and the second design batch BB.2. Distribution of computational docking scores from Autodock Vina(Right) of the \textit{de novo} designed $\beta$-barrels with starting template 6czh.A and the first design batch BB1.A and the second design batch BB.2 C) Example overlay of the computationally docked DFHBI with the designed $\beta$-barrels. D) Scatter plot of the computationally docked scores vs. the ligand RMSD}
    \label{fig:C2_R3}
\end{figure}

In this study, we focused on further developing and optimizing a DFHBI-activated $\beta$-barrel fluorescent protein previously designed by the Baker lab\cite{dou2018novo}. DFHBI (3,5-difluoro-4-hydroxybenzylidene imidazolinone) is a small molecule that becomes fluorescent when bound in a specific pose, making it an ideal candidate for designing controllable fluorescent systems\ref{fig:C2_R3}.A. Leveraging our iterative design framework, we set out to enhance the properties of this DFHBI-activated fluorescent protein. Our goal was to create a variant with improved fluorescence characteristics, higher stability, and more precise small molecule control, while maintaining the $\beta$-barrel scaffold that has proven effective for fluorescent proteins.
Building upon the initial DFHBI-activated $\beta$-barrel fluorescent protein designed by the Baker lab, we employed our iterative design framework to enhance its properties. The design process began with the generation of structural variants based on the initial design template. We used our structural generative model to generate an backbone library, preserving the DFHBI binding pocket and the surrounding residues that influence binding characteristics. A key component of our \textit{in silico} evaluation was the use of AutoDock Vina\cite{eberhardt2021autodock} for computational docking simulations. For each generated structure, we performed docking simulations with DFHBI to assess binding affinity and pose. The docking results were crucial in our iterative design algorithm, serving as a primary criterion for selecting promising candidates for further refinement. Candidates were first filtered for \textit{in silico} folding viability and selected using the computational docking metrics. As shown in \ref{fig:C2_R3}.B, the distribution of the first round of design compared to subsequent design shown significant improvement of both the ligand pose position and the computational affinity scores. On \ref{fig:C2_R3}.C we examined the ligand docking positions with the computationally designed proteins. We found stable and consistent functionally active docking simulated candidates. After multiple rounds of iteration and refinement, guided by the docking results and other computational analyses, we identified a promising candidate for further experimental validation. From \ref{fig:C2_R3}.D, we observed the progression of functional metric improvement through design rounds towards lower ligand post RMSD and better simulated binding affinities. For the designs we used two different structure design templates that are previously characterized experimentally(6CZI and 6CZH).
\begin{figure}[H]
    \centering
    \includegraphics[width=0.99\linewidth]{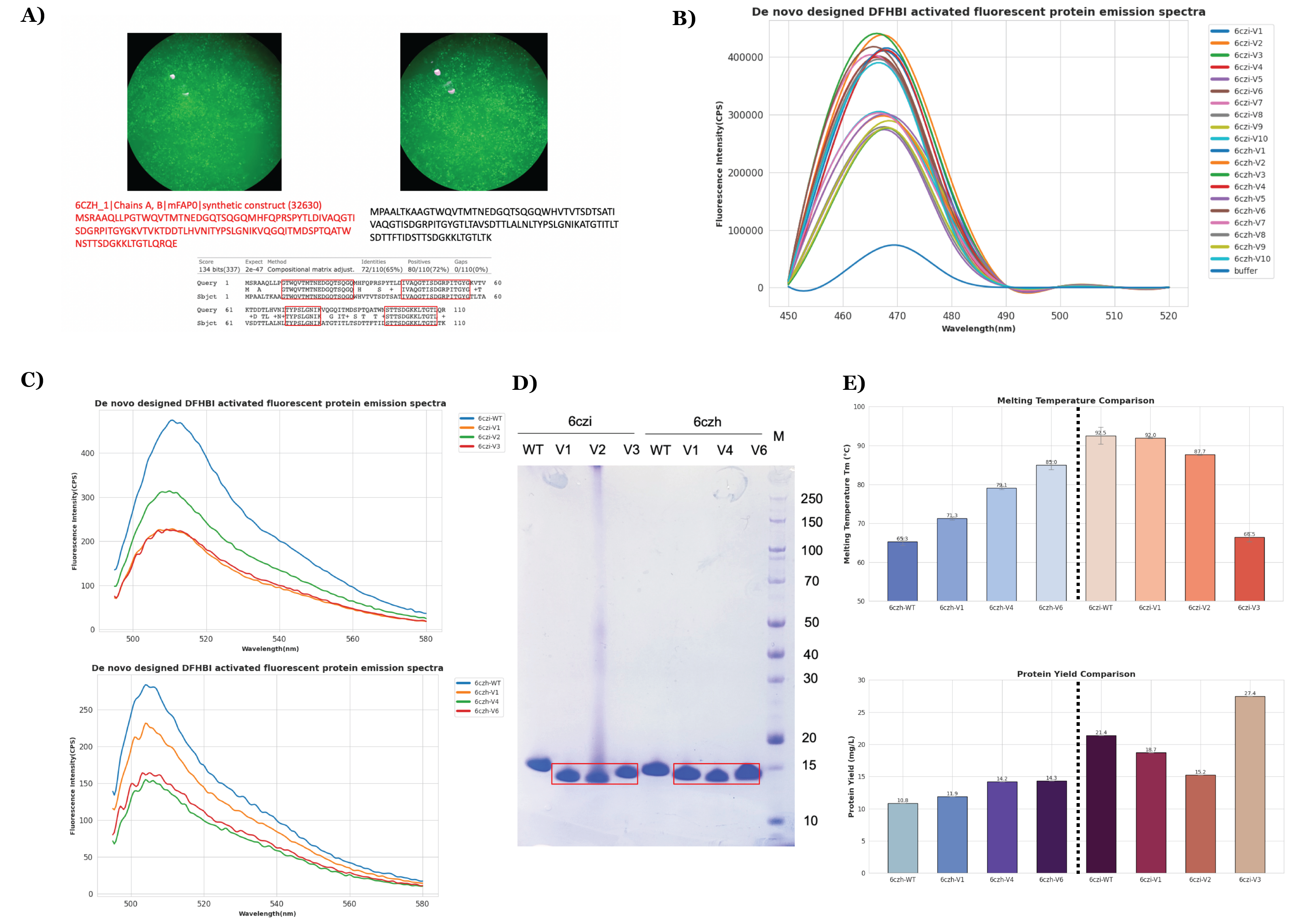}
    \caption{\textbf{Experimental validation of \textit{de novo} designed DFHBI-activated fluorescent $\beta$-barrels} A) Fluorescence microscope image of \textit{de novo} designed protein(Left) and the design template mFAP0. B) Fluorescent emission spectras from the lysates of 20 tested designs and the templates with reference buffer. All of the 20 designs were found with detectable fluorescent emission which represents a 100\% design success rate. C) Fluorescent emission spectras of purified designed proteins and their respective original templates of 6czi(up) and 6czh(down). Although we did not observe increased fluorescent peak, we saw slight shift in the emission frequency compared to the design template.  D)SDS–PAGE of purified designed protein with its respective design templates. All the designed proteins are observed to be smaller than the reference template.  E) Protein thermal stability assay results(top), and protein yield comparison results(bottom). We see improvement of thermal stability on 6czh-based designs and comparable thermal performance in 6czi-based designs. For protein yield, all of the 6czh-based designs exhibited improvement over the design reference and 6czi-V has a significant increase over the reference.}
    \label{fig:C2_R4}
\end{figure}
We selected the top 20 designs from our computational pipeline for initial experimental screening. These designs were expressed in E. coli(See Methods for details) using standard protocols. Crude cell lysates were then subjected to a fluorescence assay in the presence and absence of DFHBI. This initial screen allowed us to rapidly assess the fluorescence activation of our designs and compare their performance to the design templates. Results from this initial screen revealed that 20 out of 20 designs showed detectable DFHBI-activated fluorescence with a $100\%$ success rate(\ref{fig:C2_R4}.B). All the tested designs have sequence similarity to the design template greater than $74\%$ with the lowest candidate to have $65\%$ sequence identity to the design template as shown in \ref{fig:C2_R4}.A.\\
Based on the results of the initial screen, we selected the top 6 designs for more comprehensive evaluation. These proteins were expressed at a larger scale and purified to homogeneity using affinity chromatography followed by size exclusion chromatography. We then perform the fluorescent assay again on the purified proteins and saw very comparable emission peaks to the design templates as seen in \ref{fig:C2_R4}.C. Although we did not see increased fluorescent intensity in any of our designs, the low sequence similarity to the original templates showed a expanded functional sequence space and further wet lab optimizations such as directed evolution can be employed to improve intensity. For thermal stability, we tested the melting temperature $T_m$ of the the \textit{de novo} designed proteins, we found that for 6czh based design, all 3 purified candidates showed improved thermal stability and the best one reached $85^{o}$C which is $26\%$ improvement over the template. In the case of 6czi based design, we achieved very comparable $T_m$ despite the design template already has a high melting temperature of $92^{o}$C. For protein expression yield, we achieved increased unit expression yield in both design case as seen in \ref{fig:C2_R4}.E.\\

Building upon a previously designed DFHBI-activated $\beta$-barrel fluorescent protein, we employed our iterative computational framework, incorporating AutoDock Vina\cite{eberhardt2021autodock} for docking simulation as a fitness criterion. Our design approach yielded promising \textit{in silico} design success, as evidenced by computational docking and folding simulations. Remarkably, we achieved a 100\% design success rate from 20 computationally designed candidates tested experimentally. Our designs based on the 6czh template demonstrated improved thermal stability, with the best candidate reaching a melting temperature of 85°C, representing a significant 26\% improvement over the original design template. Furthermore, we observed increased protein expression yield for designs based on both 6czh and 6czi templates. \\ \\
Despite these achievements, our study revealed certain limitations. Most notably, the designs did not achieve increased fluorescence intensity compared to the original templates. This inability to produce higher intensity candidates could be attributed to several factors. The design process may have prioritized properties such as stability or expression yield, potentially at the expense of binding affinity. To enhance structural rigidity and stability, our designs may have inadvertently limited the flexibility of the binding pocket allosterically. Interestingly, we observed a small emission peak shift in the 6czi-based design, indicating a possible change in binding pose induced by our design. However, the general emission peak difference was maintained between the two design templates. We hypothesize that the \textit{de novo} design process has allosteric impacts on the binding pocket, which warrants further investigation. \\ \\
The differential success in improving thermal stability between the two templates is particularly noteworthy. While we saw significant improvement in thermal stability for the 6czh-based design, the 6czi template already possessed high thermal stability. Our best design based on 6czi was comparable in stability despite low sequence similarity, which is an encouraging result. Additionally, we achieved improved yield from both design templates. This outcome is not entirely surprising, as the original design was optimized for yeast display, whereas our experiments were conducted using E. coli. Our optimization pipeline did not incorporate any expression system bias, which may explain the improved yields across different hosts. \\ \\
Compared to previous design methods, our iterative design framework is fully \textit{in silico} and we leveraged DL based tools in all stages of the design pipeline. Unlike targeted mutation strategies based on predicted single mutational effects, our method enabled us to change upwards of 35\% of the total amino acids in the resulting designs. Despite this substantial sequence divergence, we successfully maintained comparable function to the original templates while simultaneously improving other properties such as thermal stability and expression yield. We see this as a promising first step towards practical \textit{in silico} protein design and optimization.\\ \\
These findings point to several directions for future research. Refining our computational model to better capture the determinants of fluorescence intensity could lead to more comprehensive improvements in protein design. Exploring ways to overcome the apparent trade-offs between different desirable properties, such as stability and fluorescence intensity, would be valuable. The study suggests that our computational approach could be particularly effective for improving proteins that are not already highly optimized, as demonstrated by the successful enhancement of the 6czh-based designs.\\ \\
To further validate our hypotheses and refine our understanding, more controlled experiments could be conducted. These might include detailed structural studies to examine the binding pocket changes, systematic exploration of the sequence-structure-function relationship in our designs, and comparative studies across different expression systems. Such investigations could provide deeper insights into the complex interaction between protein stability, expression, and function.

\subsection{de novo protein binder design via motif grounded scaffolding}
\begin{figure}[H]
    \centering
    \includegraphics[width=0.99\linewidth]{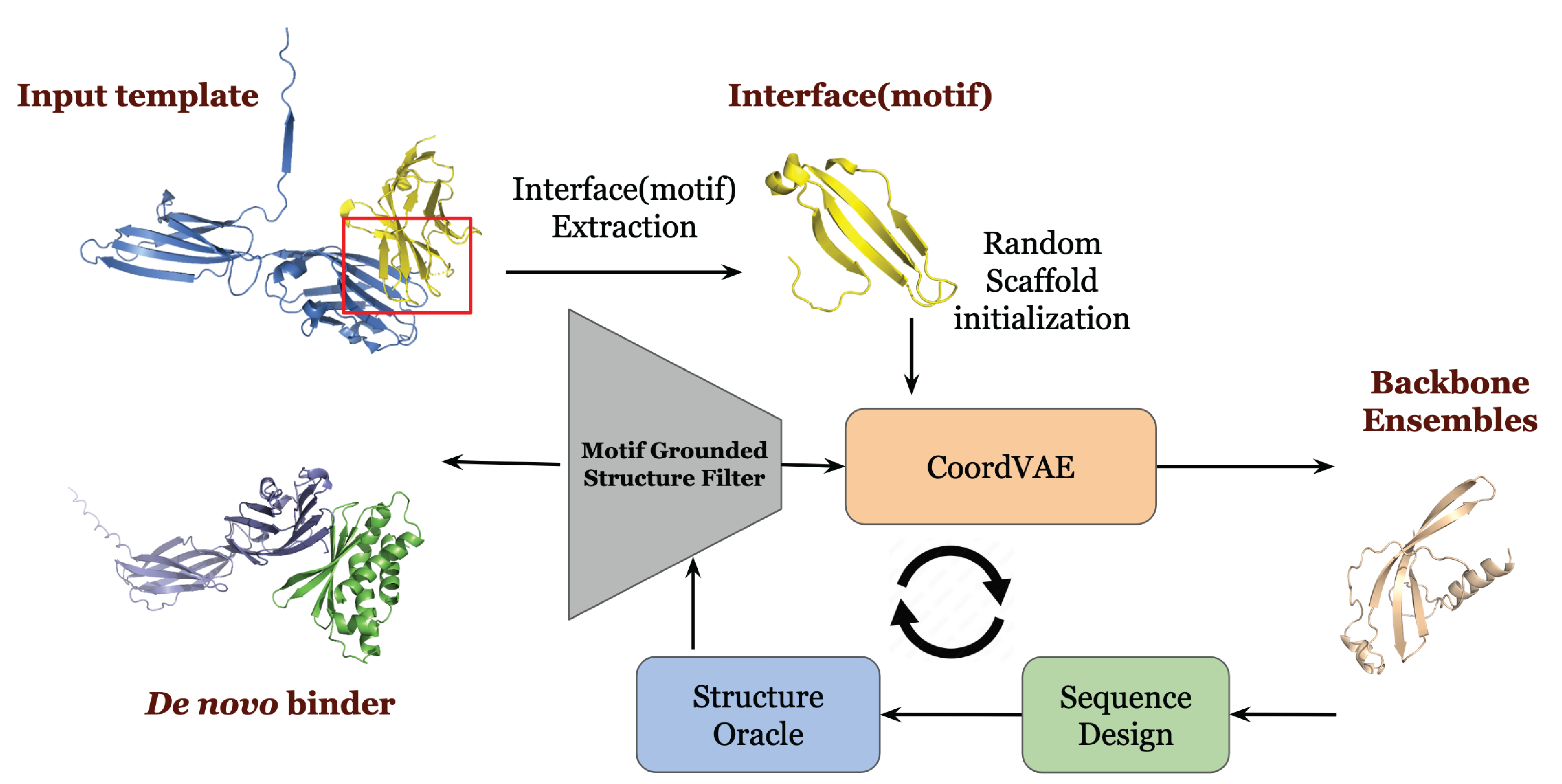}
    \caption{\textbf{Iterative design workflow for motif grounded protein scaffolding}: The design process begins with the identification of the functional motif interest, this can be done by extracting protein-protein binding surfaces or enzyme active sites. Then a randomly initialized scaffold is used as the seed template. The iterative design algorithm is then applied with a motif grounded structure fitness criterion. }
    \label{fig:C2_R5}
\end{figure}

Protein scaffolding is a crucial tool in protein engineering that involves transplanting functional motifs from one protein context to another, often more stable and versatile, protein scaffold\cite{wang2022scaffolding}. This technique has numerous applications in biotechnology and therapeutics, allowing for the creation of proteins with desired functions in optimized structural contexts. Scaffolding can improve protein stability, enhance expression levels, and even modulate the activity of the transplanted motif. To trial our proposed iterative design framework's ability to embed known functional motif into new protein scaffolds by iteratively co-optimizing structure and sequence around the dedicated motif of interest as seen in \ref{fig:C2_R5}, we performed an excise on computationally scaffold the PD1/PD-L1 complex and validated our design with \textit{in silico} binding and folding experiments. 
Programmed Cell Death Protein 1 (PD1) is a crucial immune checkpoint receptor that plays a significant role in regulating T-cell responses\cite{sharpe2018diverse}. The interaction between PD1 and its ligands (PD-L1 and PD-L2) is a key target for cancer immunotherapy\cite{sun2018regulation}. However, the use of native PD1 protein or its extracellular domain in therapeutic applications can be challenging due to stability and manufacturing consideration\cite{ganesan2019comprehensive}. In this study, we aimed to use our iterative design framework to scaffold the key binding motif of PD1 onto a more stable protein structure. Our goal was to create a novel protein that maintains PD1's binding specificity and affinity for its ligands while potentially offering improved biophysical properties.

We began by identifying the binding surface residues of PD1 involved in its interaction with PD-L1, based on available structural and mutational data \cite{pascolutti2016structure}. This binding motif served as the anchor of our scaffolding exercise. We then randomly initialize a scaffold as the starting point then employed our iterative design pipeline to design the new scaffolds for the binding surface using the motif anchored  \textit{in silico} folding confidence as the selection criteria(See Methods for details). 
\begin{figure}[h]
    \centering
    \includegraphics[width=0.99\linewidth]{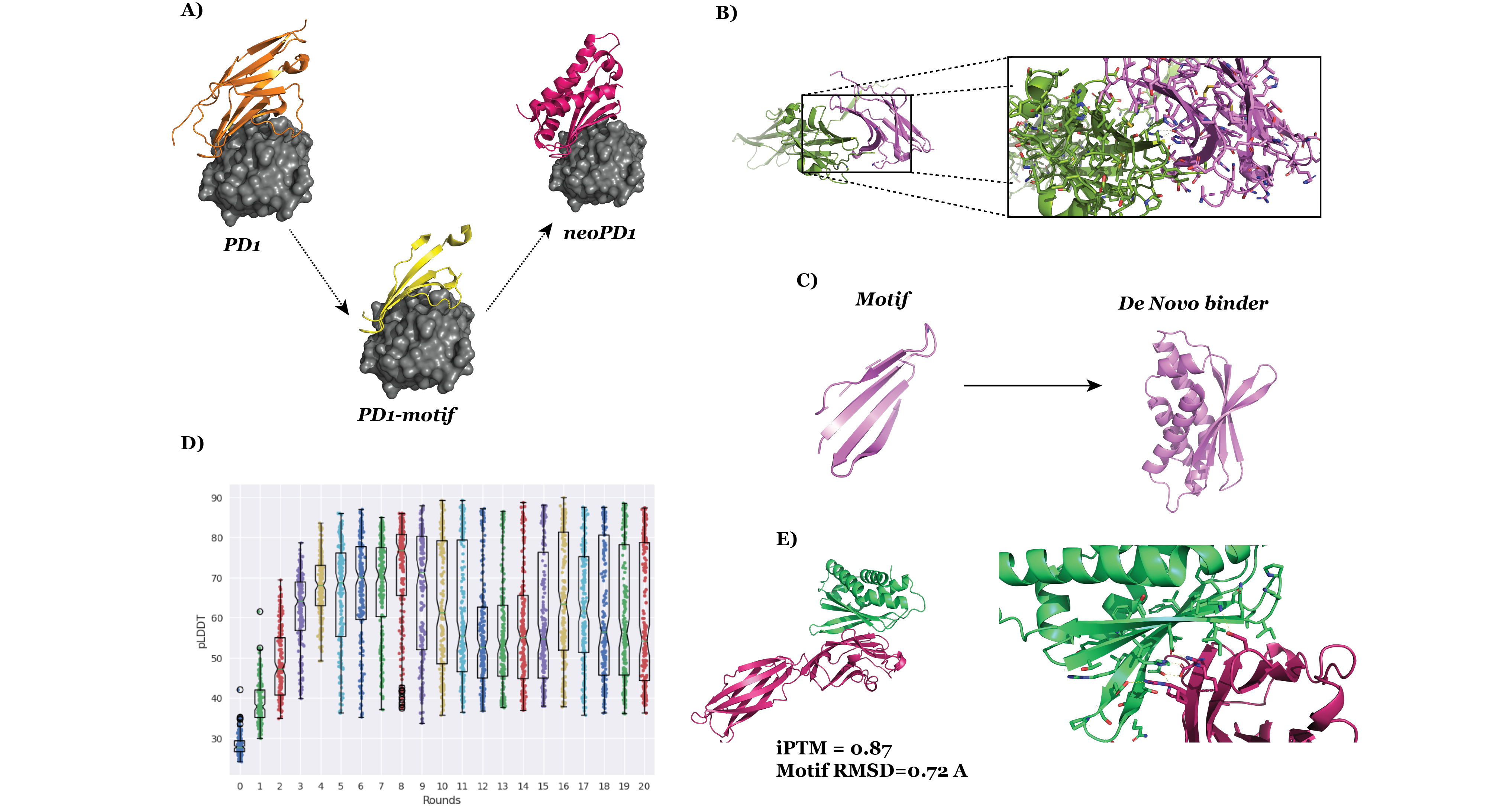}
    \caption{\textbf{Motif grounded protein binder scaffolding for PD1/PD-L1}. A) PD1 in complex with PD-L1 at different stage of the design process. PD1/PD-L1(left), PD1-motif/PD-L1(middle), and neoPD1/PD-L1(Right) B) Zoomed in image of the binding surface of the PD1/PD-L1 complex (PDB:5IUS) with polar contacts annotated. C) Starting motif and the \textit{de novo} scaffold binder comparison. D) pLDDT distribution vs. design iterations for the scaffolding process. The structures are mostly converged after 10 iterations, more rounds are conducted to obtain more viable design candidates. E) Predicted complex structure of the design scaffold binder of PD1/PD-L1. Zoomed in image of the binding surface with polar contacts annotated.}
    \label{fig:C2_R6}
\end{figure}
We performed 20 iterations in our design algorithm for scaffolding, we identified a set of promising candidates with motif centered structure RMSD and \textit{in silico} folding confidence. Our top design, is a 142-residue protein that successfully incorporates the PD1 binding motif into a novel $\alpha/\beta$ fold in contrast to the native $\beta$ only fold of PD1. We first computationally predicted the structure of the \textit{de novo} designed protein binder with PD1 binding motif and filtered with pLDDT score for \textit{in silico} folding viability and picked the top candidates for further evaluation. We showed the progression of the pLDDT distribution across all design rounds \ref{fig:C2_R6}.D.  We then computationally predicted the structure of the \textit{de novo} designed PD-1 binder with PD-L1 and most of the prediction showed high pLDDT and iPTM score from their AF-multimer models. In the example shown in \ref{fig:C2_R6}.E, the predicted complex showed a iPTM score of 0.87 and the PD1 binding surface has a RMSD of only 0.72A, we also highlighted the polar contacts in the predicted complex model.\\ \\ 

In summary,  this study employed a novel iterative design pipeline that begins with identifying the binding surface residues of PD1 involved in its interaction with PD-L1. Using this binding motif as an anchor, we randomly initialized a seed scaffold and then iteratively designed new scaffolds around the motif, using \textit{in silico} folding confidence and motif focus structure RMSD as the selection criterion. After 20 iterations, we identified a set of promising candidates, with the top design being a 142-residue protein that successfully incorporates the PD1 binding motif into a novel $\alpha/\beta$ fold, contrasting with the native $\beta$-only fold of PD1. The results of this study are promising. The designed protein showed high predicted folding stability (as indicated by pLDDT scores) and maintained the critical binding interface. Computational predictions of the designed PD-1 binder with PD-L1 showed high pLDDT and iPTM scores from AF-multimer models, with one example showing an iPTM score of 0.87 and a binding surface RMSD of only 0.72Å compared to the native structure.\\ \\
Compared to other methods that is capable of scaffolding\cite{watson2023novo,wang2022scaffolding,trippe2022diffusion}, our method generates the whole protein as a coherent entity without hard fixing any coordinates. we also demonstrates the ability to use flexible fitness selection criterion which previous models can not incorporate. This level of structural redesign is also challenging for traditional protein engineering methods that often rely on more conservative modifications.
This \textit{de novo} PD1 design showcases the power of our iterative framework in tackling complex protein engineering challenges, potentially paving the way for new approaches in cancer immunotherapy and beyond.

\subsection{Structure based Antibody design via conditional CDR inpainting}
Antibody engineering is a crucial field in biotechnology and therapeutic development, with applications ranging from cancer immunotherapy to the treatment of autoimmune diseases\cite{lu2020development}. The complementarity-determining regions (CDRs) of antibodies are primarily responsible for antigen recognition and binding, making them key targets for design and optimization. Traditional approaches to antibody design often rely on display technologies or in vivo methods, which can be time-consuming and limited in their exploration of sequence space. In recent years, computational approaches have shown promise in accelerating antibody design\cite{chungyoun2023ai}. However, many of these methods focus on sequence-based design or require extensive prior knowledge of antibody-antigen interactions. Structure-based approaches that can generate novel CDR structures while maintaining the overall antibody architecture are still quite challenging\cite{graves2020review}.
\begin{figure}[h]
    \centering
    \includegraphics[width=0.9\linewidth]{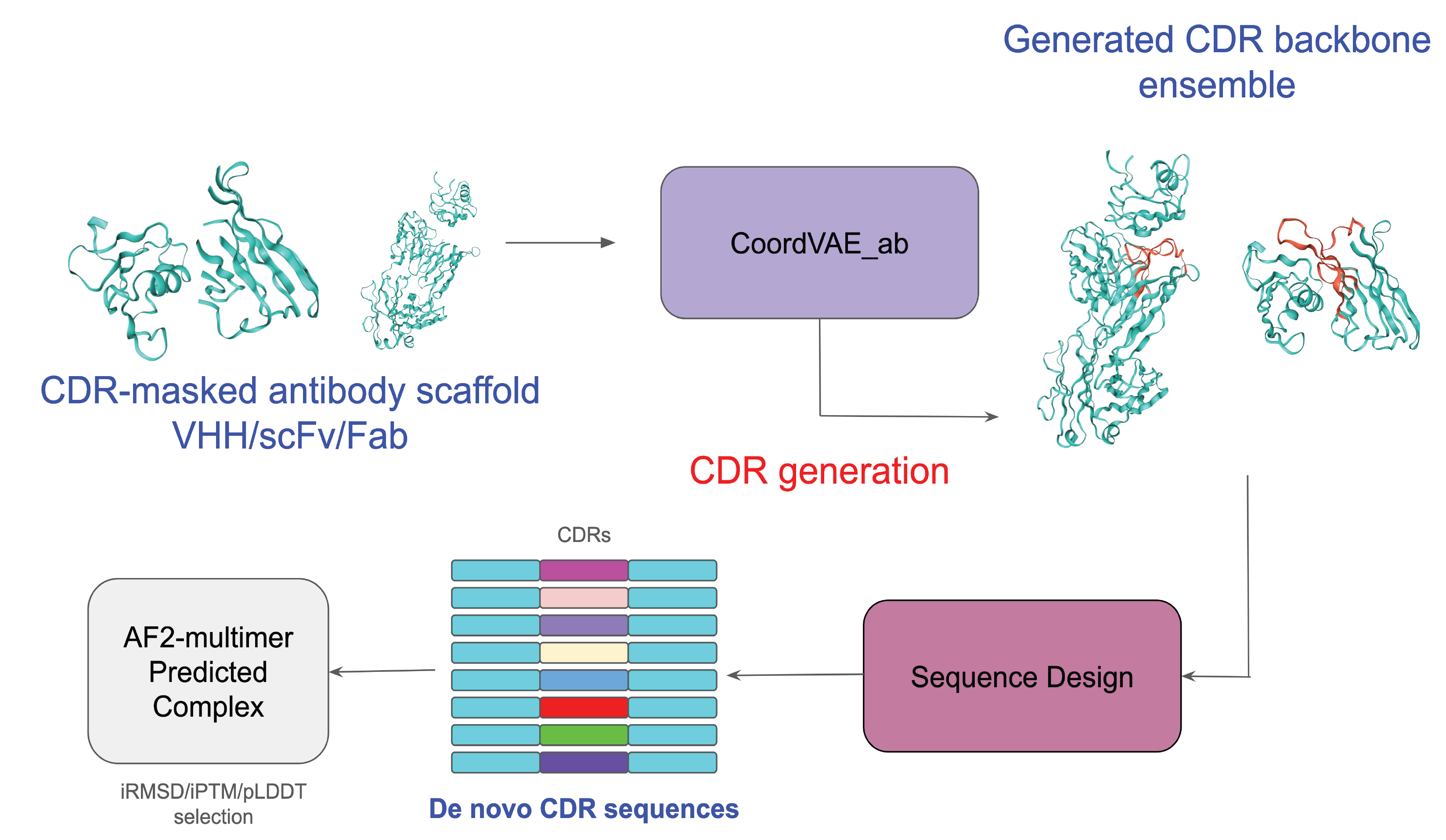}
    \caption{\textbf{Structure-based antibody design via CDR inpainting workflow}: The design process begins with a antibody-antigen pair, then the CDRs of interest are masked. CoordVAE-ab will then be used to inpaint the masked CDRs to generate a CDR backbone library. The structure library are then fed to a antibody optimized inverse folding model to recover the CDR sequence. A CDR library is then generated for downstream \textit{in silico} structure prediction tool for selection.}
    \label{fig:C2_R7}
\end{figure}
Building upon our success with the iterative design framework, we developed a specialized approach for \textit{de novo} CDR design through conditional structure inpainting. This method aims to generate novel CDR structures that are compatible with the antibody framework and optimized for target binding. To adapt our framework for CDR design, we made several key enhancements. First, we expanded our training dataset to include a comprehensive protein-protein interaction (PPI) dataset to capture diverse binding interfaces\cite{townshend2019end}, as well as a curated antibody dataset\cite{dunbar2014sabdab}. We then fine-tuned our CoordVAE model on this expanded dataset with a two-stage approach, where the first stage focused on general PPI interaction and a particular focus on antibody structures and CDR regions in the second stage. The model was modified to perform conditional inpainting, allowing us to generate new CDR structures while maintaining the rest of the antibody framework. Importantly, we incorporated antigen structure information into the design process to guide the generation of CDRs with potential binding affinity. These enhancements collectively enabled our framework to tackle the specific challenges of structure-based antibody design, leveraging both general protein interaction data and antibody-specific structural information.

Our computational design process (see \ref{fig:C2_R7}) for \textit{de novo} CDR structures begins with an antigen-antibody complex structure, where the CDRs intended for design are masked out of the antibody input. Our enhanced model, $CoordVAE_{ab}$, then generates a diverse ensemble of CDR backbone structures conditioned on this masked antibody structure and the antigen. For selected backbones from this ensemble, we perform inverse folding to obtain a CDR sequence repertoire. These designed sequences undergo rigorous \textit{in silico} folding evaluation to ensure structural integrity and stability. The most promising designs are evaluated for their potential binding interface with the antigen through computational complex structure prediction. This process is iterated, with top-scoring designs serving as seeds for subsequent rounds, allowing for progressive optimization of the CDR structures. This pipeline enables a comprehensive exploration of both CDR structure and sequence space, leveraging our $CoordVAE_{ab}$ model's ability to generate diverse yet structurally compatible CDRs while maintaining potential antigen interactions.

In Fig\ref{fig:C2_R7} we observed that the two stage fine tuning significantly improved our model's ability to accurately in paint the CDRs of interest. Our final fine tuned model $CoordVAE_{ab}$, demonstrates significant improvements across all three CDR regions compared to other models. It consistently achieves the highest TM-scores and pLDDT scores, indicating better overall structural similarity and prediction confidence. Notably, it also maintains the lowest RMSD values across all CDR types, suggesting more accurate reconstruction of the original structures. The performance gap is most significant for CDR3, which is often the most consequential CDR and the hardest to design. These results indicate that our model is exceptionally effective at generating accurate and structurally similar CDR regions, which is crucial for antibody design and engineering.

\begin{table}[h]
\centering
\resizebox{\textwidth}{!}{
\begin{tabular}{cccccccc}
\hline
 &
   &
  \multicolumn{2}{c}{CDR-H1} &
  \multicolumn{2}{c}{CDR-H2} &
  \multicolumn{2}{c}{CDR-H3} \\ \hline
\multicolumn{1}{c|}{Model} &
  \multicolumn{1}{c|}{Training Data} &
  lddt↑ &
  \multicolumn{1}{c|}{RMSD↓} &
  lddt↑ &
  \multicolumn{1}{c|}{RMSD↓} &
  lddt↑ &
  RMSD↓ \\ \hline
\multicolumn{1}{c|}{CoordVAE(10AA linear)} &
  \multicolumn{1}{c|}{} &
  0.587 &
  \multicolumn{1}{c|}{2.847} &
  0.588 &
  \multicolumn{1}{c|}{2.864} &
  0.498 &
  4.427 \\
\multicolumn{1}{c|}{CoordVAE(spatial)} &
  \multicolumn{1}{c|}{} &
  0.515 &
  \multicolumn{1}{c|}{3.944} &
  0.514 &
  \multicolumn{1}{c|}{3.914} &
  0.487 &
  4.350 \\
\multicolumn{1}{c|}{CoordVAE(Random)} &
  \multicolumn{1}{c|}{} &
  0.447 &
  \multicolumn{1}{c|}{4.216} &
  0.534 &
  \multicolumn{1}{c|}{4.152} &
  0.450 &
  5.185 \\
\multicolumn{1}{c|}{CoordVAE(mixed mask)} &
  \multicolumn{1}{c|}{\multirow{-4}{*}{CATH4.2}} &
  0.591 &
  \multicolumn{1}{c|}{2.903} &
  0.575 &
  \multicolumn{1}{c|}{3.223} &
  0.530 &
  3.889 \\ \hline
\multicolumn{1}{c|}{CoordVAE} &
  \multicolumn{1}{c|}{} &
  0.81 &
  \multicolumn{1}{c|}{1.55} &
  0.872 &
  \multicolumn{1}{c|}{1.00} &
  0.809 &
  1.55 \\
\multicolumn{1}{c|}{CoordVAE(Transfer)} &
  \multicolumn{1}{c|}{} &
  0.8 &
  \multicolumn{1}{c|}{0.81} &
   0.90 &
  \multicolumn{1}{c|}{0.85} &
   0.823 &
   1.35 \\
\multicolumn{1}{c|}{AR-GNN} &
  \multicolumn{1}{c|}{} &
  N/A &
  \multicolumn{1}{c|}{2.97} &
  N/A &
  \multicolumn{1}{c|}{2.27} &
  N/A &
  3.63 \\
\multicolumn{1}{c|}{RefineGNN} &
  \multicolumn{1}{c|}{\multirow{-4}{*}{SabDab}} &
  N/A &
  \multicolumn{1}{c|}{1.18} &
  N/A &
  \multicolumn{1}{c|}{0.87} &
  N/A &
  2.50 \\ \hline
  \multicolumn{1}{c|}{\textbf{CoordVAE-ab}} &
  \multicolumn{1}{c|}{{SabDab + CATH4.2 + DIPS}} &
  \textbf{0.95} &
  \multicolumn{1}{c|}{\textbf{0.83}} &
  \textbf{0.97} &
  \multicolumn{1}{c|}{\textbf{0.68}} &
  \textbf{0.96} &
  \textbf{0.74} \\ \hline
\end{tabular}}
\caption{Structure inpainting performance on the test monoclonal antibody dataset in CDR-H1(left), CDR-H2(middle), CDR-H3(right) across different models.}
\label{tab:table3}
\end{table}
To compare the new model with previous models, table\ref{tab:table3} demonstrates that CoordVAE-ab, trained on an extended dataset (SAbDab + CATH4.2 + DIPS) and using a two-stage fine-tuning approach, significantly outperformed all other models across all CDR regions. For CDR-H1, CoordVAE-ab achieved an LDDT score of 0.95 and an RMSD of 0.83, surpassing the next best model (CoordVAE). In CDR-H2, CoordVAE-ab's performance (LDDT: 0.97, RMSD: 0.68) exceeded the previous best (RefineGNN, RMSD: 0.87). Most notably, for the challenging CDR-H3 region, CoordVAE-ab (LDDT: 0.96, RMSD: 0.74) outperformed the next best model (CoordVAE(Transfer), LDDT: 0.823, RMSD: 1.35). These substantial improvements across all metrics and CDR regions underscore the effectiveness of CoordVAE-ab's extended dataset and refined training approach in antibody structure inpainting tasks and generate realistic and confident CDR backbone structures.

\begin{figure}[H]
    \centering
    \includegraphics[width=0.99\linewidth]{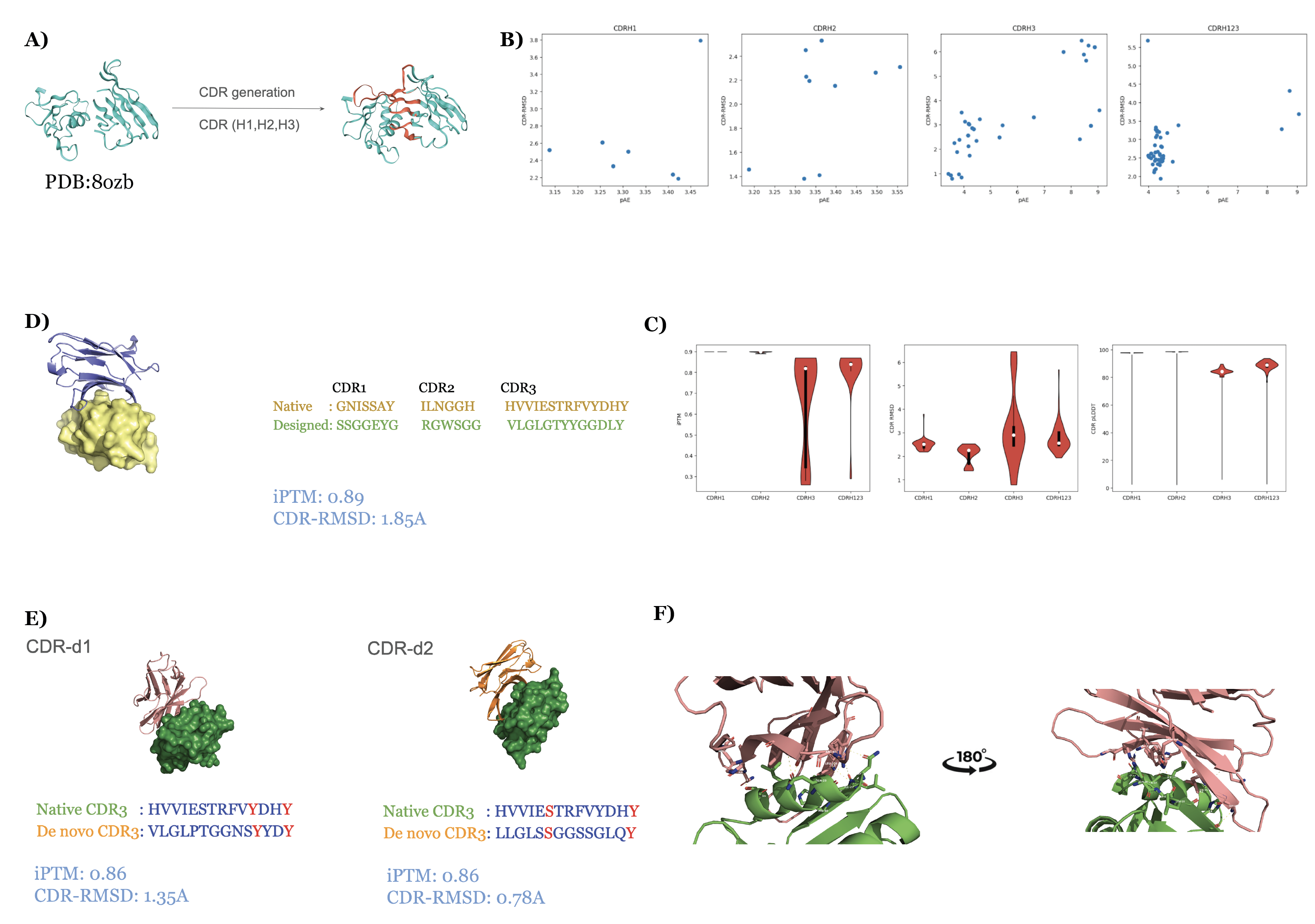}
    \caption{\textbf{Structure based conditional CDR design} A)A nanobody-antigen structure for design demonstration PDB:8ozb. B) CDR-RMSD vs. predicted pAE from computational structural models across different design modes. For CDR-H1 and CDR-H2 designs, most of the designed variants converged which is not surprising because the size of the design space is small. For CDR-H3 design, a fair share of designs pass the \textit{in silico} filter (pAE<10 and RMSD<2.5). Supricingly, we found in the case of full CDR design, the majority of the designs can pass the \textit{in silico} filter. C) Distribution of CDR-RMSD acreoss different design modes. The increased variant on CDR-H3 and full CDR design indicates that in this system the CDR-H3 is the most consequential CDR component. D) Example of full \textit{de novo} CDR design with predicted complex model. E) Example of CDR3 design with computationally predicted complex model. F) Predicted epitope-paratope binding surface of \textit{de novo} designed nanobody-antigen complex. }
    \label{fig:C2_R10}
\end{figure}

To evaluate the efficacy of our design framework, we applied it to a case study of nanobody-antigen complexes. We began by selecting a representative nanobody-antigen pair (Fig. \ref{fig:C2_R10}.A) and employed our design pipeline to varying combinations of CDRs. Our assessment of computational structure predictors' reliability in predicting the designed structures revealed promising results. For designs with predicted Aligned Error (pAE) < 10, the majority of \textit{de novo} CDR-H2 and CDR-H3 designs exhibited < 2Å RMSD compared to native CDR structures. In cases of full CDR design, most predicted structures demonstrated RMSD < 2.5Å (Fig. \ref{fig:C2_R10}.B).

Examination of the binding influence of different CDRs revealed that CDR-H3 has the most significant impact on predicted binding confidence, showing the highest variance in predicted RMSDs (Fig. \ref{fig:C2_R10}.C). This underscores the critical role of CDR-H3 in antigen recognition and binding. Examples of predicted complex structures for both full CDR design and CDR-H3-only design are illustrated in Fig. \ref{fig:C2_R10}.D and E, respectively. In both scenarios, the resulting \textit{de novo} CDRs yielded high-confidence predicted complex models, demonstrating the robustness of our design approach.

Our design framework shows strong performance in generating \textit{de novo} CDRs for nanobody-antigen complexes. The proposed designs demonstrate low predicted RMSD compared to native structures with highly diverse CDR sequences. Both full CDR and CDR-H3-only designs resulted in high-confidence predicted complexes model \textit{in silico}, with promising results on further design potentials. Compared to previous structure based antibody design methods, our approach offers several advantages. We fully leverage DL-based model in all design stages with both higher efficiency and design capacity. The method also demonstrates the ability to generate diverse yet structurally compatible CDRs while maintaining potential antigen interactions. We also offer high flexibility in the CDR design space by allowing various combination of CDR selections, this approach yield valuable insights into the systems from our computational evaluation.

\section{Discussion}
This chapter presented an innovative iterative design framework for \textit{de novo} protein engineering, demonstrating its versatility and effectiveness across four distinct applications: unconditional structure generation, functional protein design, motif-grounded scaffolding, and structure-based antibody design. By integrating deep learning-based structure generation with \textit{in silico} evaluation and refinement, our approach showcases the potential for computational methods to accelerate and enhance protein engineering efforts.

Our framework successfully generated novel protein folds with high \textit{in silico} folding confidence across a range of protein sizes. Many of these designs exhibited significant structural novelty when compared to existing structures in the Protein Data Bank, highlighting the framework's ability to explore beyond naturally occurring protein architectures. This capability opens new avenues for understanding protein structure-function relationships and potentially discovering proteins with novel functions.

For functional protein design, our iterative approach achieved remarkable success in optimizing DFHBI-activated $\beta$-barrel fluorescent proteins. The 100$\%$ success rate in experimental validation, with all 20 computationally designed candidates showing detectable DFHBI-activated fluorescence, underscores the robustness of our method. Moreover, the enhanced thermal stability and increased production yield compared to the original design reference demonstrate the practical benefits of our approach in creating improved functional proteins for biological applications.

The successful scaffolding of the key binding motif of PD1 onto a novel, potentially more stable protein structure further illustrates the framework's versatility. Computational predictions suggest that the designed protein maintains binding specificity for PD-L1 while offering a smaller, potentially more stable context. This achievement demonstrates the framework's ability to tackle complex protein engineering challenges with potential implications for cancer immunotherapy.

Our fourth application, structure-based antibody design via conditional CDR inpainting, represents a significant advancement in computational antibody engineering. The enhanced CoordVAE-ab model, trained on an extended dataset and using a two-stage fine-tuning approach, demonstrated superior performance in CDR structure prediction across all three CDR regions. This improvement was particularly notable for the challenging CDR-H3 region, which is crucial for antigen recognition and binding. The framework's ability to generate diverse, high-quality CDR designs while maintaining antigen binding specificity showcases its potential to accelerate and improve antibody design processes.

Despite these promising results, several limitations and areas for future work should be addressed. While we demonstrated experimental success with the fluorescent protein designs, the PD1 scaffolding and antibody design results are based primarily on computational predictions. Future work should include extensive experimental validation of these designs, including binding affinity measurements, structural characterization, and functional assays. The iterative nature of our framework, while powerful, can be computationally intensive. Future efforts should focus on optimizing the pipeline for increased efficiency and scalability, potentially leveraging distributed computing or more efficient sampling methods.

While our framework considers structural stability and function, future iterations could incorporate additional design objectives such as solubility, and immunogenicity. This would require the development and integration of reliable computational predictors for these properties. Our study covered a broad range of applications, but future work should explore the framework's applicability to an even wider range of protein functions, including enzymatic activity, small molecule binding, and allosteric regulation.

Developing a systematic way to incorporate experimental data into the iterative design process could further improve the success rate and efficiency of the framework. This could involve machine learning models trained on experimental outcomes to guide future design iterations. While state-of-the-art structure prediction tools were used, there is still room for improvement, especially for novel folds and antibody structures. Developing more accurate structure prediction methods tailored for \textit{de novo} designed proteins and antibodies could enhance the overall performance of the framework.

The current study focused primarily on single-domain proteins and antibody CDRs. Extending the framework to design multi-domain proteins, full antibody structures, or protein complexes could open up new possibilities for creating sophisticated molecular machines and therapeutic agents. While we demonstrated success in scaffolding existing functional motifs and designing CDRs, developing methods for \textit{de novo} functional site design within our framework could greatly expand its capabilities. This could enable the creation of entirely new protein functions not found in nature and novel antibody paratopes with enhanced binding properties.

In conclusion, our iterative design framework represents a significant advance in computational protein engineering, demonstrating success across four diverse and challenging design tasks. By addressing the limitations outlined above and expanding its capabilities, this approach has the potential to accelerate the development of novel proteins and antibodies for a wide range of biotechnological and therapeutic applications. Future work should focus on refining the computational methods, expanding the range of designable functions, improving antibody design capabilities, and integrating more closely with experimental validation to realize the full potential of this powerful approach to protein engineering.

\chapter{Concluding Remarks}
This study has introduced a novel structural generative model for proteins, termed CoordVAE, capable of directly modeling three-dimensional coordinates while addressing rotational and translational equivariance. The model demonstrates high-quality reconstruction of protein backbone structures across a wide range of sizes and folds, and generates diverse conformational ensembles. We also demonstrate both improved sequence diversity and design confidence at the same time using the generated structure ensemble compared to single backbone templates.

Building upon this generative model, we developed an iterative design framework that integrates structure generation, sequence design, and multi-faceted evaluation. This framework enables progressive optimization toward multiple design objectives, mimicking the process of directed evolution \textit{in silico}. The approach iteratively refines both structure and sequence, using computational predictions of stability and function to guide each design cycle.

The versatility and effectiveness of this iterative framework were demonstrated across four challenging applications: \textit{de novo} protein fold generation, functional protein optimization, motif-based scaffolding, and structure-guided antibody engineering. Notable successes include the generation of novel protein folds with high \textit{in silico} folding confidence, the optimization of DFHBI-activated $\beta$-barrel fluorescent proteins with $100\%$ experimental success rate and improved properties, the scaffolding of a PD1 binding motif onto a novel structure, and superior performance in antibody CDR structure prediction.

Looking ahead, several promising directions for future work emerge. Expanding experimental validation, particularly for the computationally designed binder scaffolds and antibodies, will be crucial. Incorporating additional design objectives such as solubility and immunogenicity could further improve the practical utility of designed proteins. Exploring applications to multi-domain proteins, protein-protein interfaces, and \textit{de novo} functional site design represent exciting frontiers for expanding the framework's capabilities.

Developing systematic ways to incorporate experimental feedback into the iterative process could enhance design success rates and efficiency. Improving the scalability of the pipeline will be important for tackling larger design challenges. Additionally, integrating the framework with recent advances in protein language models and emerging high-throughput experimental techniques could open new avenues for protein engineering.

In conclusion, this work represents a significant advancement in computational protein design, bridging structure generation, sequence design, and multi-faceted \textit{in silico} selection within a flexible iterative framework. By enabling the exploration of vast design spaces while maintaining a focus on both structural integrity and functional requirements, this approach provides a powerful new tool for expanding protein structure and function space. As the field of protein engineering continues to evolve, the integration and refinement of such computational methods promises to accelerate the development of novel proteins for a wide range of biotechnological and therapeutic applications.

\makebibliography
\nocite{*}


\end{document}